\documentclass[reprint, superscriptaddress, amsmath, amssymb, aps, nofootinbib, floatfix]{revtex4-2}

\usepackage{graphicx}
\usepackage{caption}
\usepackage{subcaption}
\usepackage{color}
\usepackage{dcolumn}
\usepackage{bm}

\begin{document} 

\newcommand{\bms}[1]{\textsf{\textbf{#1}}}
\newcommand{\bb}[1]{\mathbb{#1}}
\newcommand{\mc}[1]{\mathcal{#1}}
\newcommand{\od}[2]{\frac{d{#1}}{d{#2}}}
\newcommand{\pd}[2]{\frac{\partial{#1}}{\partial{#2}}}

\title{Causal Geodesics in Cylindrically Symmetric Vacuum \\ Spacetimes Using Hamilton-Jacobi Formalism}

\author{Ashiqul Islam Dip}
 \email{dip@cpbd.ac}
 \affiliation{Division of Research, Community of Physics, Kallyanpur, Dhaka 1207, Bangladesh}
 \affiliation{Department of Physics, North Carolina State University, Raleigh, NC 27607, USA}
 
\author{Nishat Anjum}
 \email{nishat@cpbd.ac}
 \affiliation{Division of Academics, Community of Physics, Kallyanpur, Dhaka 1207, Bangladesh}
 \affiliation{Department of Physics, University of Dhaka, Dhaka 1000, Bangladesh}
 
\author{Maruf Ahmed}
 \email{maruf@cpbd.ac}
 \affiliation{Division of Research, Community of Physics, Kallyanpur, Dhaka 1207, Bangladesh}
 \affiliation{Department of Physics, University of Dhaka, Dhaka 1000, Bangladesh}
 
\author{Iffat Zumarradah}
 \email{iffat@cpbd.ac}
 \affiliation{Division of Administration, Community of Physics, Kallyanpur, Dhaka 1207, Bangladesh}
 \affiliation{Department of Physics, University of Dhaka, Dhaka 1000, Bangladesh}

\date{\today}

\begin{abstract}

Cylindrically symmetric vacuum spacetimes are of immense interest in theoretical physics due to its connection to cosmic strings hypothesized in quantum field theory. In this article, we explore the properties of such a spacetime and provide the complete, exact solution by quadrature to the timelike and lightlike geodesics in it, using the Hamilton-Jacobi formalism. In addition, we compare several properties of massive particle trajectories in relativistic cylindrically symmetric vacuum spacetimes to its non-relativistic, Newtonian gravitational counterparts. On top of that, we have devised a classification scheme and utilized it to categorize the orbits of massive particles.

\begin{description}

\item[Keywords]{Levi-Civita Spacetime, Cosmic String, Geodesics, Hamilton-Jacobi Formalism}

\item[DOI]{00.0000/XXXX.000.000000}

\end{description}

\end{abstract}

\maketitle

\section{\label{section-1}Introduction}

Since its first realization by Kibble \cite{kibble1976} in 1976, cosmic string has intrigued many cosmologists and relativists. Cosmic strings are one-dimensional topological defects, formed in the early universe due to symmetry breaking phase transition \cite{kibble1976,nielsen1973,auclair2020,cui2009}. These primordial cosmic line defects, potentially form a string network at a cosmological scale \cite{albrecht1989,fernandez2020}. Their vibrations might produce gravitational waves, radiating a stochastic gravitational wave signature across the Universe \cite{auclair2020,figueroa2013,krauss1992,cui2020}. Even though it is likely that the cosmic strings exist, probability of finding one is very thin \cite{sazhina2013,sazhina2014}. However, in recent years, the optimism to detect cosmic strings by means of gravitational waves has risen due to the direct detection of gravitational waves by LIGO and Virgo collaborations \cite{abbott2016,abbott2017,abbott2020}. Furthermore, the recently released 12.5-year data set of NANOGrav collaboration might suggest the very first detection of cosmic strings and hence the existence of them, according to some authors \cite{blasi2021,ellis2021,chigusa2020}.

An infinitely-long, straight (idealized) cosmic string gives rise to a cylindrically symmetric spacetime around it \cite{hiscock1985,linet1986,gott1985}. However, the discovery of a cylindrically symmetric vacuum solution to the Einstein's equation dates back to 1919, and is called Levi-Civita spacetime, named after its discoverer \cite{dasilva1995,herrera2001}. Later in 1958, Marder proved that the Levi-Civita solution possesses two independent parameters \cite{marder1958,herrera2001}. Considering the mostly-positive sign convention, the metric tensor in the Levi-Civita spacetime can be written as \cite{kleber1980,harvey1999-8,bratek2022}

\begin{widetext}
\begin{equation}
    \bms{ds}^2= -\xi^{4\sigma} \bms{d}t\otimes\bms{d}t + (1-2\sigma)^2 P^2 \xi^{8\sigma^2-4\sigma} \bms{d}\xi\otimes\bms{d}\xi + P^2 \xi^{2-4\sigma} \bms{d}\phi\otimes\bms{d}\phi + \xi^{8\sigma^2-4\sigma} \bms{d}z\otimes\bms{d}z
\end{equation}
\end{widetext}

where, $-\infty<t<\infty$, $0\leq\xi<\infty$, $0\leq\phi\leq2\pi$, and $-\infty<z<\infty$, with $\phi=2\pi$ identified as $\phi=0$. The above-mentioned metric tensor contains two independent parameters, $\sigma$ and $P$, where $\sigma$ is dimensionless and $P$ has a dimension of length. For an infinitely-long linear mass distribution  \cite{bonnor1991}, the parameter $\sigma\in[0, 1/2]$ determines the linear mass density $\mu$ via the relation \cite{wang1997}

\begin{equation}
    \mu = \frac{\sigma}{1-2\sigma+4\sigma^2}
\end{equation}

and is known as the mass parameter. On the other hand, the parameter $P\in[1,\infty)$ carries the information about the `deficit angle' of the Levi-Civita spacetime \cite{harvey1999-8}. For $\sigma=0$ and $P=1$, the Levi-Civita spacetime reduces to the Minkowski spacetime; for $\sigma=0$ but $P\neq1$, it represents the spacetime around an idealized cosmic string, as one easily sees.

There have been multiple attempts to find the geodetic motion in some other cylindrically symmetric spacetimes. In the late 90s, Herrera and Santos derived a circular geodetic motion of massive particles in Lewis spacetime \cite{herrera1998}. Back in 2014, Brito \textit{et al.} found the solution to the geodesic equations in Linet-Tian spacetime for few special cases \cite{brito2014}. C\'{e}l\'{e}rier \textit{et al.} explored radial, axial, and circular geodetic motions in a cylindrically symmetric translating spacetime, in 2019 \cite{celerier2019}. In 2016, Hoseini \textit{et al.} came up with an analytic solution to the geodesic equation in a  static cylindrically symmetric conformal spacetime \cite{hoseini2016}. However, the search for a complete, exact solution to the geodetic motion of massive and massless particles moving in Levi-Civita spacetime continues to this day. It is because the differential geodesic equations it gives rise to are complicated enough to solve exactly in terms of elementary\footnote{Elementary functions are the ones that can be obtained in a finite number of algebraic operations, taking exponentials and logarithms. \cite{ritt1925}} functions. [see Appendix \ref{appendix-a}]. As a preview, FIG. 1 shows two typical three-dimensional trajectories of massive particles moving in Levi-Civita spacetime.

\begin{figure}[htb]
     \centering
     \begin{subfigure}{0.25\textwidth}
         \centering
         \includegraphics[width=\textwidth]{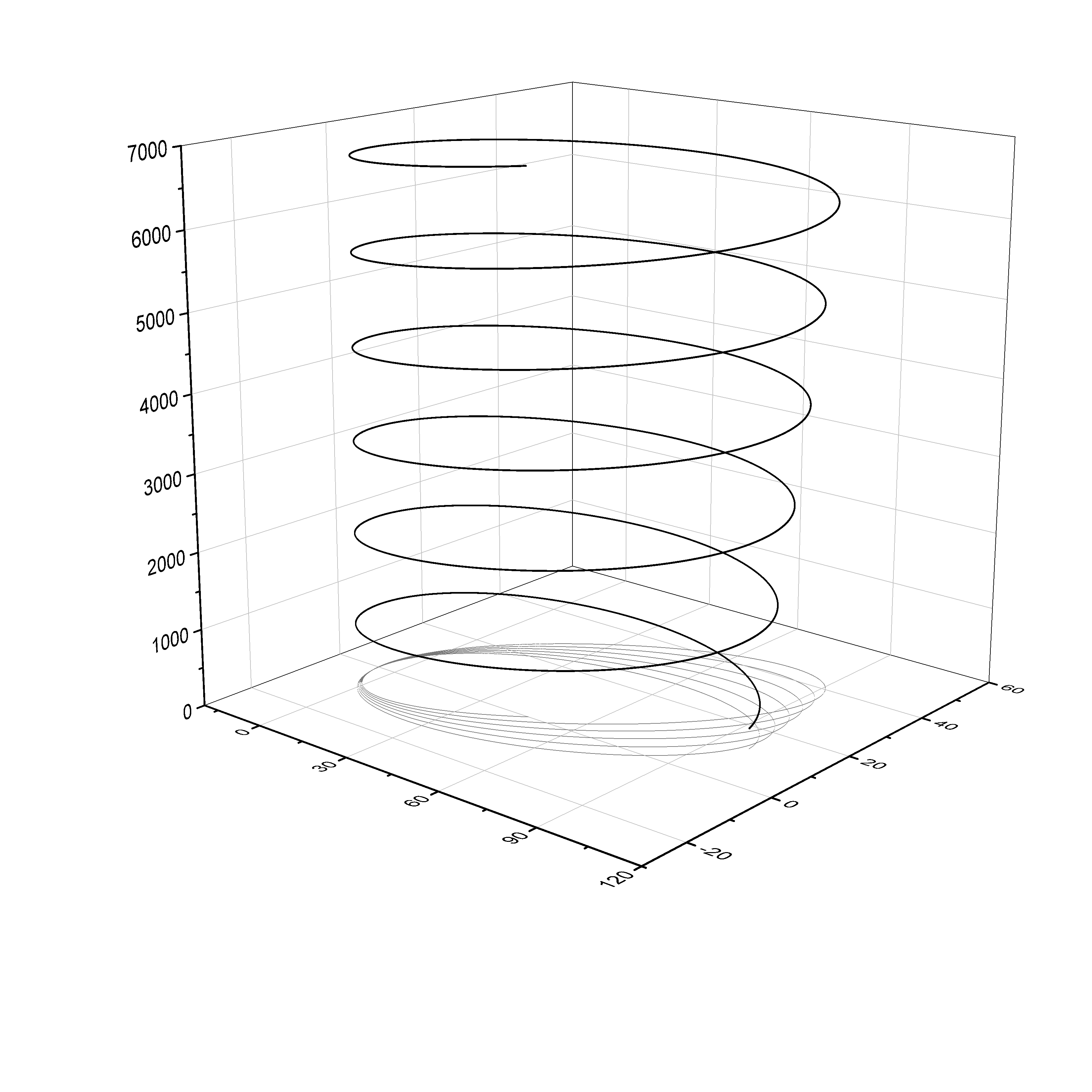}
         \caption{Trajectory in spacetime with $(\sigma,P)=(0.1,1)$}
     \end{subfigure}%
     \begin{subfigure}{0.25\textwidth}
         \centering
         \includegraphics[width=\textwidth]{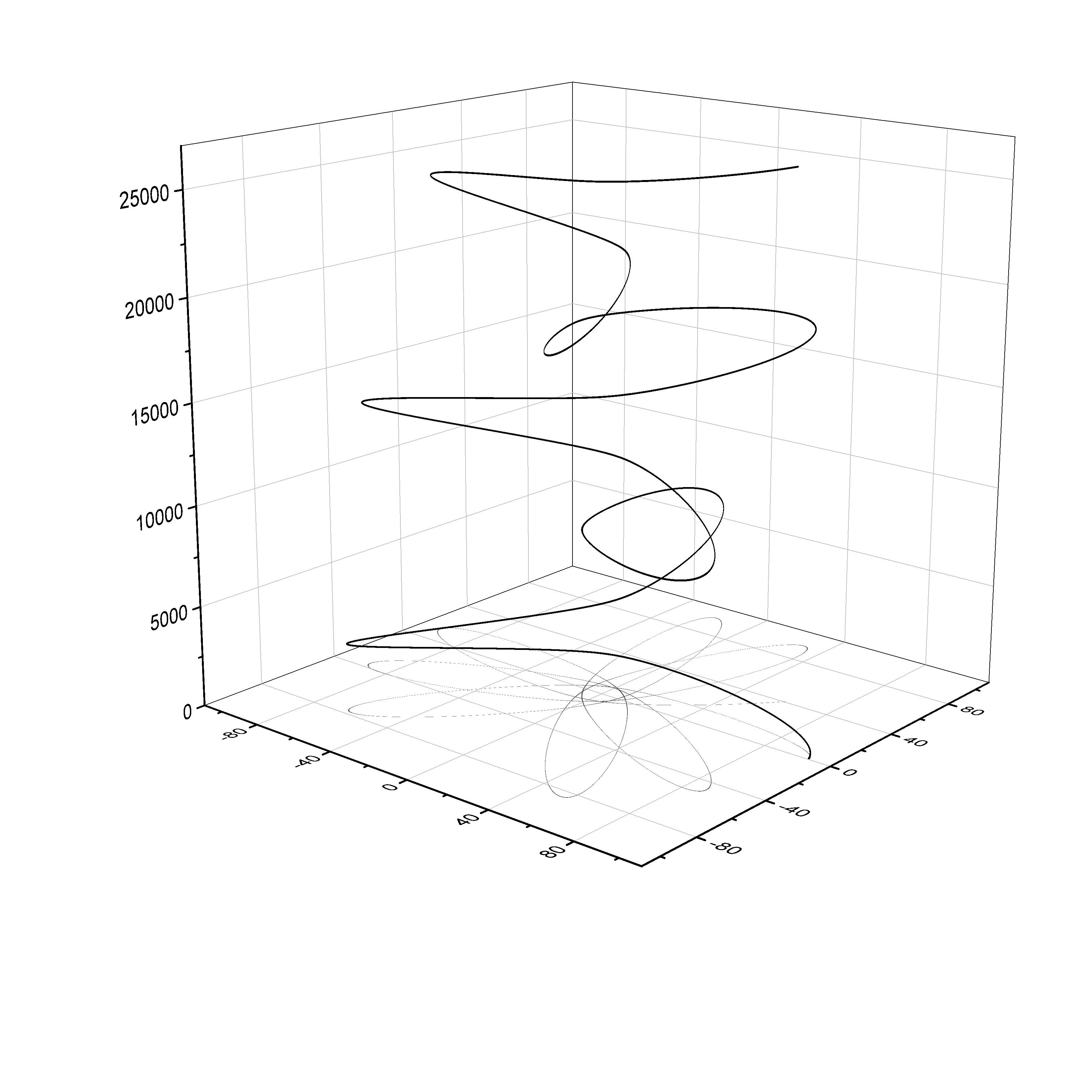}
         \caption{Trajectory in spacetime with $(\sigma,P)=(0.001,1)$}
     \end{subfigure}
     \caption{Trajectories of particles moving in Levi-Civita spacetime. The thick-black curves are the original trajectories, where the thin-grey curves are their projections on $z=0$ planes.}
\end{figure}

In this article, we present a complete, exact solution by quadrature to the differential geodesic equations in Levi-Civita spacetime using Hamilton-Jacobi formalism.\footnote{Complete solution refers to the solution to the equation of motion where the corresponding Hamilton's principal function contains the appropriate number of separation constants. \cite{goldstein2001-10} On the other hand, the exact solution means that the presented solution is neither an approximate nor a numerical one. And finally, solution by quadrature (also known as ``reduction to quadrature'') refers to the solution of a differential equation in terms of one or more definite integrals containing all the initial data explicitly. Many differential equations in physics are solved in this manner whenever a solution in terms of elementary functions is not available. \cite{kompaneyets1961-05, birkhoff1991-01}} The Hamilton-Jacobi formalism serves as a very powerful tool to solve Hamiltonian systems when the corresponding Hamilton-Jacobi equation is separable \cite{misner2017-25,deleon2017}. Interestingly enough, the Hamiltonian corresponding to the geodesic motion in Levi-Civita spacetime gives rise to a separable Hamilton-Jacobi equation, and consequently, the equation of motion is reduced to quadrature. The resulting integral solutions have also been expanded into infinite series. The resulting integral solution to the geodetic motion is presented for arbitrary parameters and initial conditions; the integrations were performed for few special cases, and the results were compared to the Newtonian counterparts. A proof of a theorem on the boundedness of the orbits for both Newtonian and relativistic cases is demonstrated. Moreover, a classification of the orbits is presented based on the orbit parameters.

\section{\label{section-2}Motion in Cylindrically Symmetric Spacetimes}

Though the metric tensor of Levi-Civita spacetime, as depicted in equation (1), represents the general cylindrically symmetric solution of Einstein's equation in vacuum, the geometric meaning of the coordinates used may not be immediately transparent. In particular, the radial coordinate $\xi$ does not represent the `circumferential radius'; that is to say, when the angular coordinate $\phi$ goes from $0$ to $2\pi$ keeping the other coordinates fixed, the proper distance covered is not equal to $2\pi\xi$. To resolve this issue, we make a coordinate transformation of the form

\begin{equation}
    \rho = P \xi^{1-2\sigma}
\end{equation}

keeping the other coordinates unchanged. Under the above transformation, the new metric tensor can be written as

\begin{widetext}
\begin{equation}
    \bms{ds}^2 = -\biggr(\frac{\rho}{P}\biggr)^{\frac{4\sigma}{1-2\sigma}} \bms{d}t\otimes\bms{d}t + \biggr(\frac{\rho}{P}\biggr)^{\frac{8\sigma^2}{1-2\sigma}} \bms{d}\rho\otimes\bms{d}\rho + \rho^2 \bms{d}\phi\otimes\bms{d}\phi + \biggr(\frac{\rho}{P}\biggr)^{-4\sigma}\bms{d}z\otimes\bms{d}z
\end{equation}
\end{widetext}

As the angular coordinate runs from $0$ to $2\pi$, the radial coordinate, $\rho$, in equation (4) can be interpreted as the `circumferential radius'. As it will turn out later, this choice of radial coordinate in the weak gravitational field limit, facilitates establishing a correspondence between the relativistic cylindrically symmetric vacuum spacetime and its Newtonian gravitational counterpart written in cylindrical polar coordinate. 

\subsection{\label{subsection-2a}Motion of a Massive Particle}

To find the solution to the geodetic motion of a point particle moving in a curved spacetime, one needs to separate the equation obtained upon substituting equation (B7) in (B5). The substitution results in

\begin{equation}
    1 + g^{\mu\nu}\pd{S}{q^\mu}\pd{S}{q^\nu} = 0
\end{equation}

Substitution of the metric tensor from equation (4) yields the Hamilton-Jacobi equation for a massive particle moving in a cylindrically symmetric vacuum spacetime.

\begin{eqnarray}
    1 & & - \biggr(\frac{\rho}{P}\biggr)^{-\frac{4\sigma}{1-2\sigma}} \biggr( \pd{S}{t}\biggr)^2 + \biggr(\frac{\rho}{P}\biggr)^{-\frac{8\sigma^2}{1-2\sigma}} \biggr( \pd{S}{\rho}\biggr)^2 
    \nonumber\\
    & & + \rho^{-2} \biggr( \pd{S}{\phi}\biggr)^2 + \biggr(\frac{\rho}{P}\biggr)^{4\sigma}\biggr( \pd{S}{z}\biggr)^2 = 0
\end{eqnarray}

The equation (6) appears to be a separable one as the coefficients are functions of $\rho$ only. To separate the equations, we try an ansatz of the form $S(t,\rho,\phi,z) = S_t(t) + S_\rho(\rho) + S_\phi(\phi) + S_z(z)$. Upon separation, we integrate the separated equations and re-construct the generation function as

\begin{widetext}
\begin{equation}
    S = -\Delta t + A\phi + B z + \int \biggr\{{\Delta^2\biggr(\frac{\rho}{P}\biggr)^{-4\sigma} - A^2\rho^{-2}\biggr(\frac{\rho}{P}\biggr)^{\frac{8\sigma^2}{1-2\sigma}} - B^2 \biggr(\frac{\rho}{P}\biggr)^{\frac{4\sigma}{1-2\sigma}} - \biggr(\frac{\rho}{P}\biggr)^{\frac{8\sigma^2}{1-2\sigma}}}\biggr\}^{\frac{1}{2}} d\rho
\end{equation}
\end{widetext}

where $\Delta$, $A$, and $B$ are separation constants and serve as the first half of the constants of motion. It is straightforward to see that $A$ is the conserved momentum conjugate to the azimuthal coordinate, whereas, $B$ can be interpreted as the conserved momentum along the $z$ coordinate, and finally, $\Delta$ is the total energy of the moving particle.

The motion of a particle in four-dimensional spacetime contains six constants of motion, three of which we have already obtained and interpreted for their physical significance. The other constants of motion are contained in the final solution of the geodesic motion obtained through applying the equation (B8). All of these constants are related to the initial data, namely the initial momenta and the initial coordinates. In this article, we have chosen the initial coordinates, $\rho_0$, $\phi_0$, and $z_0$, the initial momenta $A$ and $B$ and the total energy $\Delta$ to appear explicitly in the final integral solution. 

\begin{widetext}
\begin{subequations}
\begin{eqnarray}
    t & = & \Delta \int_{\rho_0}^\rho \biggr(\frac{\rho}{P} \biggr)^{-4\sigma} \biggr\{\Delta^2 \biggr(\frac{\rho}{P} \biggr)^{-4\sigma} -A^2\rho^{-2} \biggr(\frac{\rho}{P} \biggr)^{\frac{8\sigma^2}{1-2\sigma}}-B^2\biggr(\frac{\rho}{P} \biggr)^{\frac{4\sigma}{1-2\sigma}} - \biggr(\frac{\rho}{P} \biggr)^{\frac{8\sigma^2}{1-2\sigma}}\biggr\}^{-\frac{1}{2}} d\rho
    \\
    \phi & = & \phi_0 + A \int_{\rho_0}^\rho \rho^{-2}\biggr(\frac{\rho}{P} \biggr)^{\frac{8\sigma^2}{1-2\sigma}} \biggr\{\Delta^2 \biggr(\frac{\rho}{P} \biggr)^{-4\sigma} -A^2\rho^{-2} \biggr(\frac{\rho}{P} \biggr)^{\frac{8\sigma^2}{1-2\sigma}}-B^2\biggr(\frac{\rho}{P} \biggr)^{\frac{4\sigma}{1-2\sigma}} - \biggr(\frac{\rho}{P} \biggr)^{\frac{8\sigma^2}{1-2\sigma}}\biggr\}^{-\frac{1}{2}} d\rho
    \\
    z & = & z_0 + B \int_{\rho_0}^\rho \biggr(\frac{\rho}{P} \biggr)^{\frac{4\sigma}{1-2\sigma}} \biggr\{\Delta^2 \biggr(\frac{\rho}{P} \biggr)^{-4\sigma} -A^2\rho^{-2} \biggr(\frac{\rho}{P} \biggr)^{\frac{8\sigma^2}{1-2\sigma}}-B^2\biggr(\frac{\rho}{P} \biggr)^{\frac{4\sigma}{1-2\sigma}} - \biggr(\frac{\rho}{P} \biggr)^{\frac{8\sigma^2}{1-2\sigma}}\biggr\}^{-\frac{1}{2}} d\rho
\end{eqnarray}
\end{subequations}
\end{widetext}

However, one might choose to write the solution by quadrature above in terms of $\Gamma$, the initial value of the momentum conjugate to the radial coordinate, using the relation

\begin{eqnarray}
    \Gamma =  \biggr\{\Delta^2\biggr(\frac{\rho_0}{P}\biggr)^{-4\sigma} & - & A^2\rho_0^{-2}\biggr(\frac{\rho_0}{P}\biggr)^{\frac{8\sigma^2}{1-2\sigma}}
    \nonumber\\
    & - & B^2\biggr(\frac{\rho_0}{P}\biggr)^{\frac{4\sigma}{1-2\sigma}}
    -\biggr(\frac{\rho_0}{P}\biggr)^{\frac{8\sigma^2}{1-2\sigma}}
    \biggr\}^\frac{1}{2}
\end{eqnarray}

instead of $\Delta$.

In that case, $\Delta$ should be written in terms of $\Gamma$ and the other constants of motion.

The solution by quadrature given in equation-array (8) includes non-elementary integrals, which can be integrated numerically for the general cases if needed. Nevertheless, for the case of $A=0$ along with $B=0$, the angular and the axial equations reduce to $\phi=\phi_0$ and $z=z_0$ respectively, implying a purely radial motion. The only non-trivial integral is that of $t$, which can be performed using Euler-Gauss hypergeometric function ${}_2F_1$ {\cite{aomoto2011-01}}, giving $t$ as a function of $\rho$ in a closed form.\footnote{Using the identity $\int \frac{1}{\sqrt{1-x^n}} dx = x \cdot  {}_2F_1(\frac{1}{2},\frac{1}{n};1+\frac{1}{n};x^n)$.}

\begin{widetext}
\begin{equation}
    t = \frac{P}{1-2\sigma}\biggr[\biggr(\frac{\rho}{P}\biggr)^{1-2\sigma} {}_2F_1\biggr(\frac{1}{2}, \frac{1-4\sigma+4\sigma^2}{4\sigma}; \frac{1+4\sigma^2}{4\sigma}; \frac{1}{\Delta^2} \biggr(\frac{\rho}P\biggr)^\frac{4\sigma}{1-2\sigma}\biggr)\biggr]_{\rho_0}^{\rho}
\end{equation}
\end{widetext}

In addition, it can easily be checked that, in the limit $\sigma \to 0$ and $P = 1$, the equation-array (8) boils down to the geodesic equations of a massive particle moving in a flat spacetime written in cylindrical polar coordinates. Moreover, in the case of $\sigma \to 0$ but $P \neq 1$, the geodetic motion reduces to the motion on a `hypercone', i.e., the three-dimensional analog of a conical surface.

\subsection{\label{subsection-2b}Null Geodesics}

Not unlike a massive particle, a massless particle parallel transports its four-momentum along with its worldline. This, upon integration, leads to the constraint equation (B6), which determines the trajectory of the massless particle. The set of all possible trajectories through a point constitute a hypersurface in four-dimensional spacetime, known as the light-cone. The light-cone imposes a physical limit on the trajectories of all particles such that no causal geodesic can reside outside of it. Since a massive cosmic string might be able to strongly bend lights that are sufficiently intense to observe, obtaining the photon trajectories is therefore of particular interest. The same procedure used for the massive particles (Section IIA), when deployed for the massless ones, leads to a generating function for the Hamilton-Jacobi equation, giving rise to the massless particle trajectories. Like the massive particle case, the motion of a massless particle includes six constants of integration, three of which are $\Delta$, $A$, and $B$, and the other three are the initial coordinates $\rho_0$, $\phi_0$, and $z_0$.

\begin{widetext}
\begin{subequations}
\begin{eqnarray}
    t & = & \Delta \int_{\rho_0}^\rho \biggr(\frac{\rho}{P} \biggr)^{-4\sigma} \biggr\{\Delta^2 \biggr(\frac{\rho}{P} \biggr)^{-4\sigma} -A^2\rho^{-2} \biggr(\frac{\rho}{P} \biggr)^{\frac{8\sigma^2}{1-2\sigma}}-B^2\biggr(\frac{\rho}{P} \biggr)^{\frac{4\sigma}{1-2\sigma}} \biggr\}^{-\frac{1}{2}} d\rho
    \\
    \phi & = & \phi_0 + A \int_{\rho_0}^\rho \rho^{-2}\biggr(\frac{\rho}{P} \biggr)^{\frac{8\sigma^2}{1-2\sigma}} \biggr\{\Delta^2 \biggr(\frac{\rho}{P} \biggr)^{-4\sigma} -A^2\rho^{-2} \biggr(\frac{\rho}{P} \biggr)^{\frac{8\sigma^2}{1-2\sigma}}-B^2\biggr(\frac{\rho}{P} \biggr)^{\frac{4\sigma}{1-2\sigma}} \biggr\}^{-\frac{1}{2}} d\rho
    \\
    z & = & z_0 + B \int_{\rho_0}^\rho \biggr(\frac{\rho}{P} \biggr)^{\frac{4\sigma}{1-2\sigma}} \biggr\{\Delta^2 \biggr(\frac{\rho}{P} \biggr)^{-4\sigma} -A^2\rho^{-2} \biggr(\frac{\rho}{P} \biggr)^{\frac{8\sigma^2}{1-2\sigma}}-B^2\biggr(\frac{\rho}{P} \biggr)^{\frac{4\sigma}{1-2\sigma}} \biggr\}^{-\frac{1}{2}} d\rho
\end{eqnarray}
\end{subequations}
\end{widetext}

Similar to the massive particle case, the integral solution given in equation-array (11) involves non-elementary integrals and numerical methods can be employed to perform the integrals. However, there exist a couple of special yet physically interesting cases, where the integrals can be carried out using the Euler-Gauss hypergeometric function ${}_2F_1$, introduced in Section IIA. One of the special cases, where the integrations can be performed, is $B=0$. It involves a motion confined to $z=z_0$ plane, which is of significant importance due to its potential for aiding direct observation of cosmic strings. On performing the integrals, $t$ and $\phi$ can be expressed as functions of $\rho$ in closed forms through the Euler-Gauss hypergeometric function.

\begin{widetext}
\begin{subequations}
\begin{eqnarray}
    t &=& \frac{P}{1-2\sigma}\biggr[\biggr(\frac{\rho}{P}\biggr)^{1-2\sigma} {}_2F_1\biggr(\frac{1}{2}, -\frac{1-4\sigma+4\sigma^2}{2-8\sigma}; \frac{1-4\sigma-4\sigma^2}{2-8\sigma}; \frac{A^2}{\Delta^2P^2} \biggr(\frac{\rho}{P}\biggr)^{-\frac{2-8\sigma}{1-2\sigma}}\biggr)\biggr]_{\rho_0}^{\rho}
    \\
    \phi &=&  \phi_0 - \frac{(1-2\sigma)A}{(1-4\sigma-4\sigma^2)P\Delta}\biggr[\biggr(\frac{\rho}{P}\biggr)^{-\frac{1-4\sigma-4\sigma^2}{1-2\sigma}} {}_2F_1\biggr(\frac{1}{2},\frac{1-4\sigma-4\sigma^2}{2-8\sigma}; \frac{3-12\sigma-4\sigma^2}{2-8\sigma}; \frac{A^2}{\Delta^2P^2} \biggr(\frac{\rho}{P}\biggr)^{-\frac{2-8\sigma}{1-2\sigma}}  \biggr)\biggr]_{\rho_0}^{\rho}
\end{eqnarray}
\end{subequations}
\end{widetext}

Another important case resulting in a solution in terms of a well-known function is that of zero angular momentum with a non-zero axial momentum, leading to a motion of the massless particle confined to the plane $\phi=\phi_0$ entirely. This solution provides a graphic connection between the radial and the axial motion of the particle, which, in turn, provides insights into the gravitational strength of the cosmic string. This might allow us to measure the gravitational strength of an observable cosmic string by observing the accreting photons in its local vicinity moving through gas clouds. Similar to the case of zero axial momentum, solution to this one can also be written in a closed-form using the Euler-Gauss hypergeometric function. As one might expect, both solutions to the special cases $A=0$, given in (12), and $B=0$, given below, reduce to the same expression when these conditions hold simultaneously.\footnote{Due to the fact that ${}_2F_1(p,q;r;0)=1$ for all $p,q,r\in \bb{N}$.}

\begin{widetext}
\begin{subequations}
\begin{eqnarray}
    t &=& \frac{P}{1-2\sigma}\biggr[\biggr(\frac{\rho}{P}\biggr)^{1-2\sigma} {}_2F_1\biggr(\frac{1}{2},\frac{1-4\sigma+4\sigma^2}{8\sigma-8\sigma^2};\frac{1+4\sigma-4\sigma^2}{8\sigma-8\sigma^2};\frac{B^2}{\Delta^2}\biggr(\frac{\rho}{P}\biggr)^\frac{8\sigma-8\sigma^2}{1-2\sigma}\biggr)\biggr ]_{\rho_0}^{\rho}
    \\
    z &=&  z_0 +\frac{(1-2\sigma)PB}{(1+4\sigma-4\sigma^2)\Delta}\biggr[\biggr(\frac{\rho}{P}\biggr)^\frac{1+4\sigma-4\sigma^2}{1-2\sigma} {}_2F_1\biggr(\frac{1}{2},\frac{1+4\sigma-4\sigma^2}{8\sigma-8\sigma^2};\frac{1+12\sigma-12\sigma^2}{8\sigma-8\sigma^2};\frac{B^2}{\Delta^2}\biggr(\frac{\rho}{P}\biggr)^\frac{8\sigma-8\sigma^2}{1-2\sigma}\biggr)\biggr]_{\rho_0}^{\rho}
\end{eqnarray}
\end{subequations}
\end{widetext}

\subsection{\label{subsection-2C}Series Solution to the Geodesic Equations}

The complete, exact solution by quadrature to the massive particle geodesics presented in equation (8) and to the massless particle geodesics presented in equation (11) is enough to extract all the necessary information about the motion in Levi-Civita spacetime. However, one might want to explore the nature of the functions presented in equations (8) and (11) to a greater extent. It will allow to understand not only the motion in Levi-Civita spacetime but also its connection to similar motions in other settings.

In order to explore the nature of the functions in equations (8) and (11), we need to expand the functions into a series. To achieve that goal, we have reduced the integrals in equation (8) and (11) into a common format and defined it as a function $\kappa(a, b; p, q, r; x)$ as shown below.

\begin{widetext}
\begin{eqnarray}
    \kappa(a,\ b;\ p,\ q,\ r;\ x)& = & \int_0^x \Big(1 - x^p - ax^q - bx^r\Big)^{-\frac{1}{2}} dx
    \nonumber\\
    & = & \sum_{k=0}^\infty \sum_{k_1, k_2, k_3} \binom{-\frac{1}{2}} {k} \frac{(-1)^k\  \delta^k_{\ k_1 + k_2 + k_3}\ a^{k_2}\  b^{k_3}\ k!} {(1 + pk_1 + qk_2 + rk_3)\ k_1!\ k_2!\ k_3!} x^{1 + pk_1 + qk_2 + rk_3}
\end{eqnarray}
\end{widetext}

In equation (14), we have expanded the integrand using multinomial expansion and we have performed the integration term-wise obtaining a series expansion for the function $\kappa(a, b; p, q, r; x)$. In the expansion, $k, k_1, k_2$ and $k_3$ are all positive integers and range from $0$ to $\infty$. Furthermore, $\delta^{k}_{k_1+k_2+k_3}$ ensures that the relation $k=k_1+k_2+k_3$ is satisfied for each of those terms in the sum. Using the integral definition and the series expansion of $\kappa(a, b; p, q, r; x)$, one can find relevant properties of the function.

The solutions by quadrature for the massive particle geodesics in equation-array (8) can easily be expressed using the definition of the $\kappa(a, b; p, q, r; x)$ function, as we have shown below. It is worth mentioning that, both the integral expressions and the series expansions are equivalent when pursuing numerical methods to generate trajectories, and ultimately choosing one of these methods boils down to a matter of preference.

\begin{widetext}
\begin{subequations}
\begin{eqnarray}
    t & = & \frac{P}{1-2\sigma} 
    \biggr(\frac{A}{\Delta P}\biggr)^{\frac{1-4\sigma+4\sigma^2}{1-4\sigma}} \biggr[ \kappa\biggr(\frac{B^2}{\Delta^2}\biggr(\frac{A}{\Delta P}\biggr)^{\frac{8\sigma-8\sigma^2}{1-4\sigma}}, \frac{1}{\Delta^2}\biggr(\frac{A}{\Delta P}\biggr)^{\frac{4\sigma}{1-4\sigma}};\cdots
    \nonumber\\
    & & \hspace{25mm} -\frac{2-8\sigma}{1-4\sigma+4\sigma^2},\frac{8\sigma-8\sigma^2}{1-4\sigma+4\sigma^2},\frac{4\sigma}{1-4\sigma+4\sigma^2}; \biggr(\frac{A}{\Delta P} \biggr)^{-\frac{1-4\sigma+4\sigma^2}{1-4\sigma}} \biggr( \frac{\rho}{P} \biggr)^{1-2\sigma} \biggr)\biggr]_{\rho_0}^{\rho}
    \\
    \phi & = & \phi_0 - \frac{PB(1-2\sigma)}{\Delta(1-4\sigma -4\sigma^2)}\biggr({\frac{A}{P\Delta}}\biggr)^{\frac{1+4\sigma-4\sigma^2}{1-4\sigma}}  \biggr[ \kappa\biggr(\frac{B^2}{\Delta^2}\biggr(\frac{A}{\Delta P}\biggr)^{\frac{8\sigma-8\sigma^2}{1-4\sigma}}, \frac{1}{\Delta^2}\biggr(\frac{A}{\Delta P}\biggr)^{\frac{4\sigma}{1-4\sigma}};\cdots 
    \nonumber\\
    & &\hspace{15mm} -\frac{2-8\sigma}{1-4\sigma-4\sigma^2},-\frac{8\sigma-8\sigma^2}{1-4\sigma-4\sigma^2},-\frac{4\sigma}{1-4\sigma-4\sigma^2}; \biggr(\frac{A}{\Delta P}\biggr)^{\frac{1-4\sigma-4\sigma^2}{1 - 4\sigma}}\biggr(\frac{\rho}{P}\biggr)^{-\frac{1-4\sigma-4\sigma^2}{1 - 2\sigma}} \biggr)\biggr]_{\rho_0}^{\rho}
    \\ 
    z & = & z_0 + \frac{PB(1-2\sigma)}{\Delta(1+4\sigma -4\sigma^2)}\biggr({\frac{A}{P\Delta}}\biggr)^{\frac{1+4\sigma-4\sigma^2}{1-4\sigma}} \biggr[ \kappa\biggr(\frac{B^2}{\Delta^2}\biggr(\frac{A}{\Delta P}\biggr)^{\frac{8\sigma-8\sigma^2}{1-4\sigma}}, \frac{1}{\Delta^2}\biggr(\frac{A}{\Delta P}\biggr)^{\frac{4\sigma}{1-4\sigma}};\cdots
    \nonumber\\
    & &\hspace{20mm} -\frac{2-8\sigma}{1+4\sigma-4\sigma^2},\frac{8\sigma-8\sigma^2}{1+4\sigma-4\sigma^2},\frac{4\sigma}{1+4\sigma-4\sigma^2}; \biggr(\frac{A}{\Delta P}\biggr)^{-\frac{1+4\sigma-4\sigma^2}{1 - 4\sigma}}\biggr(\frac{\rho}{P}\biggr)^{\frac{1+4\sigma-4\sigma^2}{1 - 2\sigma}} \biggr)\biggr]_{\rho_0}^{\rho}
\end{eqnarray} 
\end{subequations}
\end{widetext}

On the other hand, to find the series expansion of the solution for the massless particle in equation-array (11), one can use the same set of equations as for the massive particles in equation-array (15) except substituting $b=0$ and $r=0$ in $\kappa(a, b; p, q, r; x)$ appearing in the equations for $t$, $\phi$, and $z$.

\subsection{\label{subsection-2D}Weak Gravitational Field Limit}

Another interesting scenario one might want to explore is that when the mass parameter $\sigma$ is sufficiently small, leading to a weak gravitational field. For all causal geodesics having $\sigma \ll 1$, the mass density $\mu$ in the equation (2) can be approximated as $\mu\approx\frac{\sigma}{1-2\sigma}$, since the parameter $\sigma$ is bounded by $1/2$ from the above in such cases. As a result, the Newtonian gravitational potential $\Phi$ for an infinitely-long linear mass distribution can be written as

\begin{equation}
    \Phi = \frac{2\sigma}{1-2\sigma}\ln{\biggr(\frac{\rho}{R}\biggr)}
\end{equation}

where the linear mass density $\mu$ is approximated as mentioned above and $\rho=R$ is the surface on which the gravitational potential is taken to be zero, for an arbitrarily chosen $R$.

On the other hand, the metric tensor itself can be interpreted as the potential for the general relativistic case. To see a connection between the Newtonian gravitational potential and the metric tensor in general relativity, one might expand $\bms{ds}^2(\bms{e}_t,\bms{e}_t)$ considering the weak gravitational field limit, where $\bms{e}_t$ is the coordinate basis vector along the coordinate $t$. Upon expansion, $\bms{ds}^2(\bms{e}_t,\bms{e}_t)\approx -(1+2\Phi)$ is expected, ignoring the second and higher order corrections \cite{weinberg1972-03}. Considering the metric tensor in equation (4), we write

\begin{eqnarray}
    \bms{ds}^2(\bms{e}_t,\bms{e}_t) = -\biggr[1 &+& \frac{4\sigma}{1-2\sigma}\ln{\biggr(\frac{\rho}{P}\biggr)} 
    \nonumber\\
    &+& \frac{1}{2}\biggr\{\frac{4\sigma}{1-2\sigma}\ln{\biggr(\frac{\rho}{P}\biggr)}\biggr\}^2+\cdots\biggr]
\end{eqnarray}

The quadratic and higher order terms can be ignored as $\frac{4\sigma}{1-2\sigma} \approx 4\sigma$ in the limit $\sigma \to 0$. Indeed, it can be seen that, $\bms{ds}^2(\bms{e}_t,\bms{e}_t)\approx -(1+2\Phi)$ if we choose $R=P$. This correspondence allows us to interpret $P$, the parameter related to the `deficit angle' of the Levi-Civita spacetime, to be the radius of the cylindrical surface on which the Newtonian gravitational potential is taken to be zero.

As demonstrated above, for small $\sigma$, there exists a correspondence between the Levi-Civita spacetime and the Newtonian gravitational field generated by an infinitely-long straight mass distribution. Consequently, the respective motions of a massive particle coincide as well, in the appropriate limit, i.e., $\sigma\ll 1$.

\section{\label{section-3}Boundedness of Orbits}

Long before the birth of general relativity, Newton found the exact solution to the motion of a bounded point particle moving in spherically symmetric Newtonian gravity to be an ellipse, consistent with the Kepler's empirical laws of planetary motion. A key insight for all the bounded orbits, including the ones in Schwarzschild geometry, is that the radial coordinates describing the motions are always bounded between a minimum and a maximum value. Extending similar investigation for the geodetic motion in Levi-Civita spacetime, we have seen that the radial coordinates, $\rho$, describing the motion is bounded between a lower limit $\underline{\rho}$ and an upper limit $\bar{\rho}$ as well. To show this, we rewrite the equation (6) in terms of conserved momenta and set $B=0$ to get the equation on $z=z_0$ plane. In addition, we set the radial velocity to zero and obtain

\begin{equation}
    \rho^2 - \Delta^2\rho^2\biggr(\frac{\rho}{P}\biggr)^{-\frac{4\sigma}{1-2\sigma}}+A^2=0
\end{equation}

which determines the extrema of the radial coordinate. This is a transcendental equation which can be solved either by graphically or numerically. To graphically see that the radial coordinate $\rho$ is bounded between a lower limit and an upper limit, we plot the function $f$, defined by $f(\rho; A, \Delta, \sigma, P) := \rho^2 - \Delta^2\rho^2(\rho/P)^{-\frac{4\sigma}{1-2\sigma}}+A^2$, against $\rho$ with physically attainable parameter values, and a horizontal line defined by $h(\rho)=0$. 

In FIG. 2 below, apparently the equation (18) has two distinct solutions given by the intersection points between the plots of $f(\rho; A, \Delta, \sigma, P)$ and $h(\rho)$. 

On the other hand, it is instructive to find numerical solutions to the equation (18) and obtain $\underline{\rho}$ and $\bar{\rho}$. To obtain the lower limit $\underline{\rho}$, we rewrite the equation (18) as $\rho = \frac{A}{\sqrt{\Delta^2(\rho/P)^{-\frac{4\sigma}{1-2\sigma}}-1}}$ and apply the fixed-point iteration technique to obtain an approximate solution for $\underline{\rho}$, using $\rho_0$ as the initial value solution. This is because, for fixed-point iteration technique, $\underline{\rho}$ is a fixed point, i.e. a stable point and any initial value that lies in between $\underline{\rho}$ and $\bar{\rho}$ the iteration converges to $\underline{\rho}$. However, for $\bar{\rho}$, this method fails as the iteration diverges away from the point. In such a case, other iteration techniques, such as the secant method, can be employed.

\begin{figure}[htb]
     \centering
     \begin{subfigure}{0.25\textwidth}
         \centering
         \includegraphics[width=\textwidth]{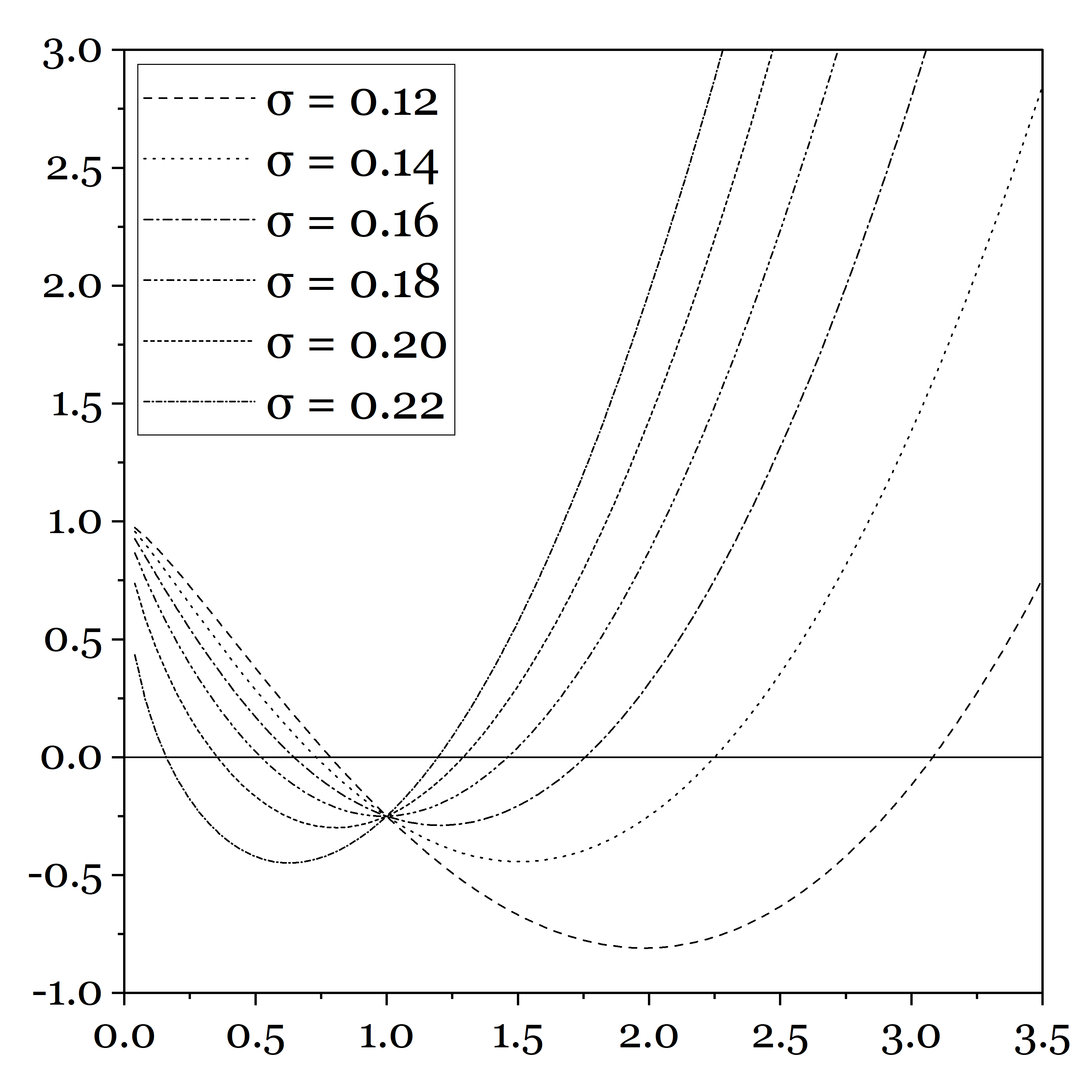}
         \caption{Varying $\sigma$ keeping $A=1$, $\Delta=1.5$, and $P=1$}
     \end{subfigure}%
     \begin{subfigure}{0.25\textwidth}
         \centering
         \includegraphics[width=\textwidth]{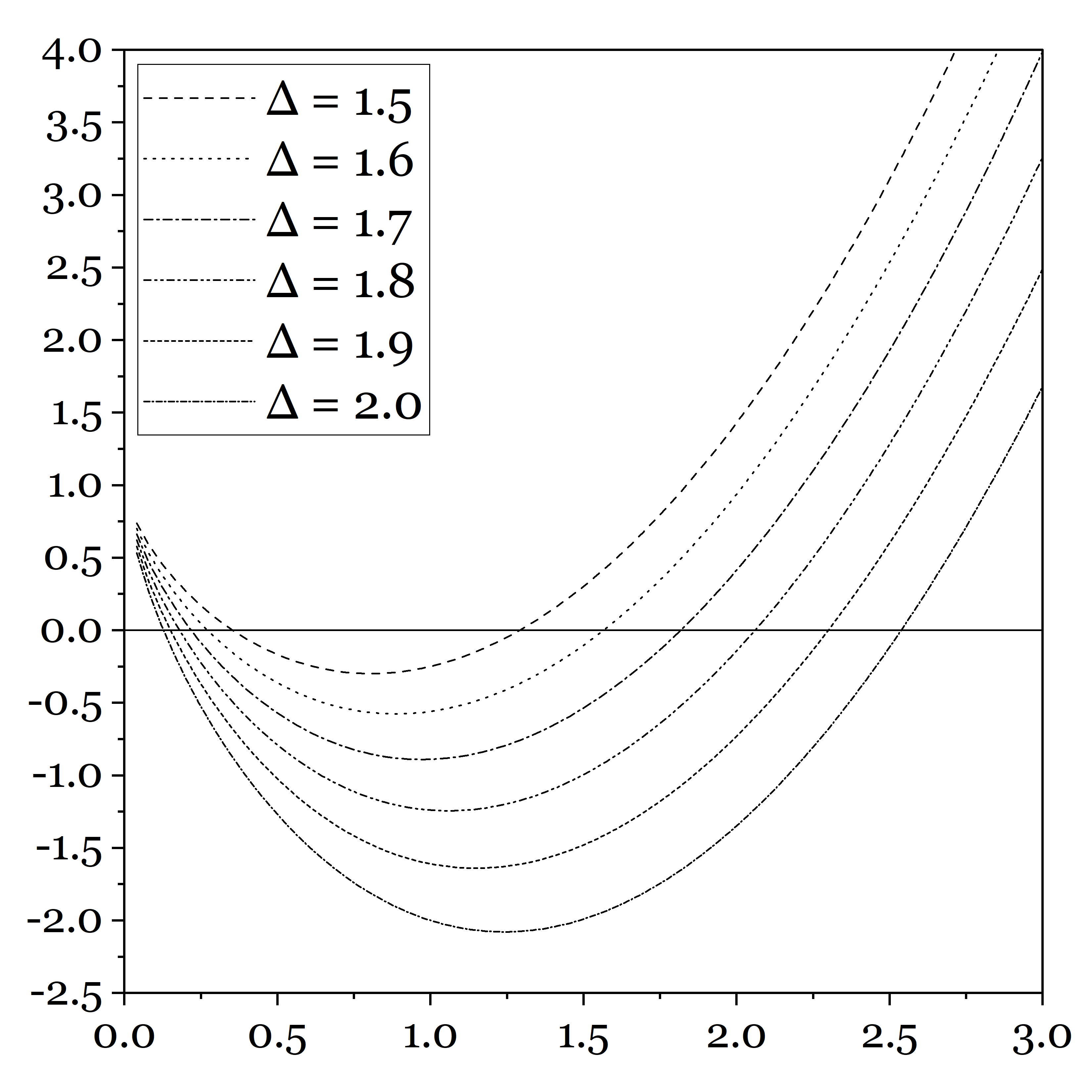}
         \caption{Varying $\Delta$ keeping $A=1$, $\sigma=0.2$, and $P=1$}
     \end{subfigure}
          \centering
     \begin{subfigure}{0.25\textwidth}
         \centering
         \includegraphics[width=\textwidth]{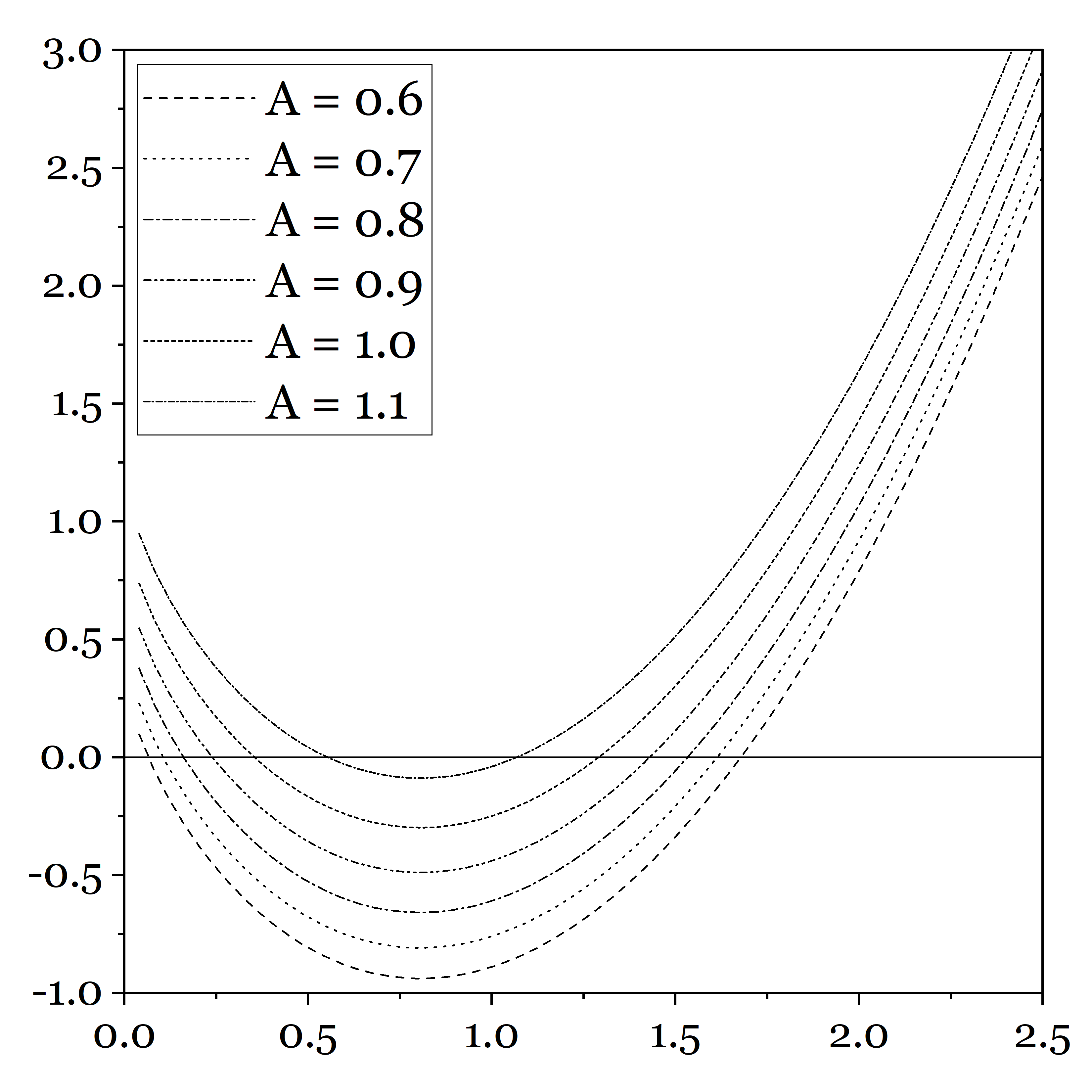}
         \caption{Varying $A$ keeping $\Delta=1.5$, $\sigma=0.2$ and $P=1$}
     \end{subfigure}%
     \begin{subfigure}{0.25\textwidth}
         \centering
         \includegraphics[width=\textwidth]{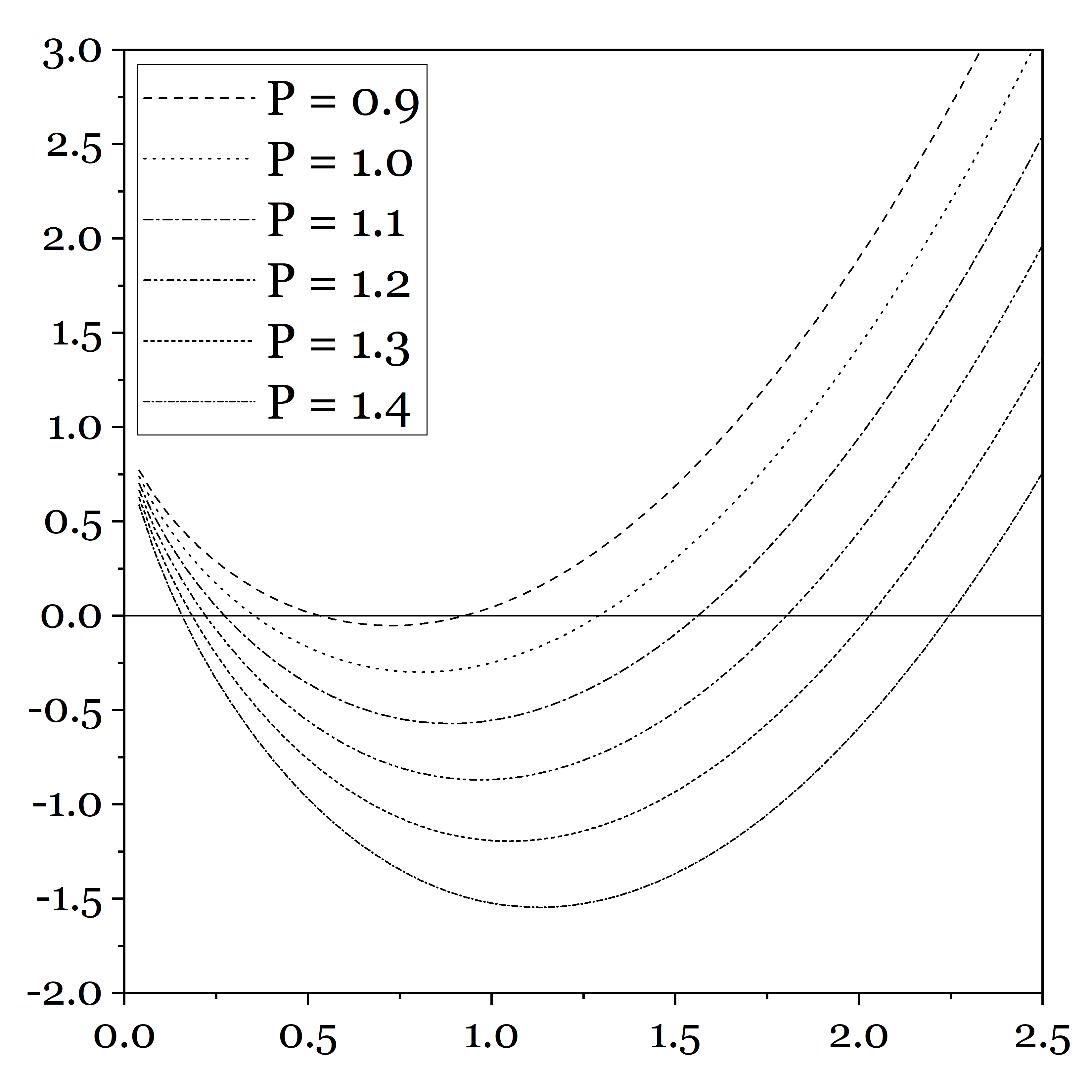}
         \caption{Varying $P$ keeping $A=1$, $\Delta=1.5$, and $\sigma=0.2$}
     \end{subfigure}
     \caption{$f(\rho; A, \Delta, \sigma, P)$ plots intersect $h(\rho)=0$ line twice, showing the existence of a lower limit $\underline{\rho}$ and an upper limit $\bar{\rho}$ of the radial coordinate}
\end{figure}

\subsection{\label{section-3A}Absence of Unbounded Solution}

A key feature to the Levi-Civita spacetime is that, an orbit of a massive point particle is always bounded, unlike the trajectory in Schwarzschild spacetime. This can be proven quite generally from the fact that the function $f(\rho; A, \Delta, \sigma, P)$ increases unboundedly as the radial coordinate increases and has only one minimum; while no maximum occurs except at $\rho=0$. Therefore, for all physically valid values of $A$ and $\Delta$, the equation (18) always has two distinct solutions, $\underline{\rho}$ and $\bar{\rho}$ (excepting the circular orbits where $\underline{\rho}=\bar{\rho}$) and the radial coordinate oscillates between in between these two points. This proves that the radial coordinate is always bounded from above by $\bar{\rho}$, leading to a bounded orbit of a massive particle for all physically valid initial conditions.

This behaviour persists in the orbit of a massive particle moving in a Newtonian gravitational field, generated by an infinitely long linear mass distribution. As can be seen from equation (16), the potential $\Phi$ also increases unboundedly due to its logarithmic nature. For this reason, a particle moving in it would require an infinite amount of energy to escape from the gravitational pull. 

\section{\label{section-4}Classification of Orbits}

Even though, the motion of a massive particle confined to the plane $z=z_0$ is of significant importance for astrophysical observations, the solution for this case in terms of elementary functions is yet to be found. That is why, it is worth resorting to numerical procedures. However, numerical solutions are difficult to extract general characteristics of orbits from, which could have been readily found from a complete, exact solution expressed in terms of well-known functions if those were available. To mitigate the situation, we need to categorize the orbits depending on their properties. The goal of this section is to categorize different orbits based on their visual appearances and lay out a system that assigns a set of unique identification parameters to each types of orbits. To identify an orbit uniquely, only three real numbers, $(C,\epsilon,\gamma)$, are required, which we will demonstrate in the next section (i.e., section IVA). Based on those parameters, we will lay out a detailed categorization scheme in section IVB.

\subsection{\label{subsection-4a}Background of the Classification Scheme}

Point particles moving in Schwarzschild spacetime follow precessing elliptical orbits \cite{synge1960-08}. It is formally shown in the Appendix C as well. As proven earlier, orbits of massive point particles in Levi-Civita spacetime are also bounded between the periastron at $\rho = \underline{\rho}$ and the apastron at $\rho = \bar{\rho}$, one might hypothesise that their planar geodetic motion can be approximated to be precessing ellipses as well. In order to see that, we make a number of apt approximations to the equation (8b) considering $B=0$, $\sigma \ll 1$, and $\lvert{\underline{\rho}/\rho}\rvert < 1$ obtaining a closed form formula of trajectory on $z=z_0$ plane.  

\begin{equation}
    \biggr(\frac{\rho}{\underline{\rho}}\biggr)^{1-\eta} = \frac{C}{1-\epsilon \cos \big[\gamma(\phi-\delta)\big]}
\end{equation}

where, $C$, $\epsilon$, $\gamma$, and $\delta$ are dimensionless quantities, given by the formulae below, carry information of the size, eccentricity, precession rate, and phase shift of the orbit. These parameters are dependent on the initial conditions. On the other hand, the parameter $\eta$ is dependent on the spacetime itself. For Levi-Civita spacetime, it is $\eta = 2\sigma + 8 \sigma^2$.

\begin{subequations}
\begin{eqnarray}
    C = & & \biggr[ 4\sigma{\underline{\rho}^2} + 8\sigma^2 A^2 + 24\sigma^2{\underline{\rho}^2} + 16\sigma^2{\underline{\rho}^2}\ln\biggr(\frac{\underline{\rho}}{P}\biggr)\biggr]^{-1}
    \nonumber \\
    & & \times \biggr[A^2 + 2\sigma{\underline{\rho}^2} + 4\sigma A^2 \ln\biggr(\frac{\underline{\rho}}{P}\biggr) + 12 \sigma^2 A^2 
    \nonumber \\
    & &  + 16\sigma^2{\underline{\rho}^2} + 8\sigma^2 A^2 \ln\biggr(\frac{\underline{\rho}}{P}\biggr)  
    \nonumber \\ 
    & & + 8\sigma^2 {\underline{\rho}^2}\ln\biggr(\frac{\underline{\rho}}{P}\biggr) + 8\sigma^2 A^2 \ln^2\biggr(\frac{\underline{\rho}}{P}\biggr) \biggr]
    \\
    \epsilon = & & \biggr[4\sigma {\underline\rho^2} + 24\sigma^2 {\underline\rho^2} + 8\sigma^2 A^2
    + 16 \sigma^2 {\underline\rho^2}\ln(\frac{\underline{\rho}}{P})\biggr]^{-1}  \nonumber \\
    & & \times \biggr[\Delta^2A^2 {\underline\rho^2} -  A^2 {\underline\rho^2} + 2\sigma\Delta^2{\underline{\rho}^4} -2\sigma{\underline{\rho}^4} \nonumber \\
    & &  - 6 \sigma A^2 \underline{\rho}^2+ 4 \sigma \Delta^2A^2 \underline\rho^2 \ln\biggr(\frac{\underline{\rho}}{P}\biggr) \nonumber \\
    & & + 4 \sigma \Delta^2A^2 \underline\rho^2 \ln\biggr(\frac{\underline{\rho}}{P}\biggr) - 8 \sigma A^2 \underline\rho^2 \ln\biggr(\frac{\underline{\rho}}{P}\biggr) \nonumber \\
    & & - 4 \sigma^2 A^4 + 16\Delta^2{\underline\rho^4}\sigma^2 - 12{\underline\rho^4}\sigma^2 \nonumber \\
    & & + 12\Delta^2\sigma^2 A^2 \underline\rho^2- 44\sigma^2 A^2 \underline\rho^2  \nonumber \\
    & & + 8 \Delta^2\sigma^2 {\underline\rho^4}\ln\biggr(\frac{\underline{\rho}}{P}\biggr) - 16 \sigma^2 {\underline\rho^4}\ln\biggr(\frac{\underline{\rho}}{P}\biggr)  \nonumber \\
    & & + 8\Delta^2 \sigma^2 A^2 \underline\rho^2\ln\biggr(\frac{\underline{\rho}}{P}\biggr) - 64 \sigma^2 A^2 \underline\rho^2\ln\biggr(\frac{\underline{\rho}}{P}\biggr)  \nonumber \\
    & & + 8  \Delta^2\sigma^2 A^2 \underline\rho^2\ln^2\biggr(\frac{\underline{\rho}}{P}\biggr) \nonumber \\
    & & - 32 \sigma^2 A^2 \underline\rho^2\ln\biggr(\frac{\underline{\rho}}{P}\biggr)\biggr]^\frac{1}{2}
    \\
    \gamma = & & \biggr[\frac{1-4\sigma -4\sigma^2}{1-2\sigma}\biggr]\times\biggr[1 + 2\sigma\frac{\underline{\rho}^2}{A^2} + 12\sigma^2 
    \nonumber \\   
    & & + 16\sigma^2\frac{\underline{\rho}^2}{A^2} - 8\sigma^2\ln\biggr(\frac{\underline{\rho}}{P}\biggr)\biggr]^\frac{1}{2}
    \\
    \delta = & &\phi_0-\frac{1}{\gamma}\arccos\biggr[\frac{1}{\epsilon} - \frac{C}{\epsilon} \biggr( \frac{\underline{\rho}}{\rho_0} \biggr)^{1-\eta}\biggr]
\end{eqnarray}
\end{subequations}

It is evident from equation (19) that the trajectory approximately follows a precessing elliptical orbit. We also notice that the shape of a trajectory is uniquely determined by three of the parameters only, namely $C$, $\epsilon$, and $\gamma$. This is because, the parameter $\delta$ can always be set to zero by a mere rotation with respect to the $z$-axis.

The parameter $C$ determines the size of an orbit. An important thing to notice in the equation (20a) is that, $C$ goes to infinity as $\sigma$ approaches to zero. This is reasonable, since $\sigma=0$ represents the complete absence of the matter distribution, leading to a straight line trajectory. Another important parameter characterizing an orbit is its eccentricity $\epsilon$, such that $0\leq\epsilon<1$ for spacetime with a non-zero $\sigma$. In such a spacetime, the orbit of a massive particle is always bounded between $\rho=\underline{\rho}$ and $\rho=\bar{\rho}$, as we proved in section IIIA. From this, one concludes that the orbit must have an eccentricity which is always less than $1$.

Out of the four parameters of the trajectory, $\gamma$ is the most important one for our purpose, since it determines the rate of precession and consequently the number of ``petals''. However, for a closed orbit, $\gamma$ must be a rational number. For an irrational $\gamma$, the orbit never closes and the trajectory keeps filling out a circular area as time increases. Hence, simply counting the number of ``petals'' loses its meaning. In the following section (IVB), we lay out a classification scheme that is both practical and incorporates all possible orbits.

\begin{figure}[htb]
     \centering
     \begin{subfigure}{0.25\textwidth}
         \centering
         \includegraphics[width=\textwidth]{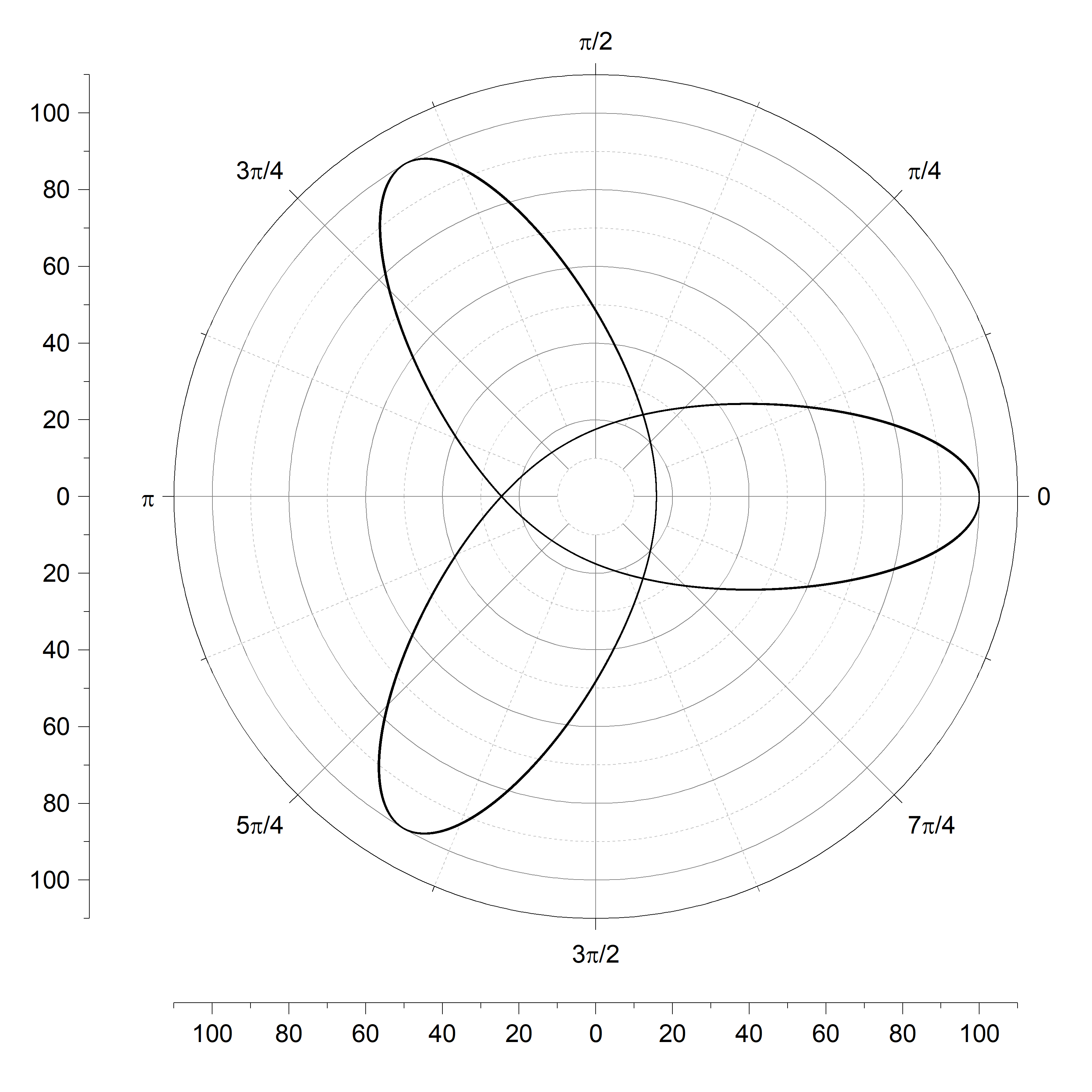}
         \caption{A closed orbit with three petals in spacetime with $(\sigma,P)=(0.001,1)$}
     \end{subfigure}%
     \begin{subfigure}{0.25\textwidth}
         \centering
         \includegraphics[width=\textwidth]{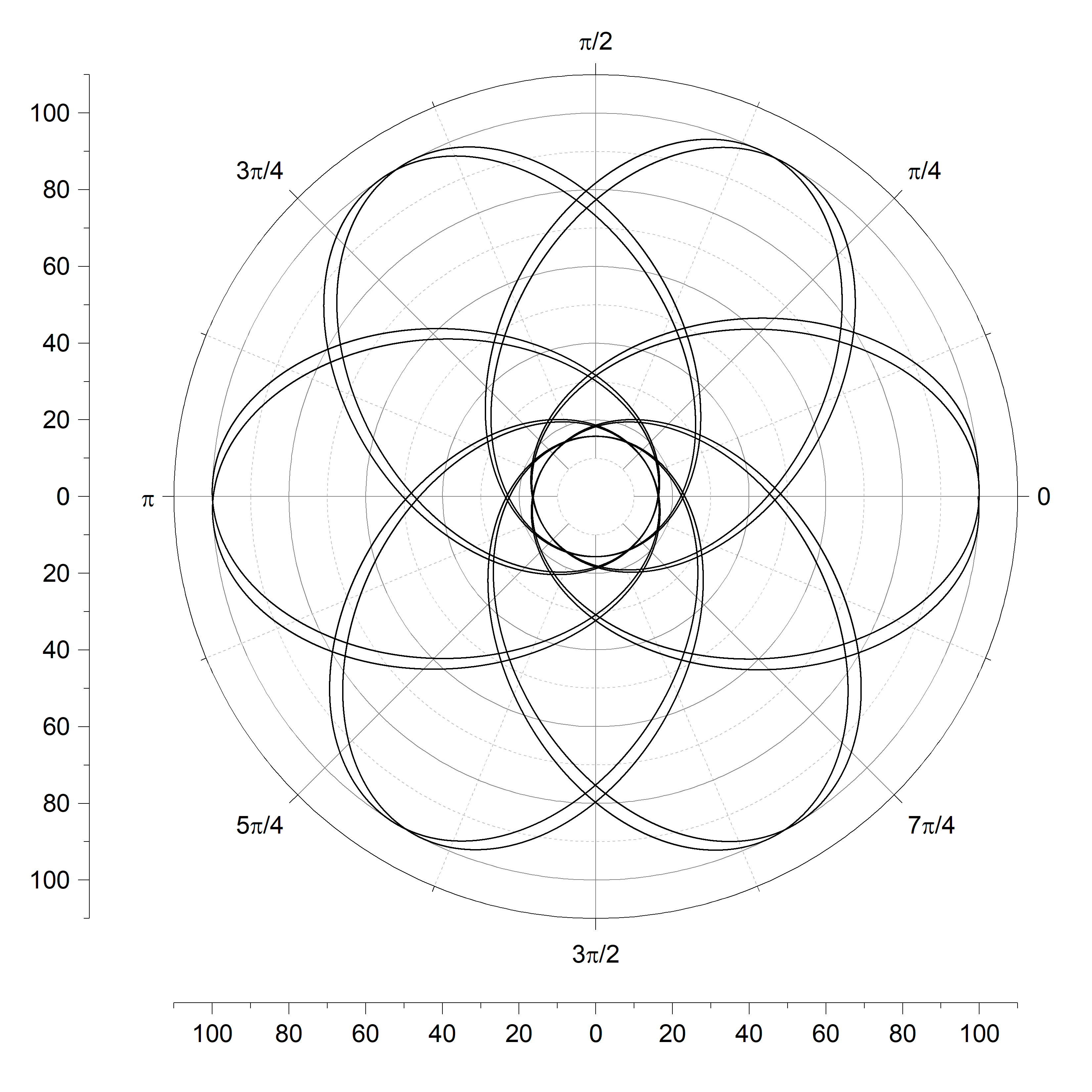}
         \caption{A precessing orbit with six petals in spacetime with $(\sigma,P)=(0.1,1)$}
     \end{subfigure}
     \caption{Particle moving in Levi-Civita spacetime exhibiting closed and precessing trajectories}
\end{figure}

\subsection{\label{subsection-4b}The Classification Scheme}

It is apparent from the equation (20) that the parameters $C$, $\epsilon$ and $\gamma$ contain all the information regarding an orbit on $z=z_0$ plane. The parameter $C$ defines the size, whereas, the parameter $\epsilon$ defines the eccentricity of the orbit. A rather interesting one is the parameter $\gamma$. Depending on its value, an orbit can be a closed or an ever-precessing one. Tweaking its value, one can even change the number of ``petals'' in a closed orbit. FIG. 3 shows two examples of possible orbits of massive particles in Levi-Civita spacetime. Among two, the first one is a closed orbit with three petals and the other second one is a precessing orbit with six petals. Therefore, given enough time ($t\approx 28300$), the trajectory will start to fill the localized region in between $\underline{\rho}$ and $\bar{\rho}$. Before discussing the categorization scheme, we need to define few terminologies. 

\begin{figure}[htb]
     \centering
     \begin{subfigure}{0.25\textwidth}
         \centering
         \includegraphics[width=\textwidth]{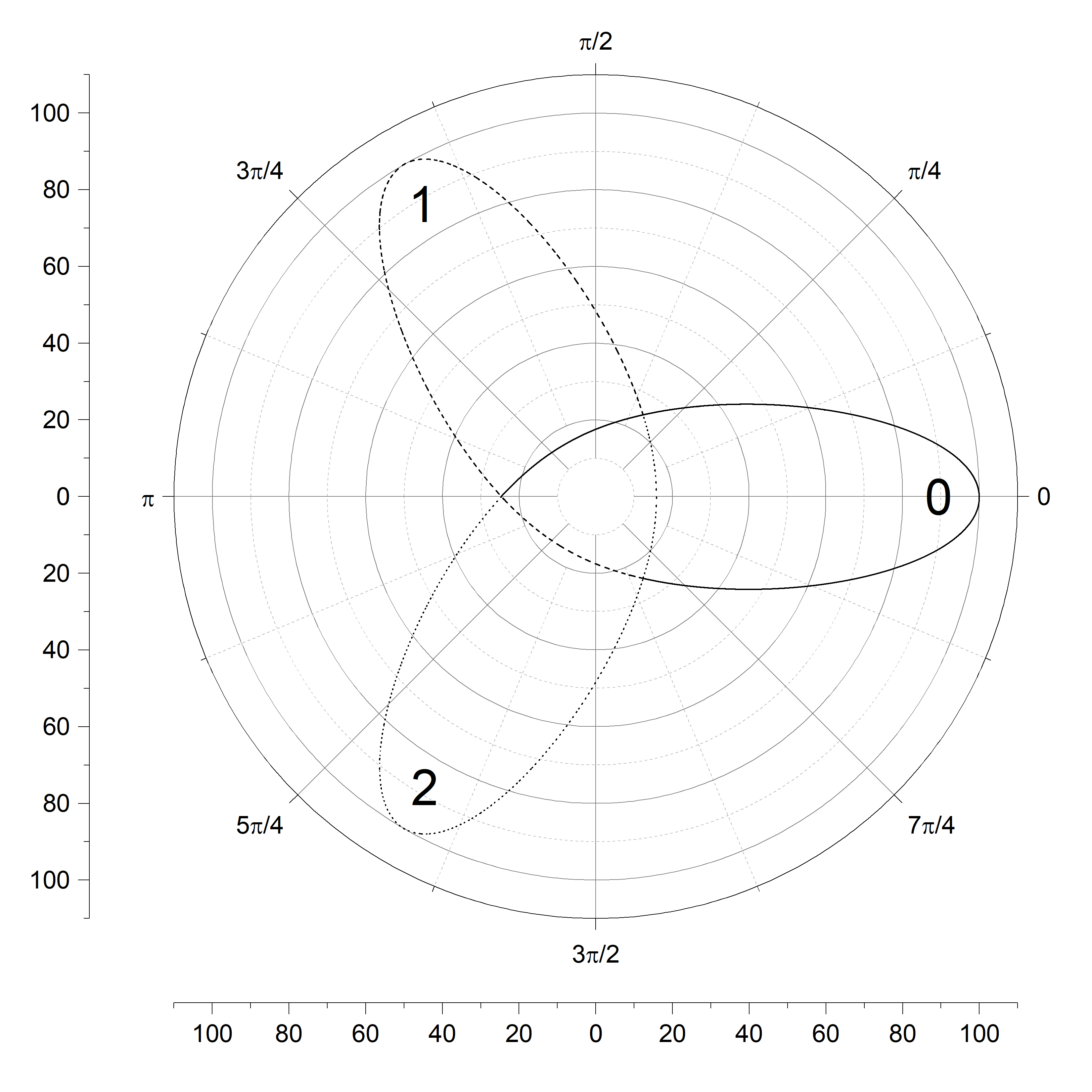}
         \caption{A corolla with three petals of a closed orbit}
     \end{subfigure}%
     \begin{subfigure}{0.25\textwidth}
         \centering
         \includegraphics[width=\textwidth]{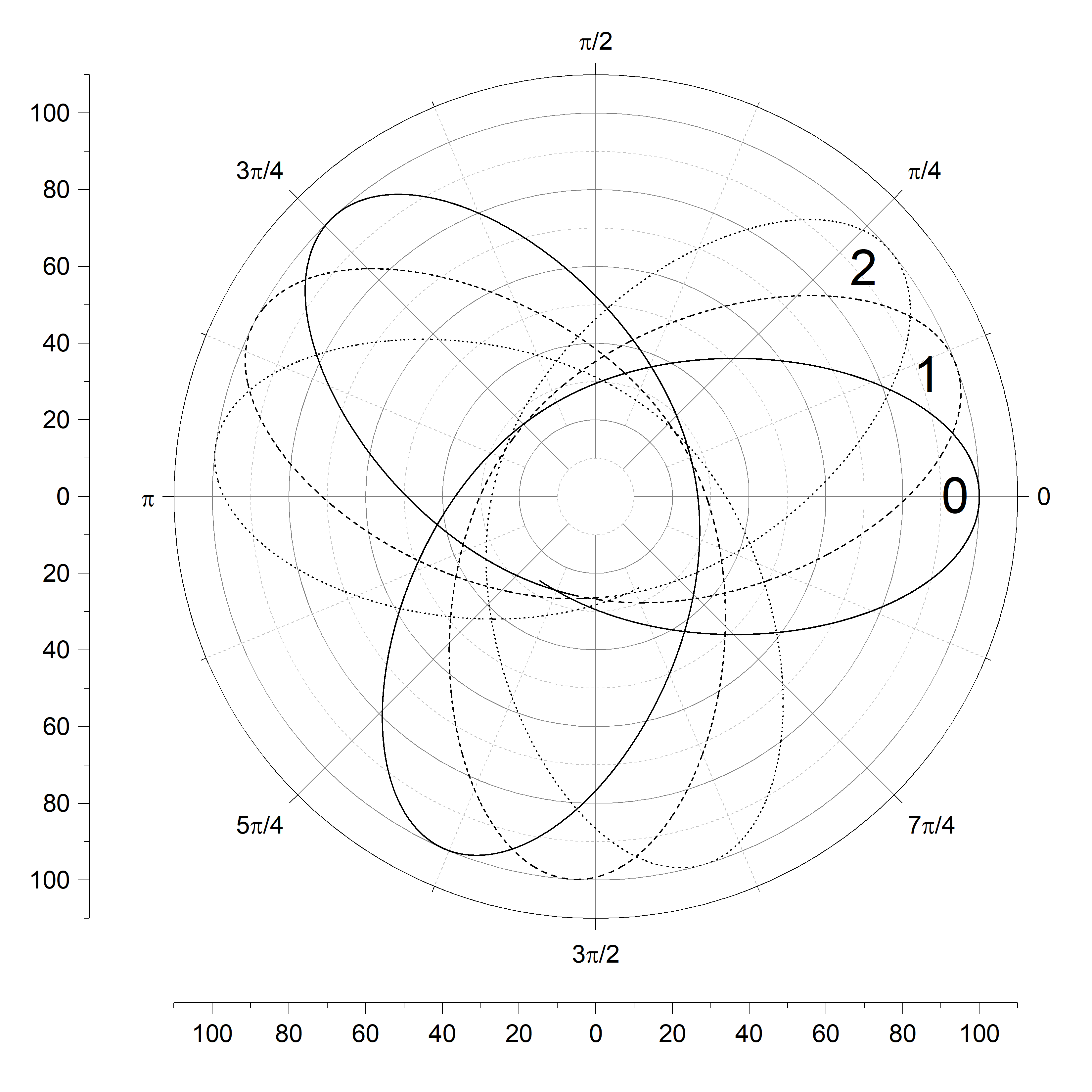}
         \caption{Three corollas of a precessing orbit}
     \end{subfigure}
     \caption{A closed and a precessing orbits with three petals}
\end{figure}

Let us use the orbits shown in FIG. 4 as specimens to aid our discussion. Both orbits have what we shall call three ``leaves’’ or ``petals’’. To demonstrate, we number the petals of the closed orbit in FIG. 4(a), starting from $0$ in the counter-clockwise direction. In FIG. 4(b), a three-petal orbit is precessing and we have plotted three complete traces of the orbit. We shall call each trace a ``corolla’’ and have numbered them starting from $0$, based on their chronology. In the diagram, the solid-line, the dashed-line, and the dotted-line represent corolla $0$, corolla $1$, and corolla $2$ respectively.  This procedure of assigning an ordered pair of numbers to identify a petal of a corolla can be generalized to any such orbits, as has been done in FIG. 5(a). The first number of the pair denotes the corolla and second one denotes the petal in that corolla. We call the petals that share the same corolla (e.g., $(1,0)$ and $(1,2)$) as ``sisters'' and petals that share the same second number (e.g., $(1,1)$ and $(2,1)$) as ``clones'’ of each other.

\begin{figure}[htb]
     \centering
     \begin{subfigure}{0.25\textwidth}
         \centering
         \includegraphics[width=\textwidth]{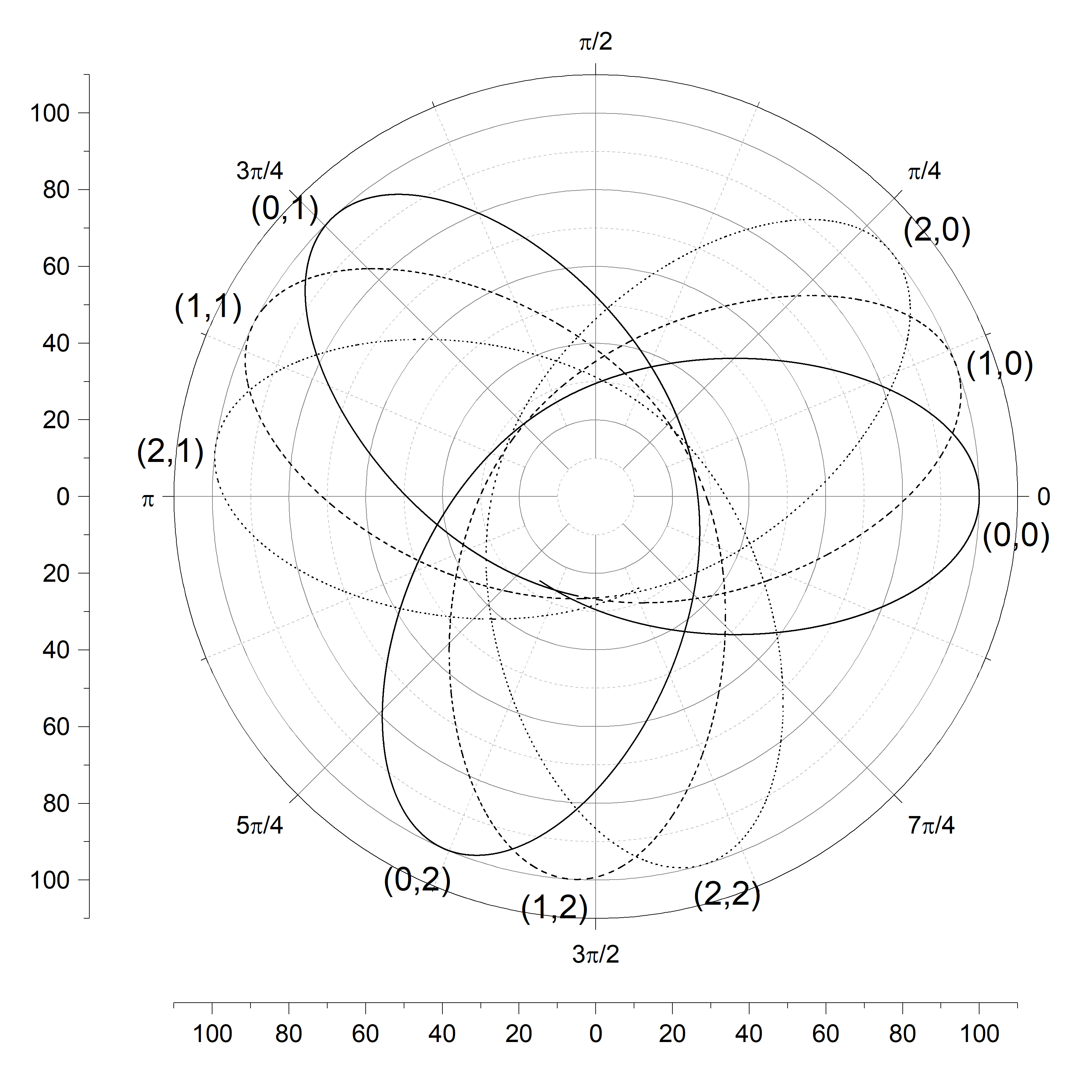}
         \caption{Scheme for numbering petals in an orbit}
     \end{subfigure}%
     \begin{subfigure}{0.25\textwidth}
         \centering
         \includegraphics[width=\textwidth]{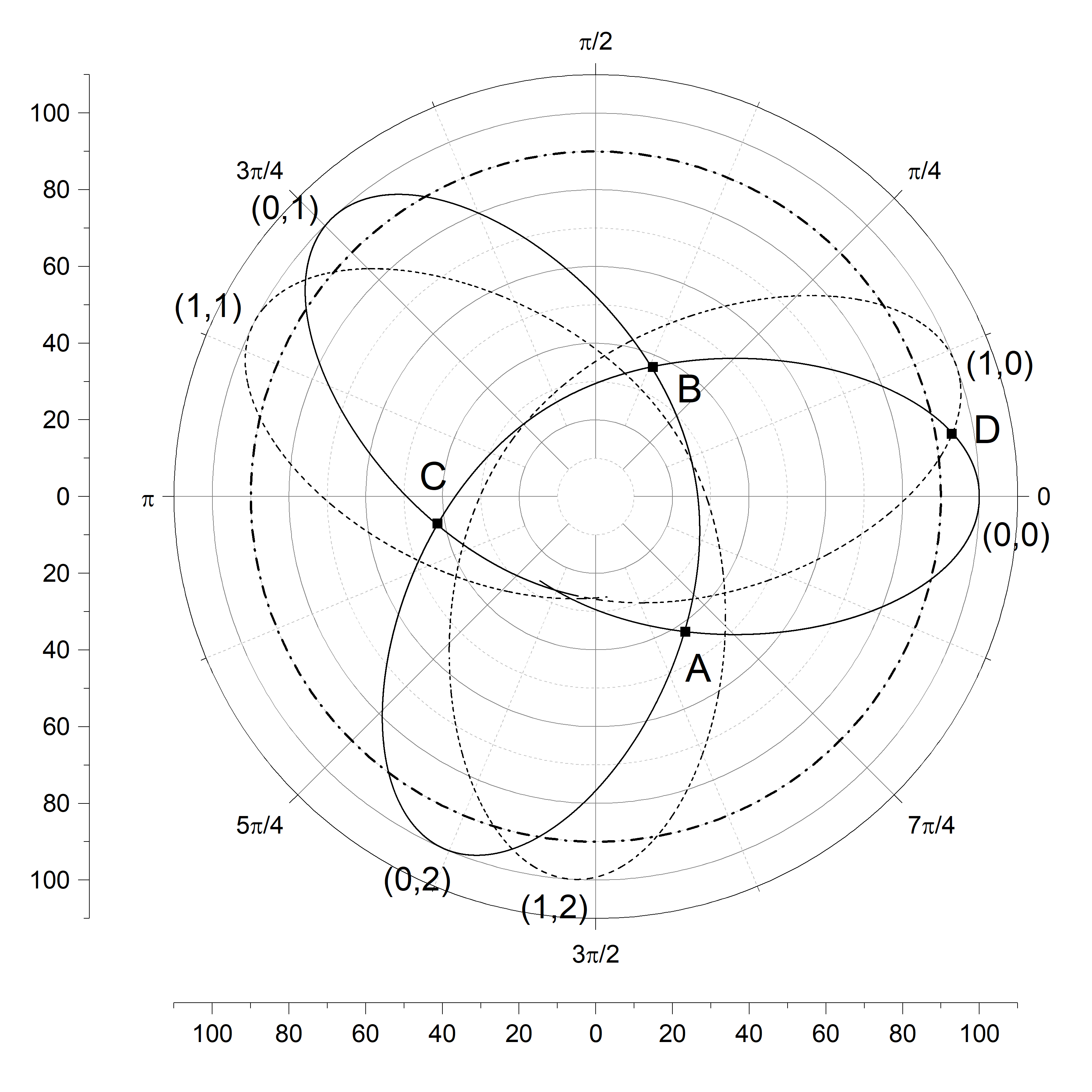}
         \caption{Distinguishing sister petals from their clones}
     \end{subfigure}
     \caption{Identification of a corolla in a precessing orbit}
\end{figure}

At this point, it is natural to ask for a criterion that makes the distinction between sister petals and their clones. Our definition goes like this: \emph{If, for a set of chronologically consecutive petals, all the intersection points between any two petals falls inside of a circle defined by $\rho=\lambda\bar{\rho}$, for some $\lambda\in[0,1]$, then the petals are sisters for that chosen $\lambda$. The maximal set of sister petals makes a corolla. A petal in a corolla is said to be a clone of a petal in a preceding corolla, if the mentioned petals share an intersection point that lies at the farthest, among all the intersection points in the said consecutive corollas, from the origin.}

For instance, in FIG. 5(b), we have chosen $\lambda=0.9$ and therefore, the corresponding circle $\rho=\lambda\bar{\rho}$ (dash-dot curve) in the figure portrays a circle with a radius of $90\%$ of the peak radial coordinate $\bar\rho$. In the figure, the orbit starts with the petal $(0,0)$, which is then followed by the petals $(0,2)$ and $(0,1)$. This three petals are sisters, as any intersection point (for example, the points $A$, $B$, and $C$) between these petals remains inside the dash-dot circle. The set of $(0,0)$, $(0,1)$, and $(0,2)$ also make a corolla, as the chronologically fourth petal $(1,0)$ has an intersection point (point $D$) that falls outside of the dash-dot circle. Since, in FIG. 5(b), $D$ is the farthest from the origin among all the intersection points, the petal $(1,0)$ is a clone of $(0,0)$.

Finally, armed with a precise definition of corolla, we are ready to lay out the desired categorization scheme. We split the orbit parameter $\gamma$ in equation (19) such that

\begin{equation}
    \gamma=\frac{m-\chi}{n}
\end{equation}

where, $m$ and $n$ are positive coprime integers and $\chi$ is a real number on the interval $(-1, 1)$. A quick inspection reveals that $m$ is the number of petals in a corolla, $n$ is the ``winding number'', i.e., the number of complete rotations the particle massive makes around the cosmic string during the course of a single corolla, and $\chi$ is a small number that represent the ``rate of precession'' of a corolla. If $\chi=0$ for the orbit, we will refer its corolla to be ``closed'', meaning, it is not precessing. It is worth to mention that an orbit with closed corolla is always closed, but the converse is not true. Evidently, the numbers $m$, $n$, and $\chi$ depend on our choice of $\lambda$. Throughout this article, we have chosen $\lambda=0.9$. One can pick a higher or a lower value of $\lambda$ for convenience. To avoid ambiguity, one must mention the $\lambda$ parameter of their classification scheme. For example, we call our classification scheme a ``$90\%$ Level classification scheme''.

\begin{figure}[htb]
     \centering
     \begin{subfigure}{0.25\textwidth}
         \centering
         \includegraphics[width=\textwidth]{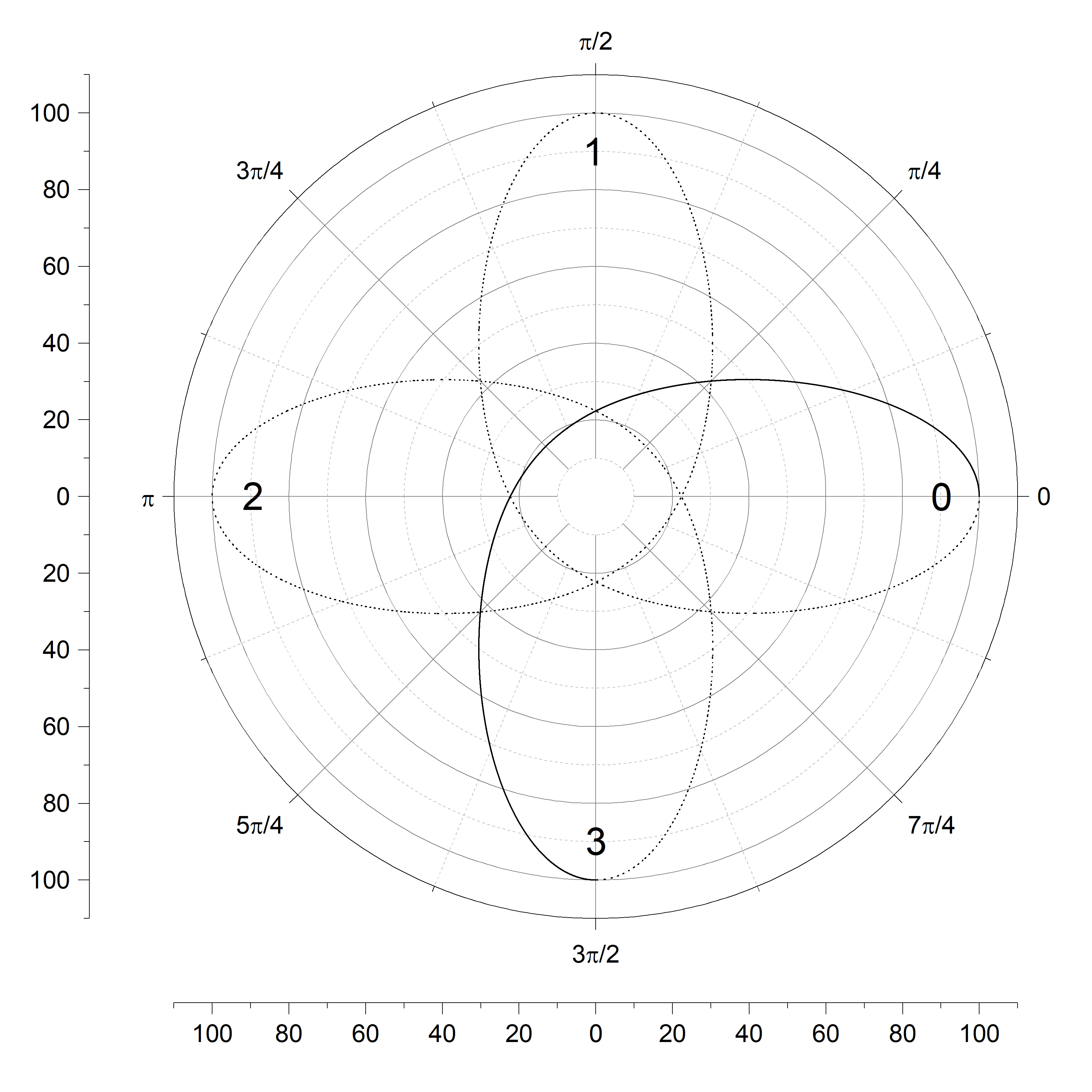}
         \caption{$\gamma=4/3$}
     \end{subfigure}%
     \begin{subfigure}{0.25\textwidth}
         \centering
         \includegraphics[width=\textwidth]{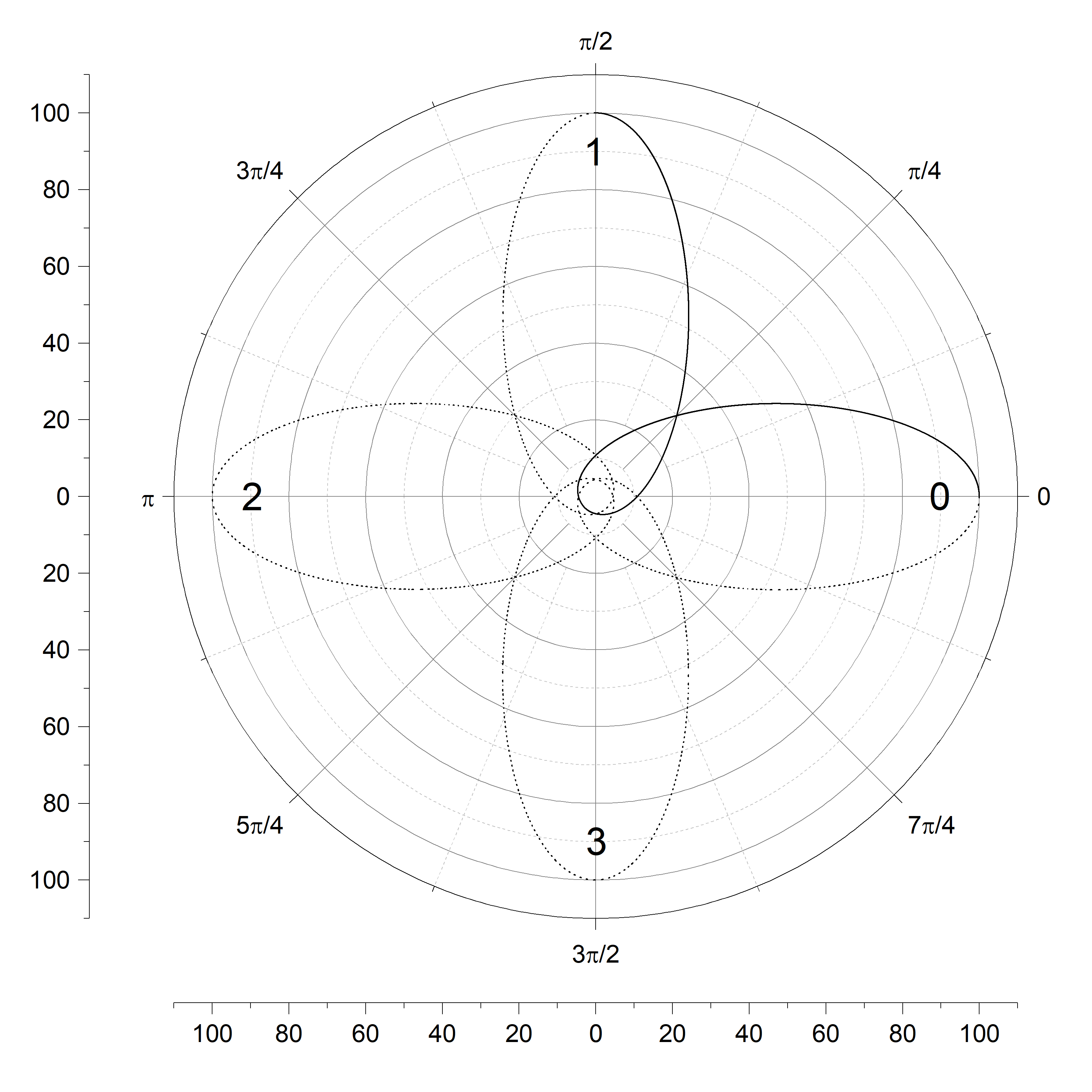}
         \caption{$\gamma=4/5$}
     \end{subfigure}
     \caption{Closed orbits with four petals}
\end{figure}

In FIG. 6, both orbits have closed corollas with four petals each, as they have $m=4$. However, the one in FIG. 6(a) makes three full rotations around the center before it reaches its initial position as it has $n=3$. Right after leaving the petal $0$, the particle in FIG 6(a) skips two petals and traces out petal $3$ directly. On the other hand, the orbit in FIG. 6(b) has $n=5$ and makes five full rotations before reaching the initial position. After leaving petal $0$, the particle makes a full rotation around the center and reaches petal $1$. In the course of going from petal $0$ to petal $1$, the particle skips four petals. It is impossible to have an orbit of four petals in which the particle skips one petal after leaving petal $0$ and enters petal $2$ directly. Because, in that situation, the particle would enter the petal $0$ after leaving $2$, which would make it a closed orbit of two petals only.

\begin{figure}[htb]
     \centering
     \begin{subfigure}{0.25\textwidth}
         \centering
         \includegraphics[width=\textwidth]{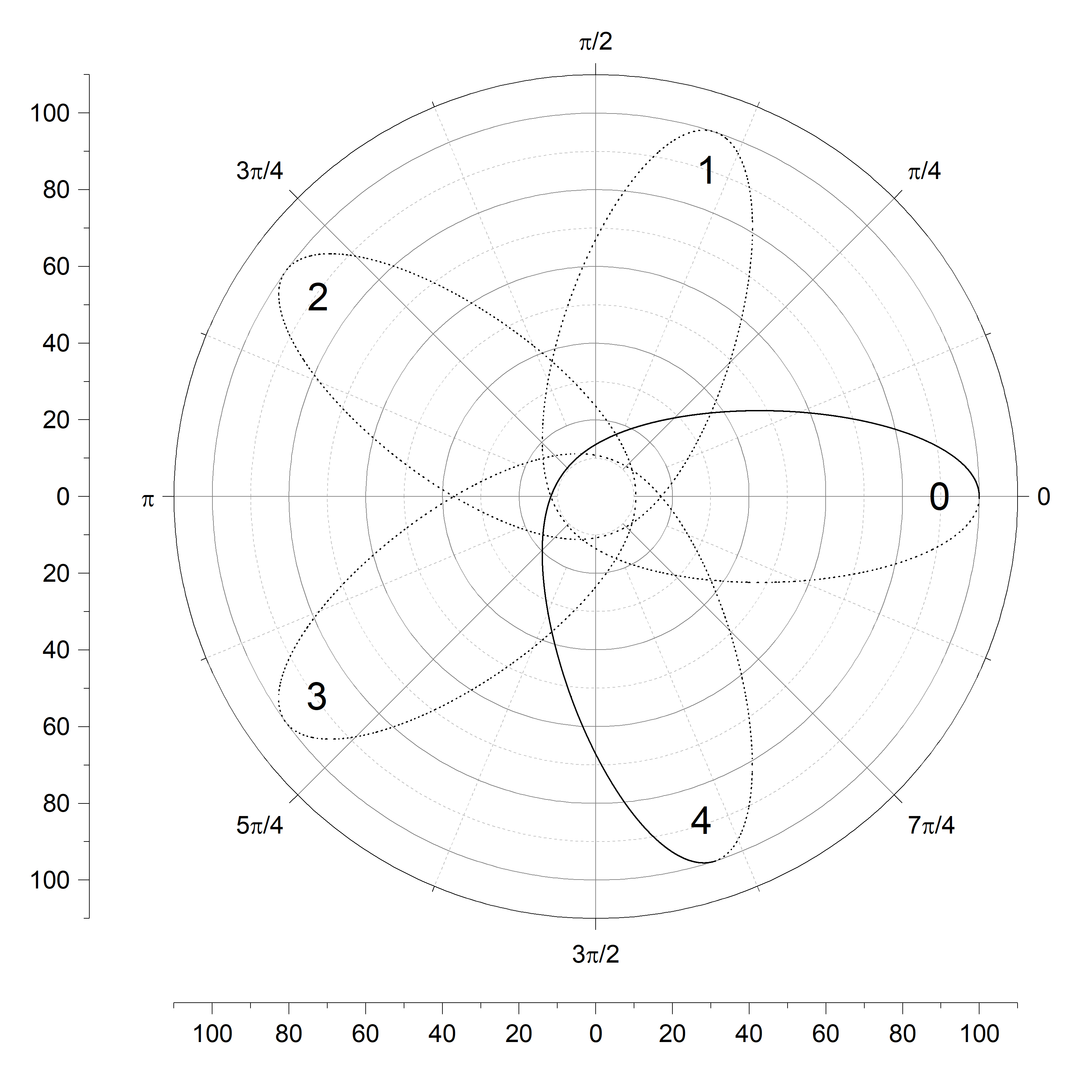}
         \caption{$\gamma=5/4$}
     \end{subfigure}%
     \begin{subfigure}{0.25\textwidth}
         \centering
         \includegraphics[width=\textwidth]{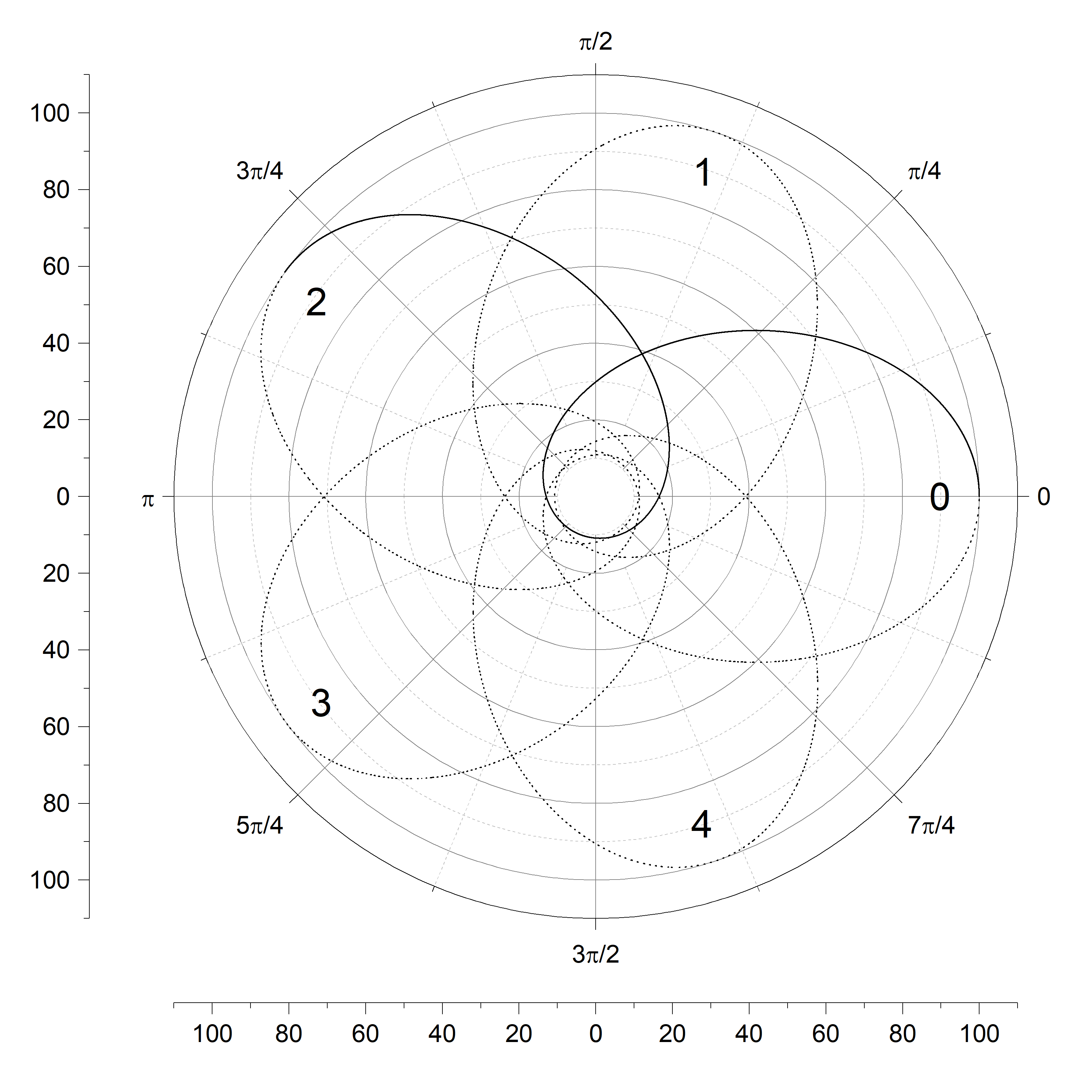}
         \caption{$\gamma=5/7$}
     \end{subfigure}
          \centering
     \begin{subfigure}{0.25\textwidth}
         \centering
         \includegraphics[width=\textwidth]{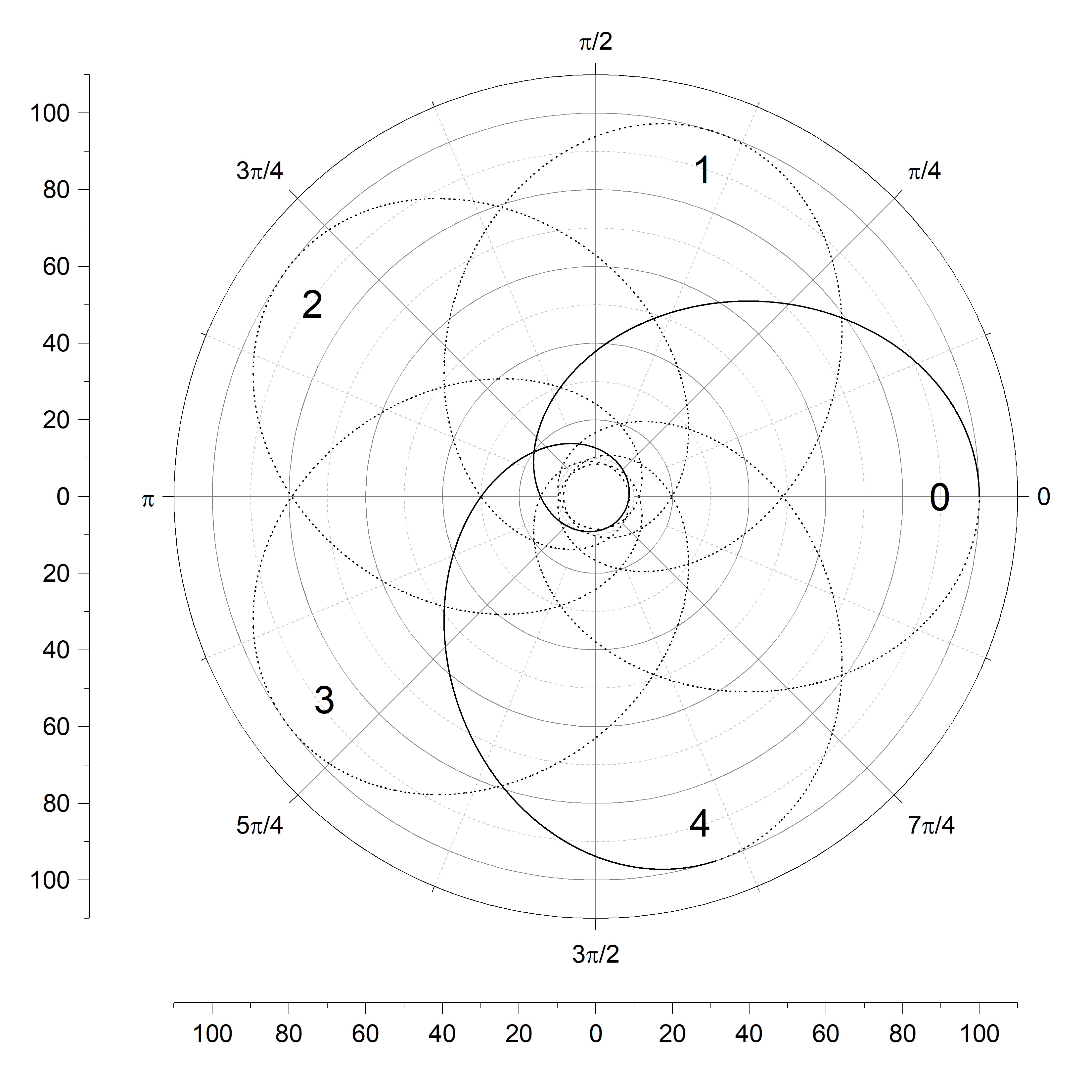}
         \caption{$\gamma=5/9$}
     \end{subfigure}%
     \begin{subfigure}{0.25\textwidth}
         \centering
         \includegraphics[width=\textwidth]{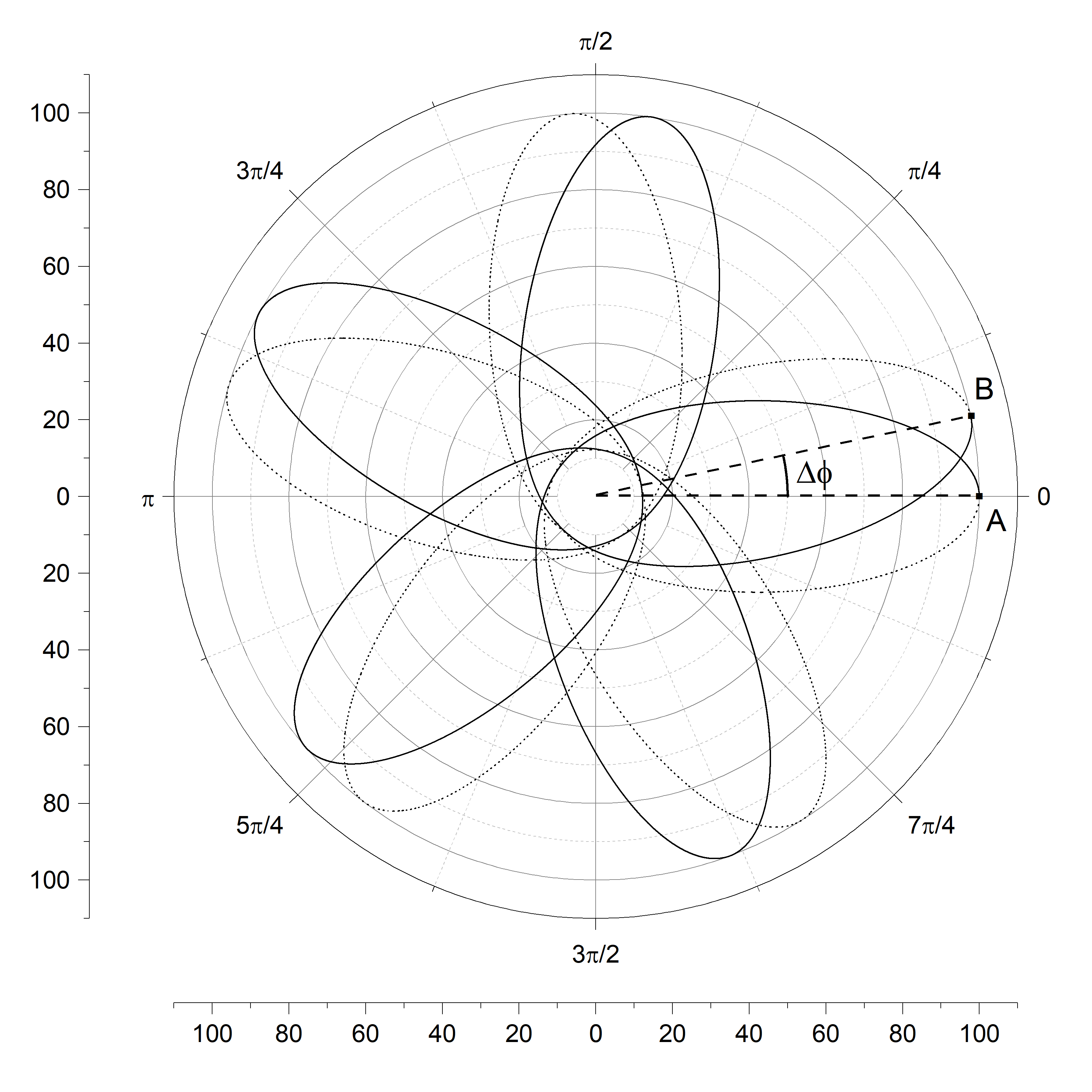}
         \caption{$\gamma\approx\frac{1}{4}(5-0.0413065)$}
     \end{subfigure}
     \caption{Orbits with five petals}
\end{figure}

Similarly, in FIG. 7, all the orbits have $m=5$ and therefore, have five petals each. However, they differ in $n$. Orbits in FIG 7(a), FIG 7(b), and FIG 7(c) have $n=4$, $n=7$, and $n=9$ respectively. We observe that, for any orbit, right after leaving the petal $0$, the particle directly enters in and traces out the petal $n\bmod m$.

There is one more property that could make two closed orbits visually different. That is the number of extra full rotations the particle makes during the transition from one petal to another. Both particles in FIG. 7(a) and FIG. 7(c) enter in petal $4$ after leaving petal $0$. But the particle in FIG. 7(c) makes an extra $2\pi$ rotation around the center before entering in petal $4$. It can be seen easily that the number of extra full rotations $w$ is related to $m$ and $n$ by the relation $n=(w+1)m+n\bmod m$.

Finally, consider the orbit in FIG 7(d). The orbit is, essentially, the one in FIG. 7(a) with a small precession. Starting at point $A$, the particle traces out five petals and ends up at point $B$, creating an angular separation, $\Delta\phi$, with its starting point $A$. For any orbit, one can show that the angular separation is related to the parameter $\chi$ by the relation

\begin{equation}
    \Delta\phi = \frac{2\pi n}{m-\chi}\chi
\end{equation}

using equation (19) and (21). For an orbit a parameter $\chi=0$ leads to a zero angular separation between the end points, leaving a closed trace. Therefore, the parameter $\chi$ measures how badly a corolla in a given orbit fails to close.

So far, we have discussed the geometric interpretations of the parameters $C$, $\epsilon$, $m$, $n$, and $\chi$. These parameters are sufficient to identify any orbit around an infinitely-long, straight cosmic string. Therefore, in our classification scheme, we assign an unique quintuple $(C,\epsilon, m, n, \chi)$ to each orbit. One might suppress the first two numbers and write the triplet $(m, n, \chi)$ only, if the size and eccentricity of the orbits are not the parameters of interest.

The classification scheme we have laid out here is inspired from a 2008 work of Levin and Perez-Giz \cite{levin2008}. In their article, Levin and Perez-Giz laid out a scheme that categorizes the orbits around a black hole according to geometric shapes of the orbits. They used three numbers, $z$, $w$, and $v$, to identify ``zoom'' or the number of petals, ``whirl'' or extra complete rotation per petal the particle makes, and the petal number that particle hits right after leaving the petal $0$ respectively. It is apparent from the discussions above that the parameters for closed orbits $(z,w,v)$, by Levin and Perez-Giz, are related to our parameters $m$ and $n$ by the relations

\begin{subequations}
\begin{eqnarray}
    z & = & m
    \\
    w & = & \frac{1}{m}(n - n\bmod m) -1
    \\
    v & = & n\bmod m
\end{eqnarray}
\end{subequations}

The equation (23) shows that only two parameters are enough to express the information contained in the parameters $z$, $w$, and $v$. Another limitation of the classification scheme developed by Levin and Perez-Giz is that it would label two visually similar orbits far from each other. For example, the orbits in FIG. 7(a) and FIG 7(d) have $\gamma\approx1.25$ and $\gamma\approx1.2396734$ respectively. The classification scheme of Levin and Perez-Giz would label the first one as $(5,0,4)$ and the second one as $(6198367,0,1198367)$. It is also prone to truncation errors. On the other hand, our classification scheme labels the orbits in FIG. 7(a) and FIG 7(d) as $(5,4,0)$ and $(5,4,0.0413065)$ respectively, when the first two parameters are suppressed. For categorizing astronomically observable ellipse-like orbits, our classification scheme might prove to be useful as it labels orbits closely that are visually similar.

\section{\label{section-5}Examples of Massive Particle Orbits}

In this section, we explore the the orbits of massive particles in Levi-Civita spacetime graphically. In three consecutive figures, FIG. 8, FIG. 9, and FIG. 10, we get to see 48 orbits in total. The list includes orbits as simple as an ellipse to orbits with many petals and whirls. In these figures, the images in the first and the third columns are of closed orbits, whereas, the rest, in third and fourth columns, are of orbits with small precession.

\begin{figure*}[htb]
     \centering
     \begin{subfigure}{0.25\textwidth}
         \centering
         \includegraphics[width=\textwidth]{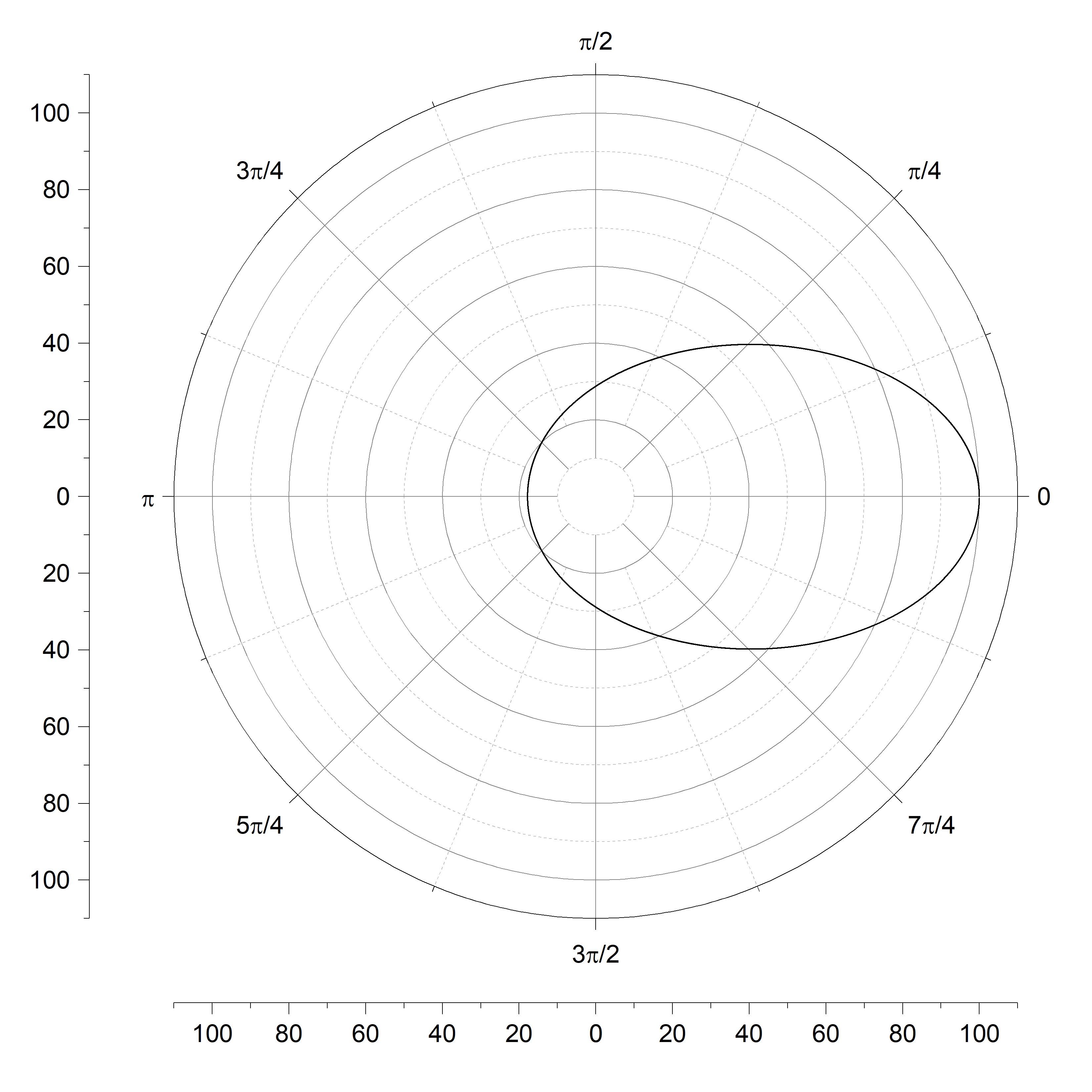}
         \caption{$(27,0.73,1,1,0)$}
     \end{subfigure}%
     \begin{subfigure}{0.25\textwidth}
         \centering
         \includegraphics[width=\textwidth]{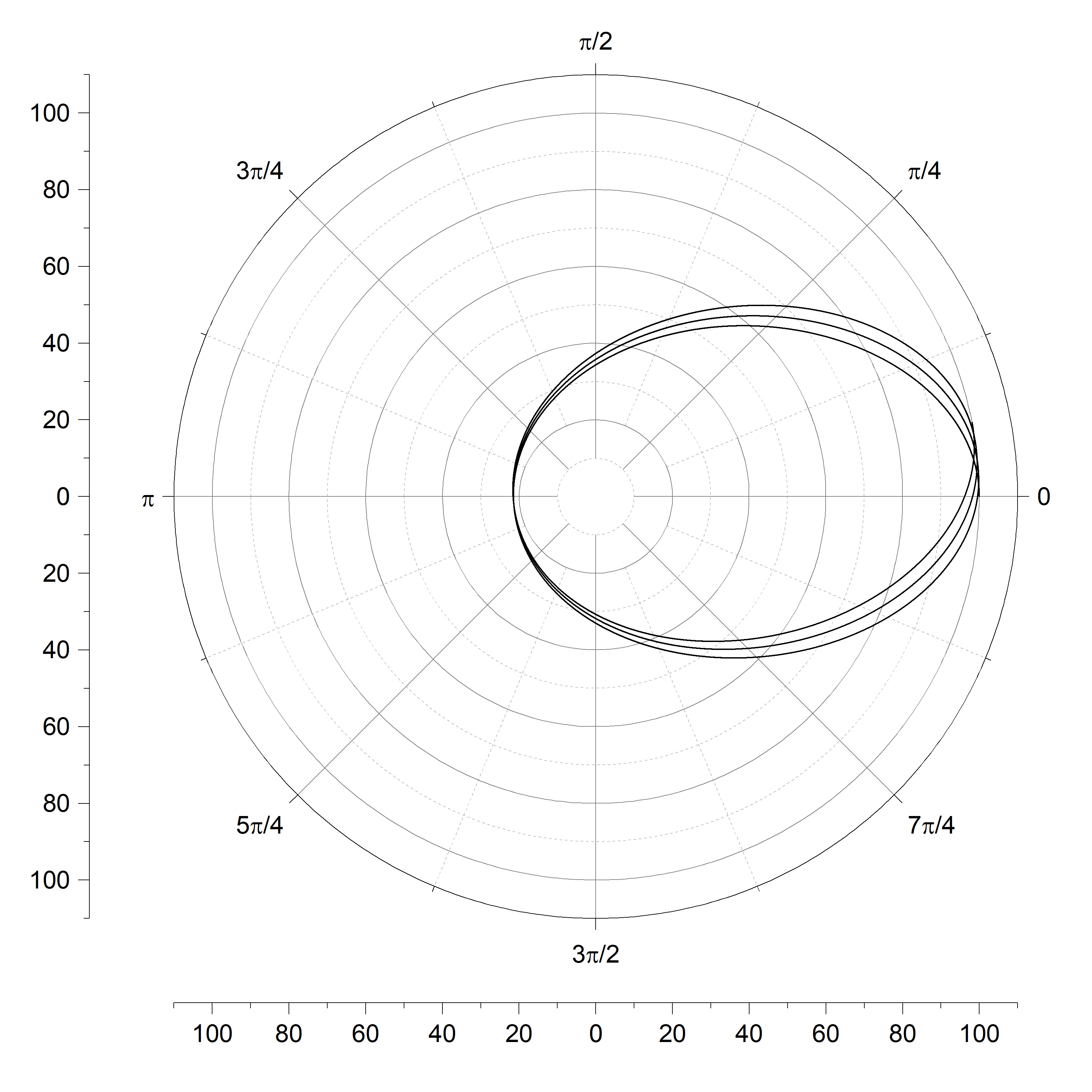}
         \caption{$(32.5,0.675,1,1,0.01024)$}
     \end{subfigure}%
          \begin{subfigure}{0.25\textwidth}
         \centering
         \includegraphics[width=\textwidth]{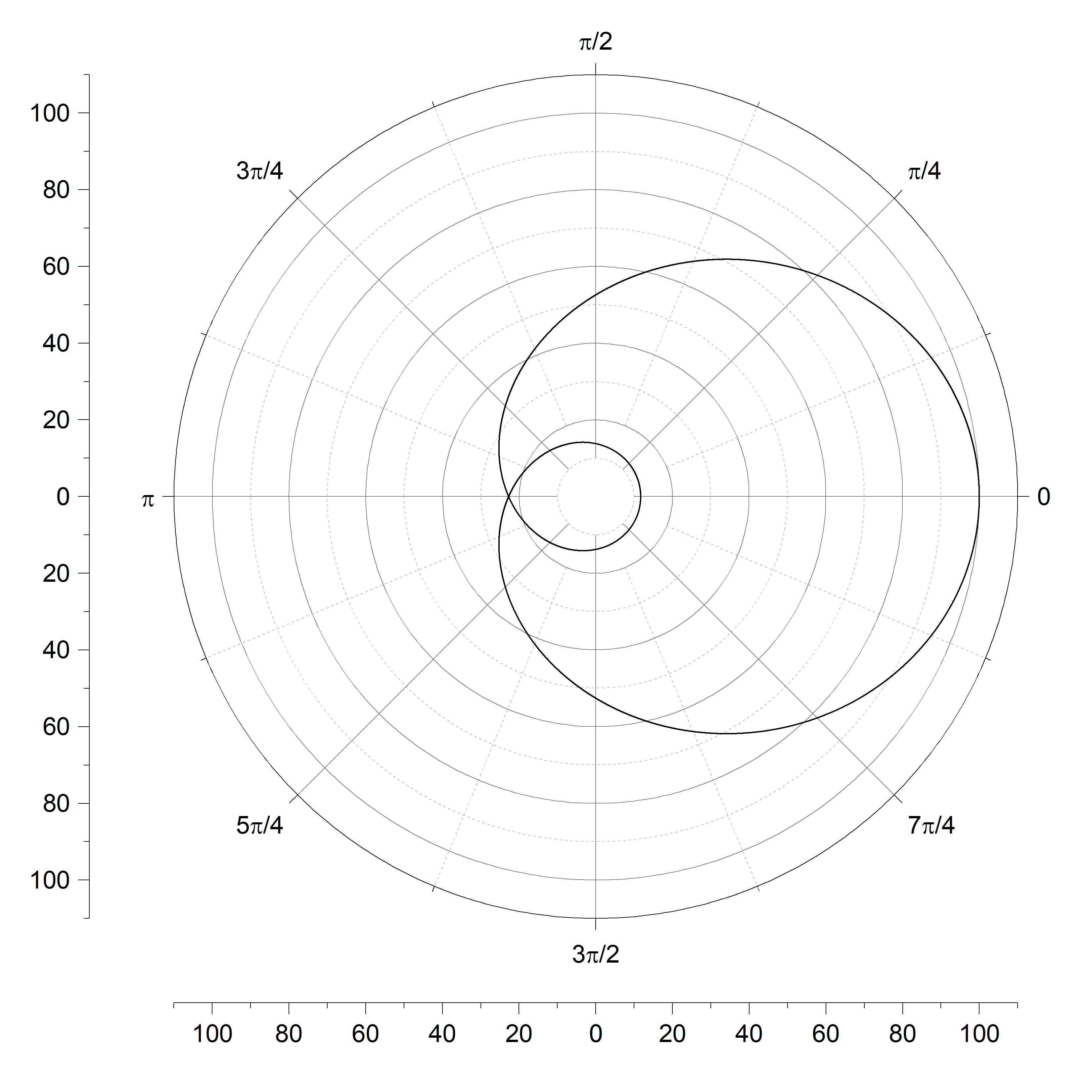}
         \caption{$(25,0.75,1,2,0)$}
     \end{subfigure}%
     \begin{subfigure}{0.25\textwidth}
         \centering
         \includegraphics[width=\textwidth]{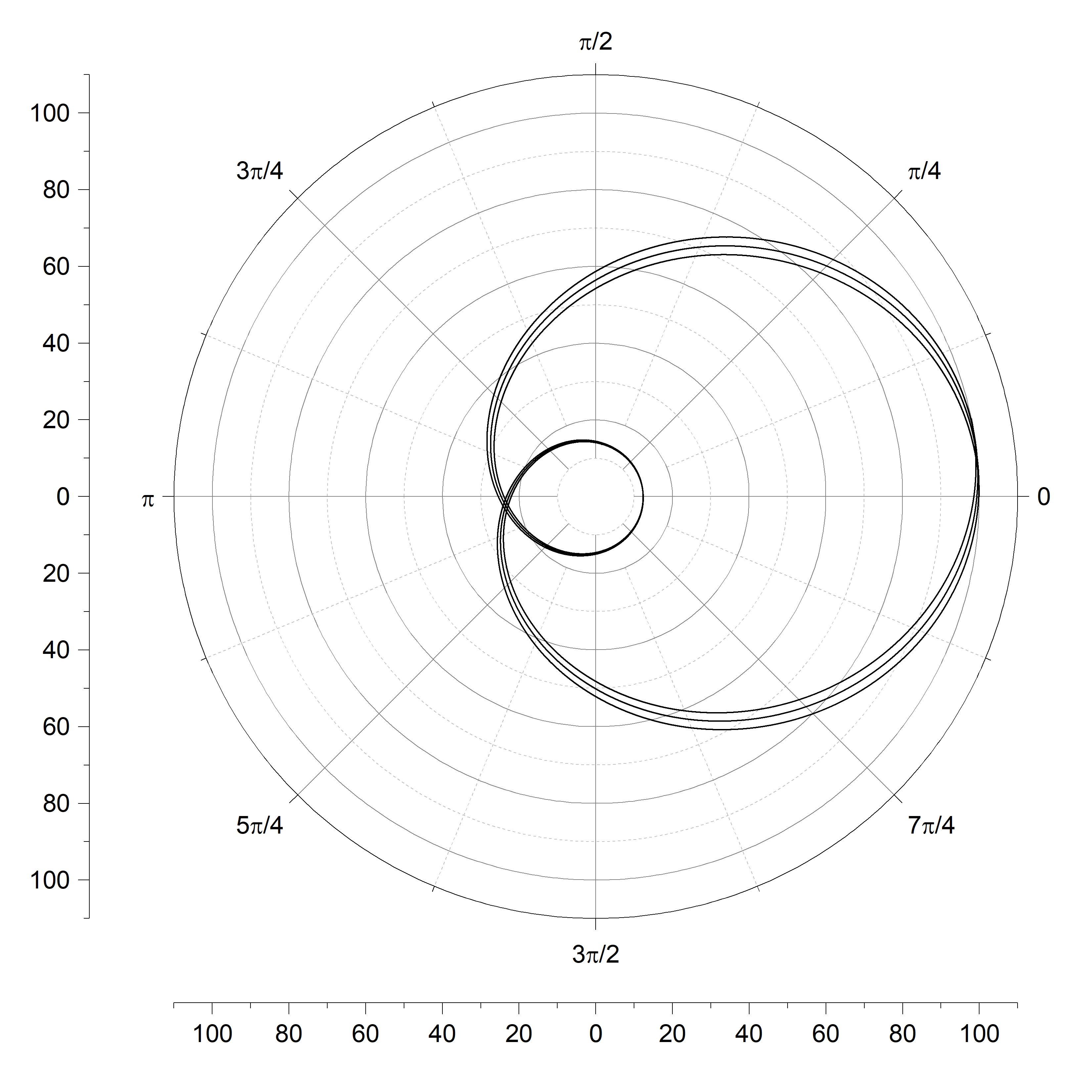}
         \caption{$(26.2,0.738,1,2,0.005392)$}
     \end{subfigure}
          \centering
     \begin{subfigure}{0.25\textwidth}
         \centering
         \includegraphics[width=\textwidth]{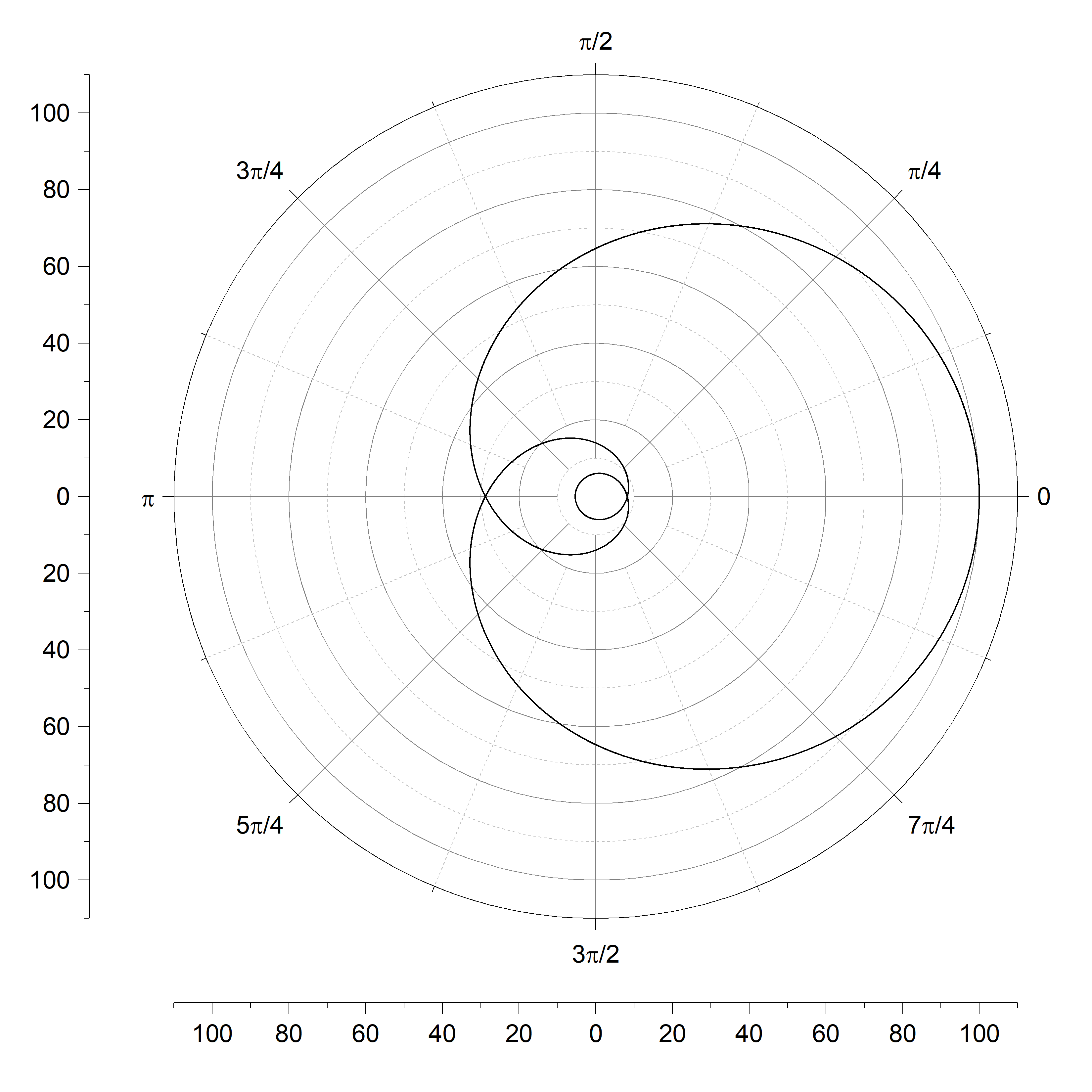}
         \caption{$(20,0.8,1,3,0)$}
     \end{subfigure}%
     \begin{subfigure}{0.25\textwidth}
         \centering
         \includegraphics[width=\textwidth]{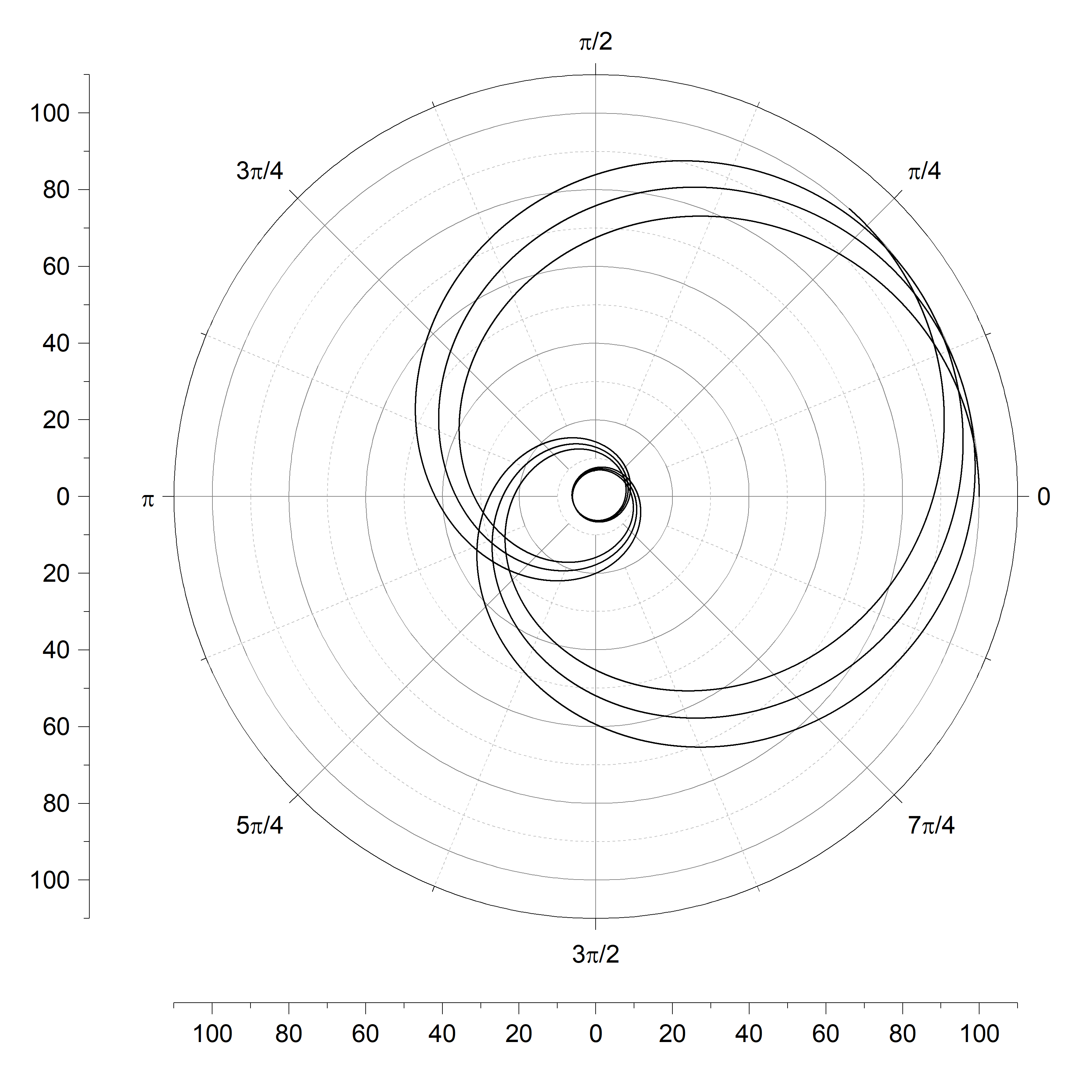}
         \caption{$(22,0.78,1,3,0.014784)$}
     \end{subfigure}%
          \begin{subfigure}{0.25\textwidth}
         \centering
         \includegraphics[width=\textwidth]{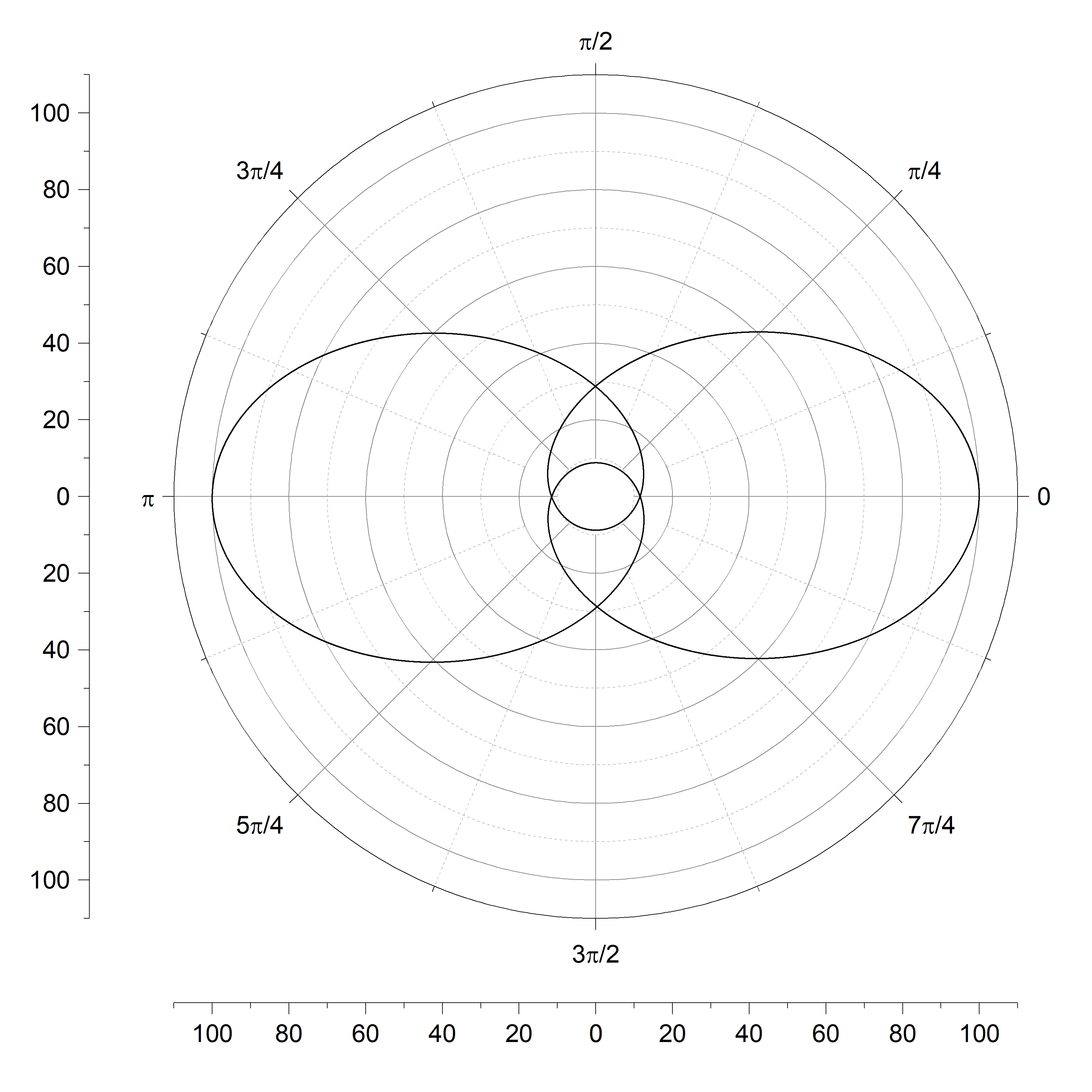}
         \caption{$(17.5,0.825,2,3,0)$}
     \end{subfigure}%
     \begin{subfigure}{0.25\textwidth}
         \centering
         \includegraphics[width=\textwidth]{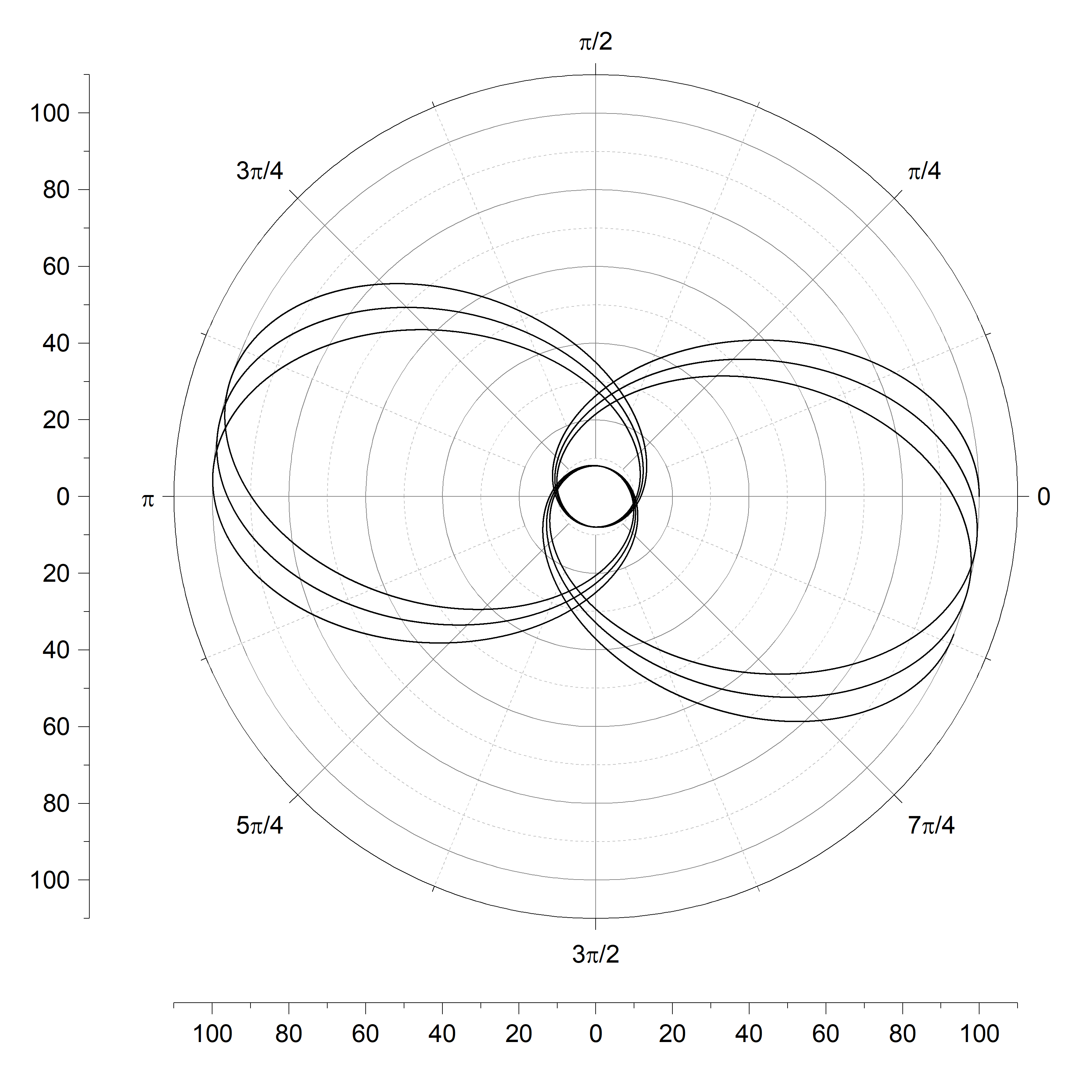}
         \caption{$(15.8,0.842,2,3, -0.013078)$}
     \end{subfigure}
          \centering
     \begin{subfigure}{0.25\textwidth}
         \centering
         \includegraphics[width=\textwidth]{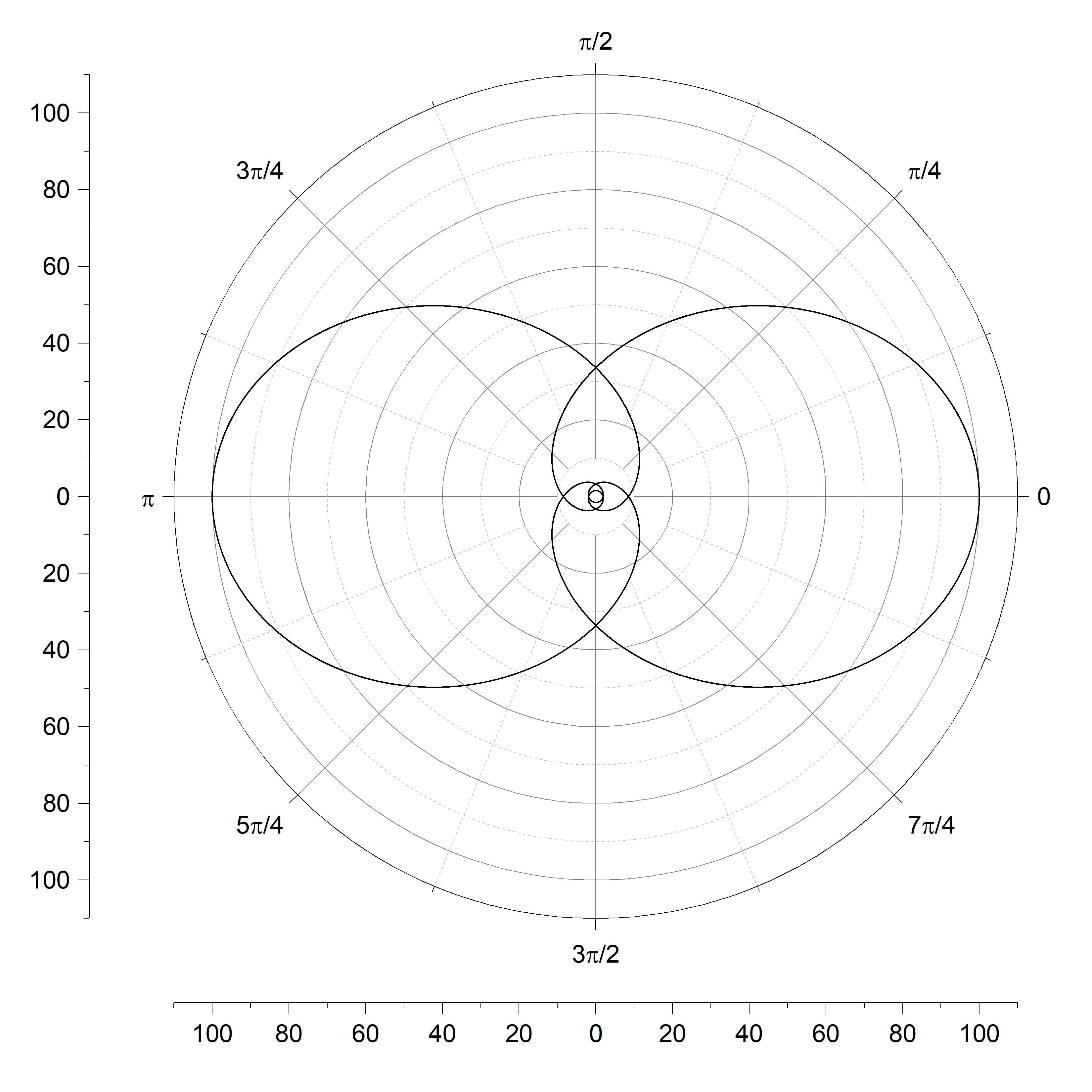}
         \caption{$(10.3,0.897,2,5,0)$}
     \end{subfigure}%
     \begin{subfigure}{0.25\textwidth}
         \centering
         \includegraphics[width=\textwidth]{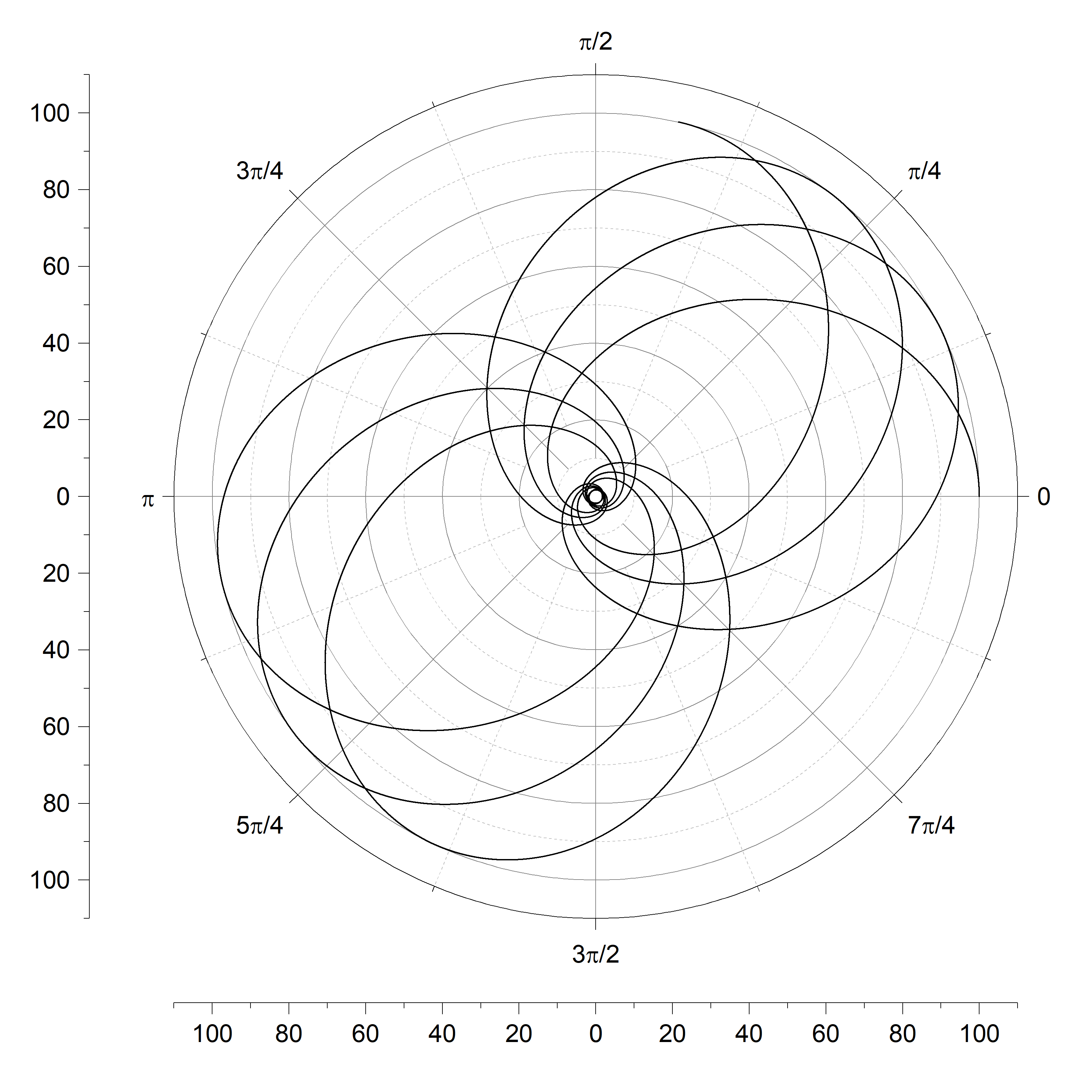}
         \caption{$(10.5,0.895,2,5,0.028316)$}
     \end{subfigure}%
          \begin{subfigure}{0.25\textwidth}
         \centering
         \includegraphics[width=\textwidth]{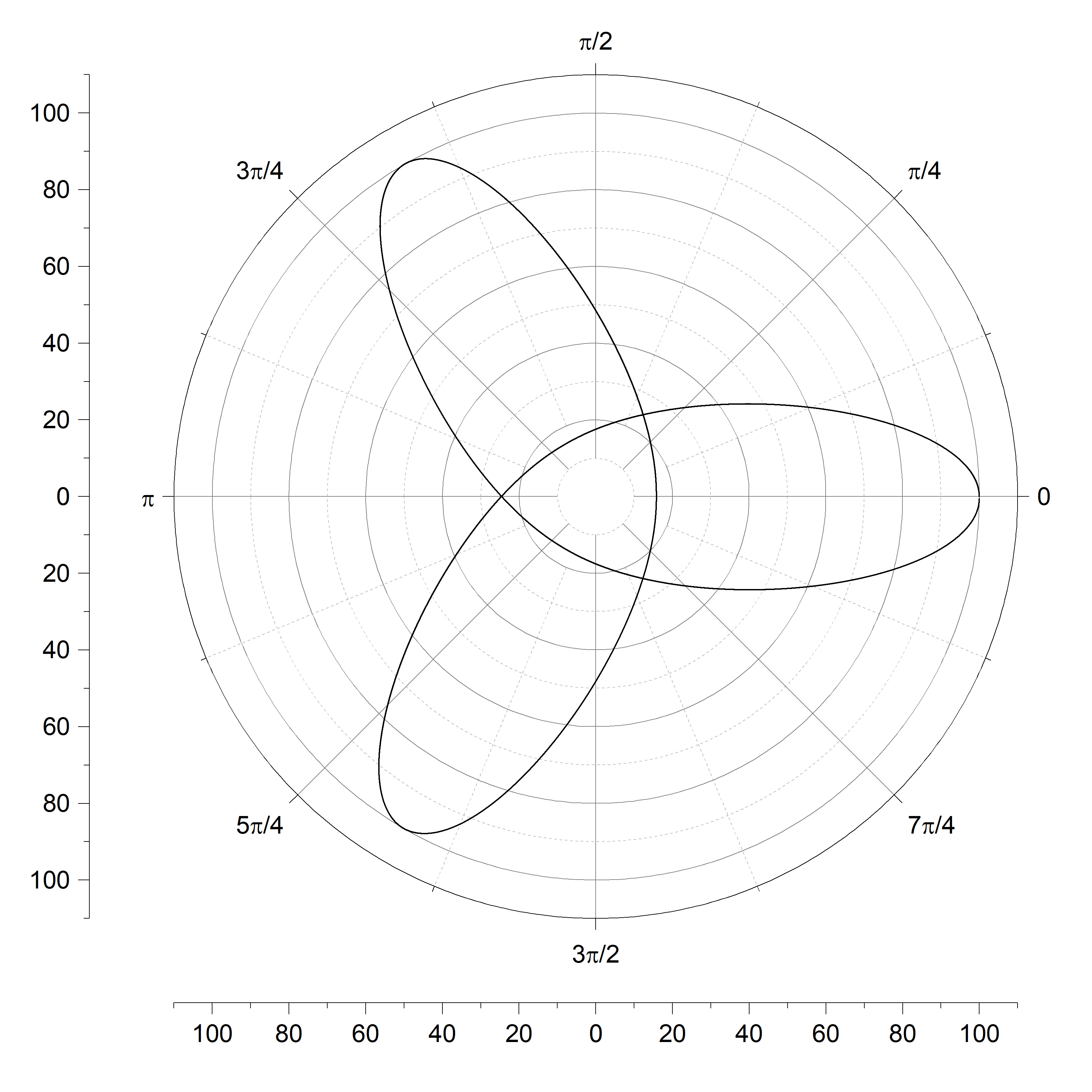}
         \caption{$(20.3,0.797,3,2,0)$}
     \end{subfigure}%
     \begin{subfigure}{0.25\textwidth}
         \centering
         \includegraphics[width=\textwidth]{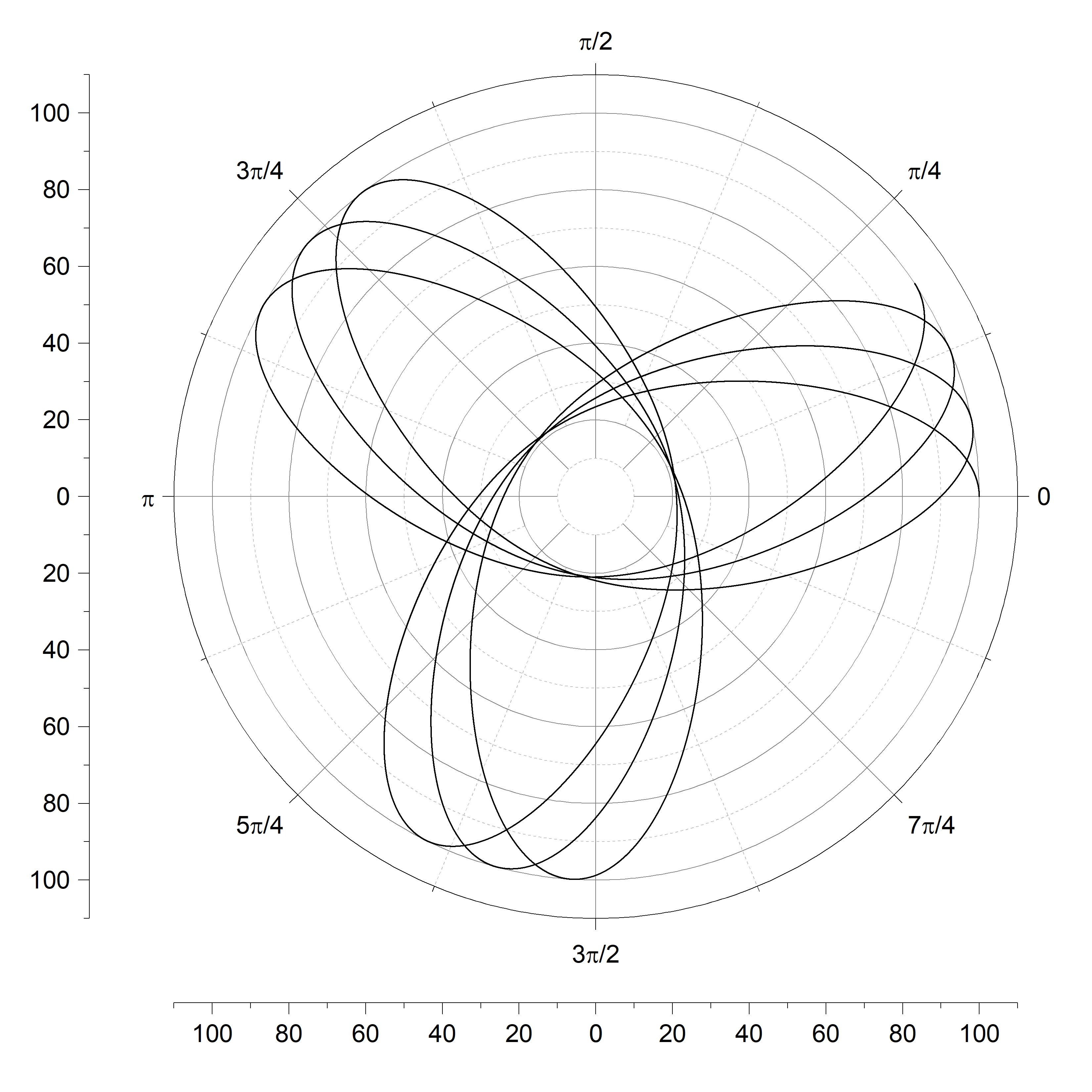}
         \caption{$(28.3,0.717,3,2,0.046172)$}
     \end{subfigure}
          \centering
     \begin{subfigure}{0.25\textwidth}
         \centering
         \includegraphics[width=\textwidth]{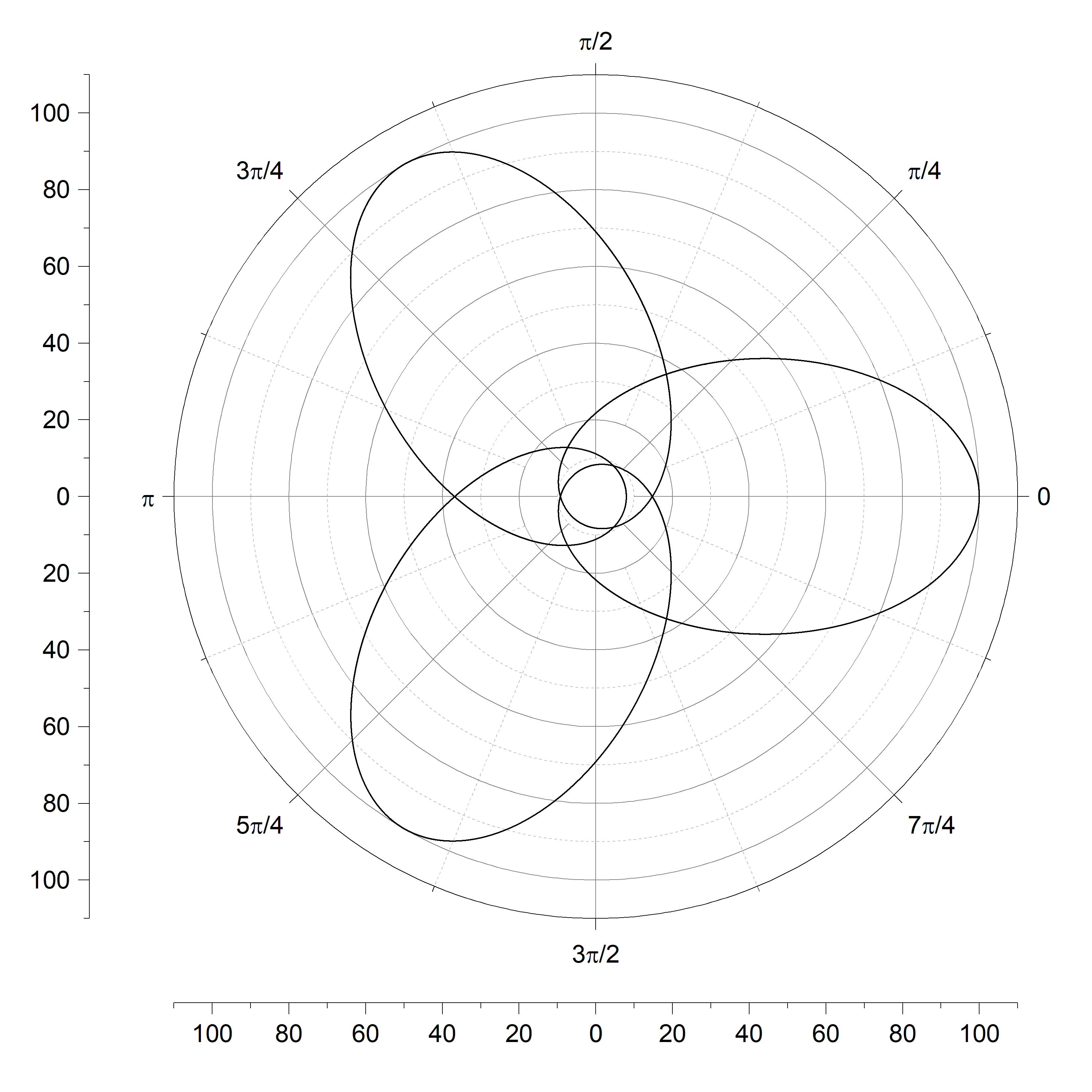}
         \caption{$(14.5,0.855,3,4,0)$}
     \end{subfigure}%
     \begin{subfigure}{0.25\textwidth}
         \centering
         \includegraphics[width=\textwidth]{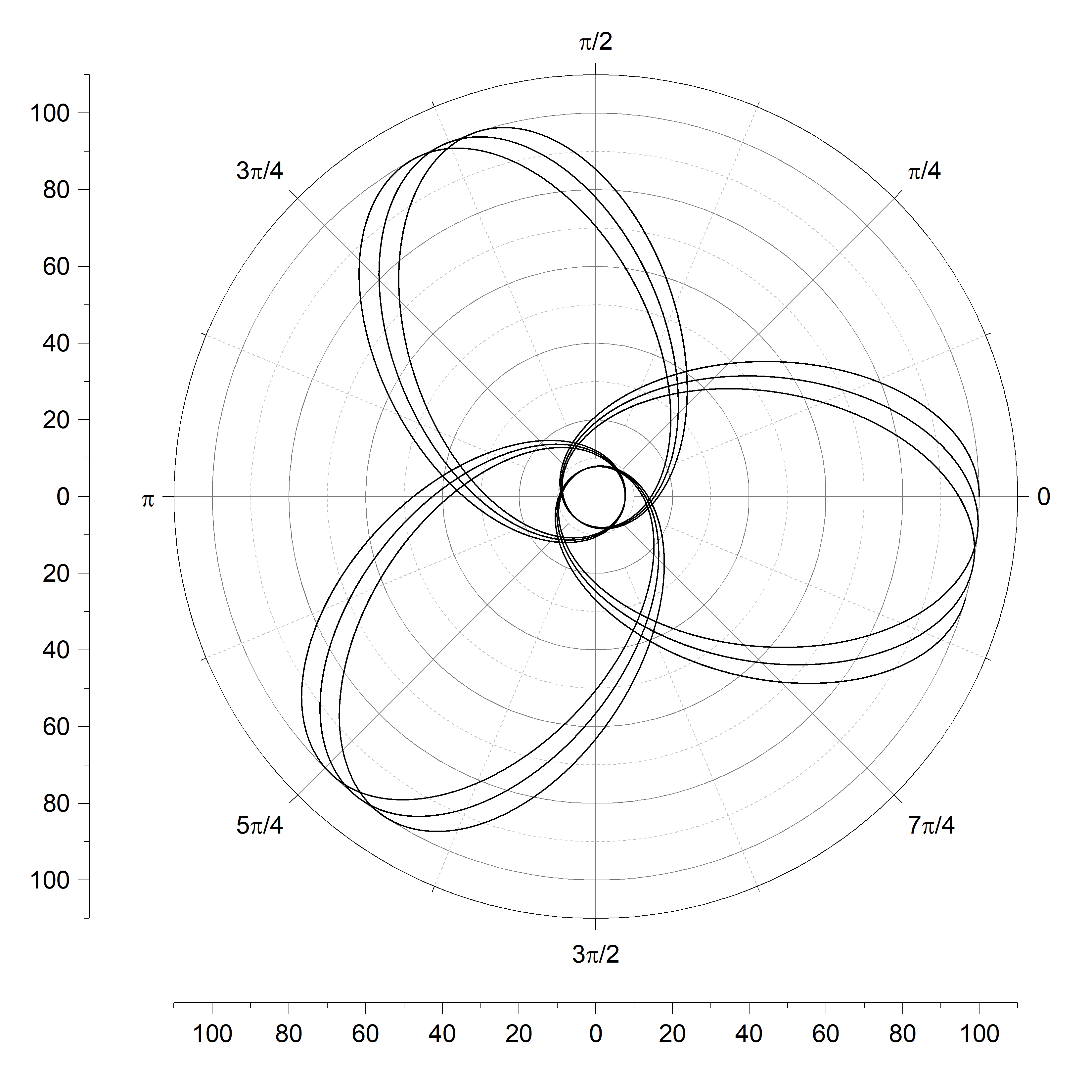}
         \caption{$(14,0.86,3,4,-0.010714)$}
     \end{subfigure}%
          \begin{subfigure}{0.25\textwidth}
         \centering
         \includegraphics[width=\textwidth]{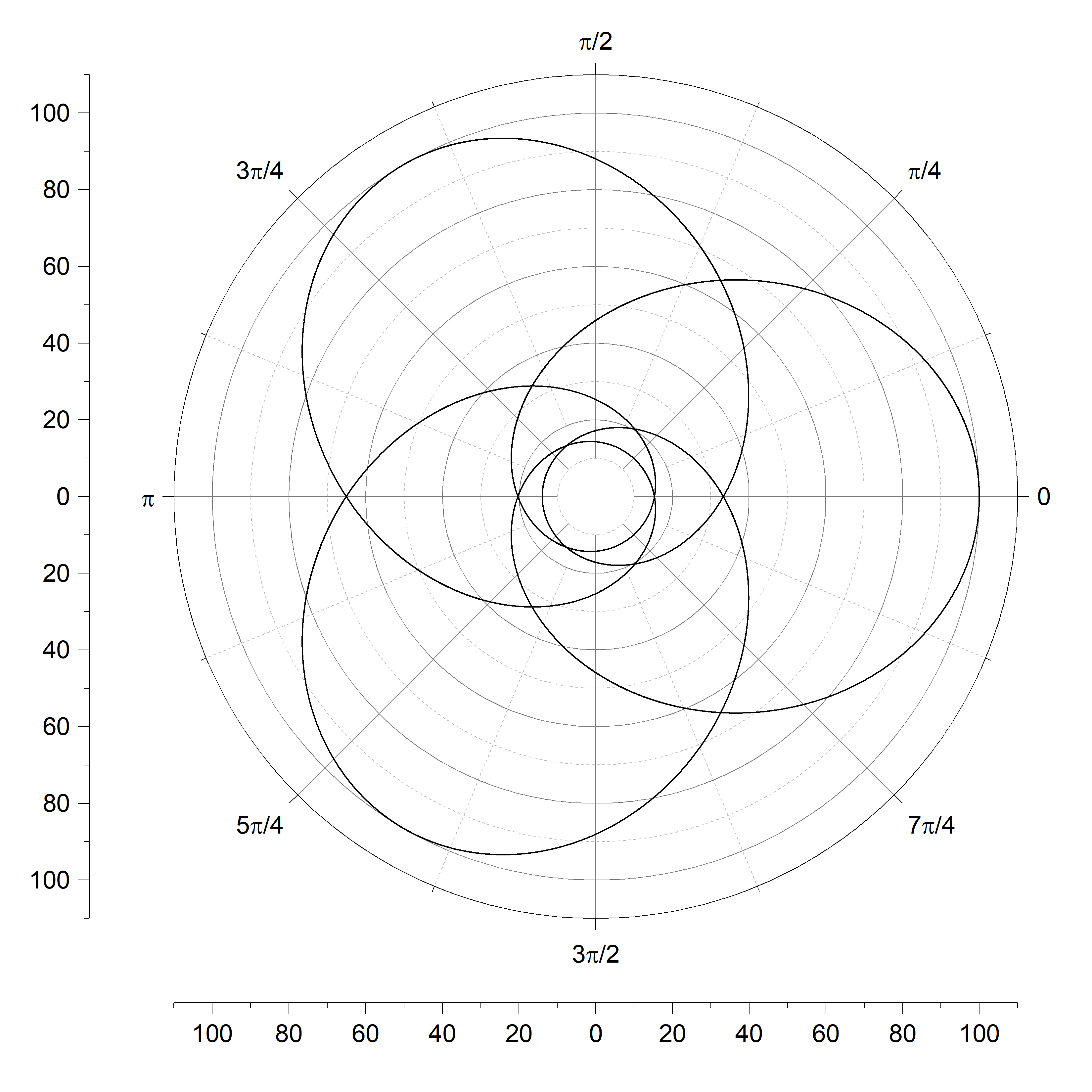}
         \caption{$(26.5,0.735,3,5,0)$}
     \end{subfigure}%
     \begin{subfigure}{0.25\textwidth}
         \centering
         \includegraphics[width=\textwidth]{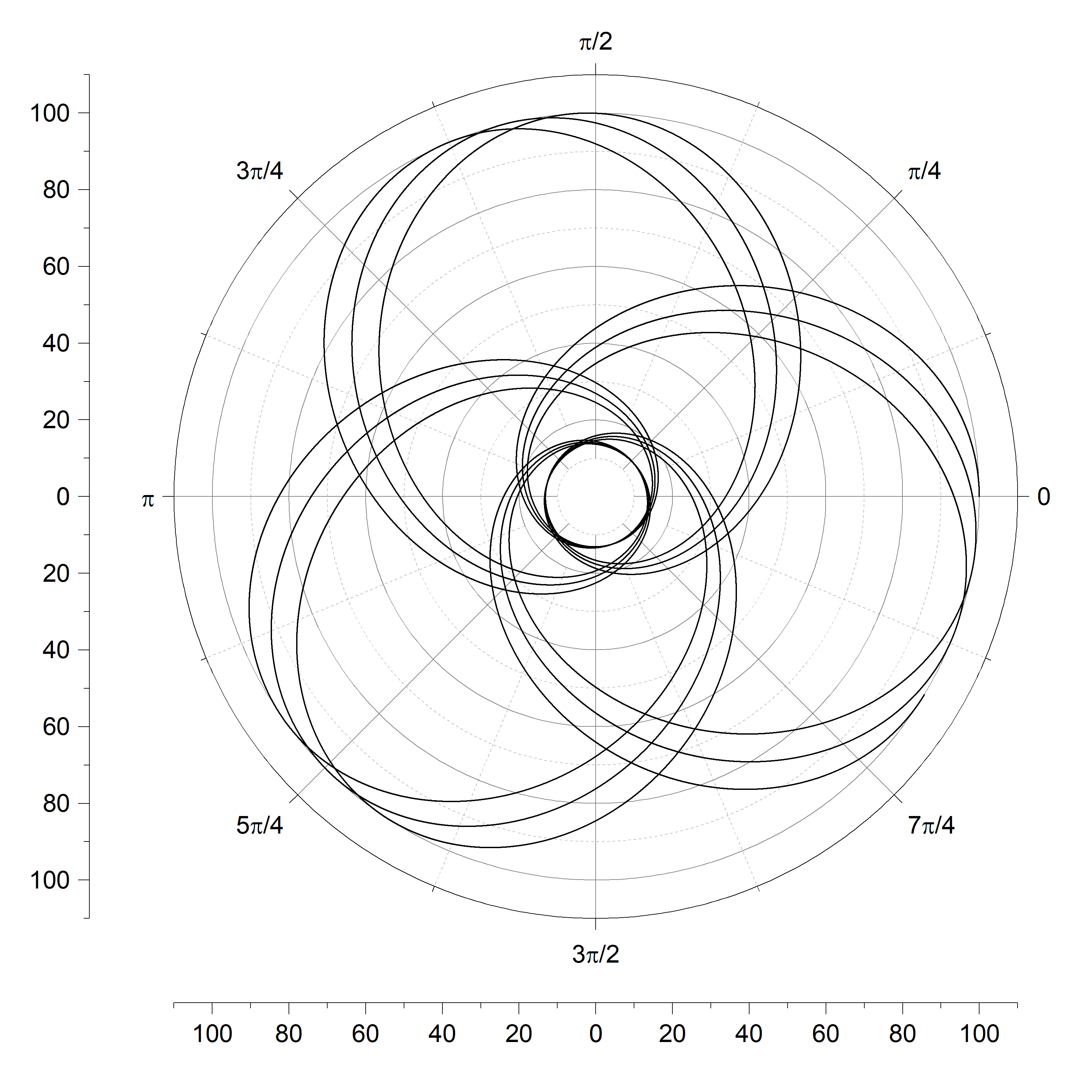}
         \caption{$(25,0.75,3,5,-0.01738)$}
     \end{subfigure}
     \caption{Examples of massive particle orbits in Levi-Civita spacetime I}
\end{figure*}

\begin{figure*}[htb]
     \centering
     \begin{subfigure}{0.25\textwidth}
         \centering
         \includegraphics[width=\textwidth]{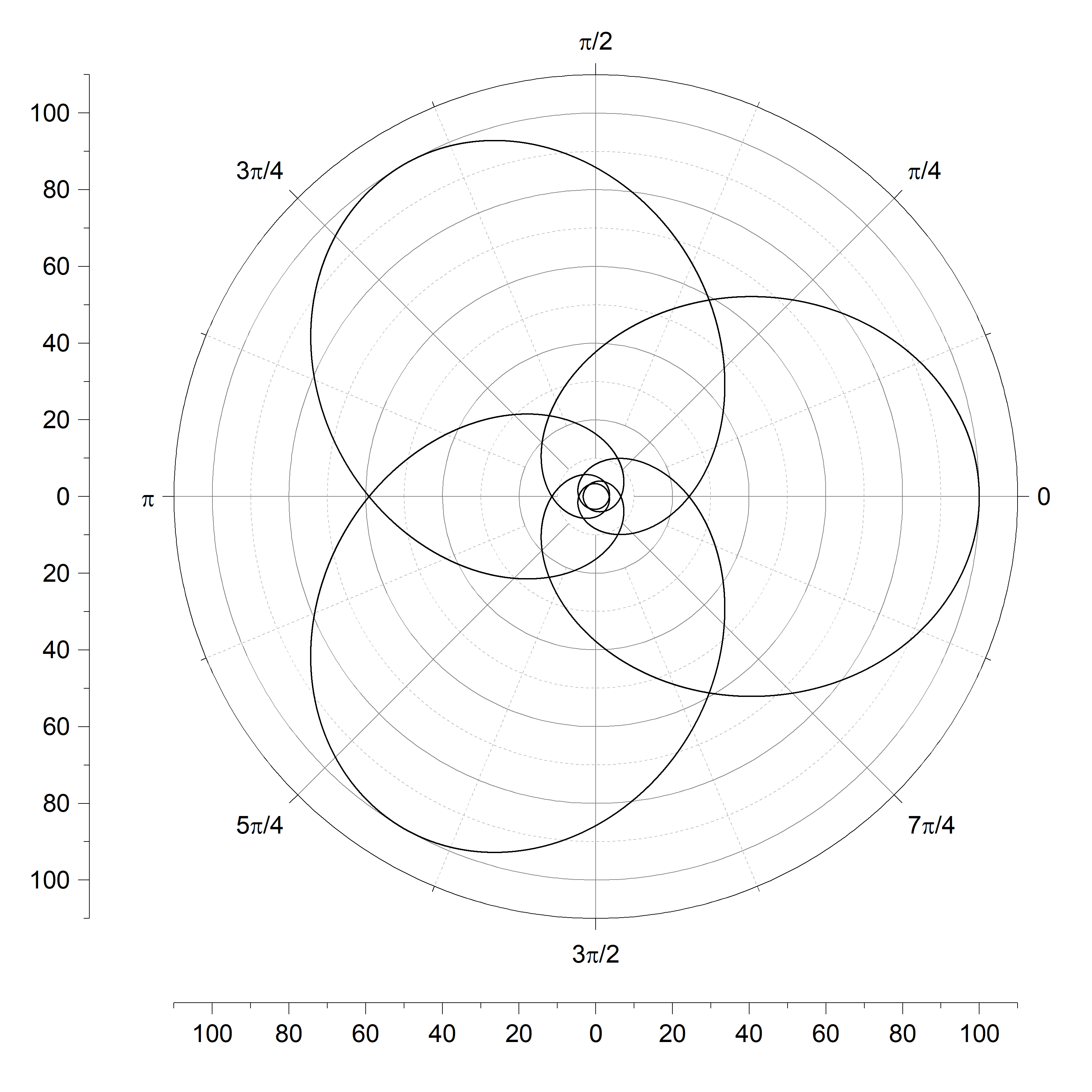}
         \caption{$(13.1,0.869,3,7,0)$}
     \end{subfigure}%
     \begin{subfigure}{0.25\textwidth}
         \centering
         \includegraphics[width=\textwidth]{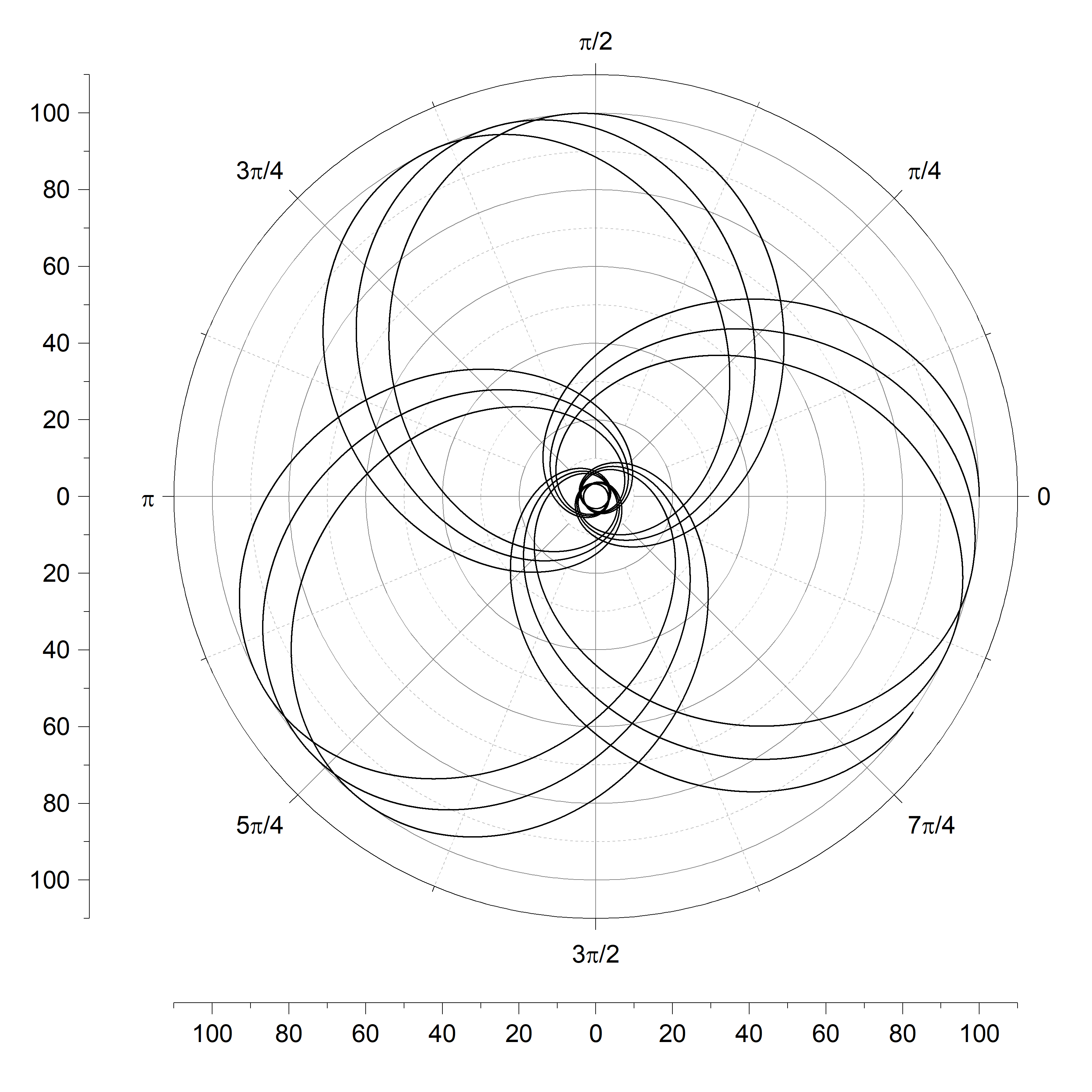}
         \caption{$(12.9,0.871,3,7,-0.01365)$}
     \end{subfigure}%
          \begin{subfigure}{0.25\textwidth}
         \centering
         \includegraphics[width=\textwidth]{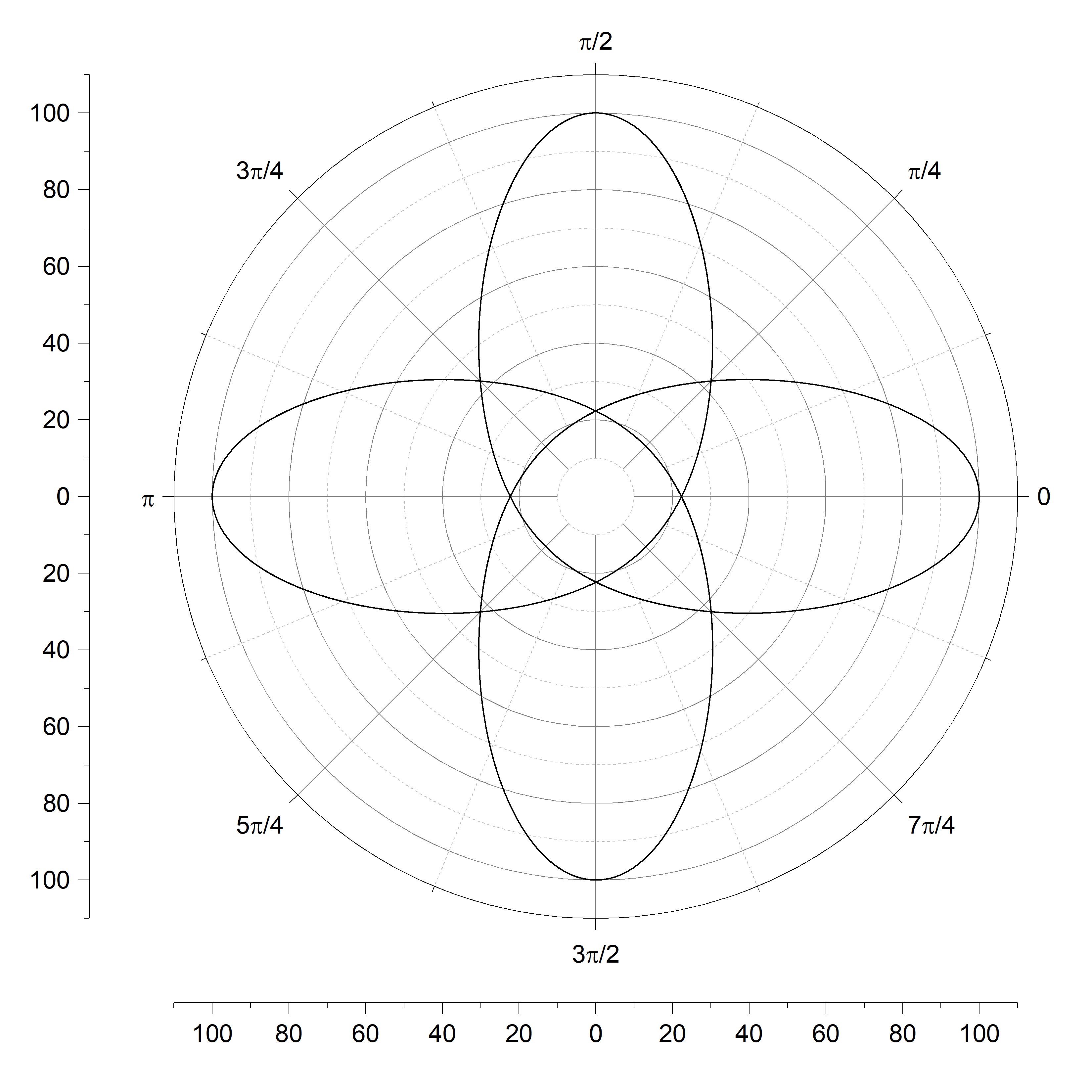}
         \caption{$(25.4,0.745,4,3,0)$}
     \end{subfigure}%
     \begin{subfigure}{0.25\textwidth}
         \centering
         \includegraphics[width=\textwidth]{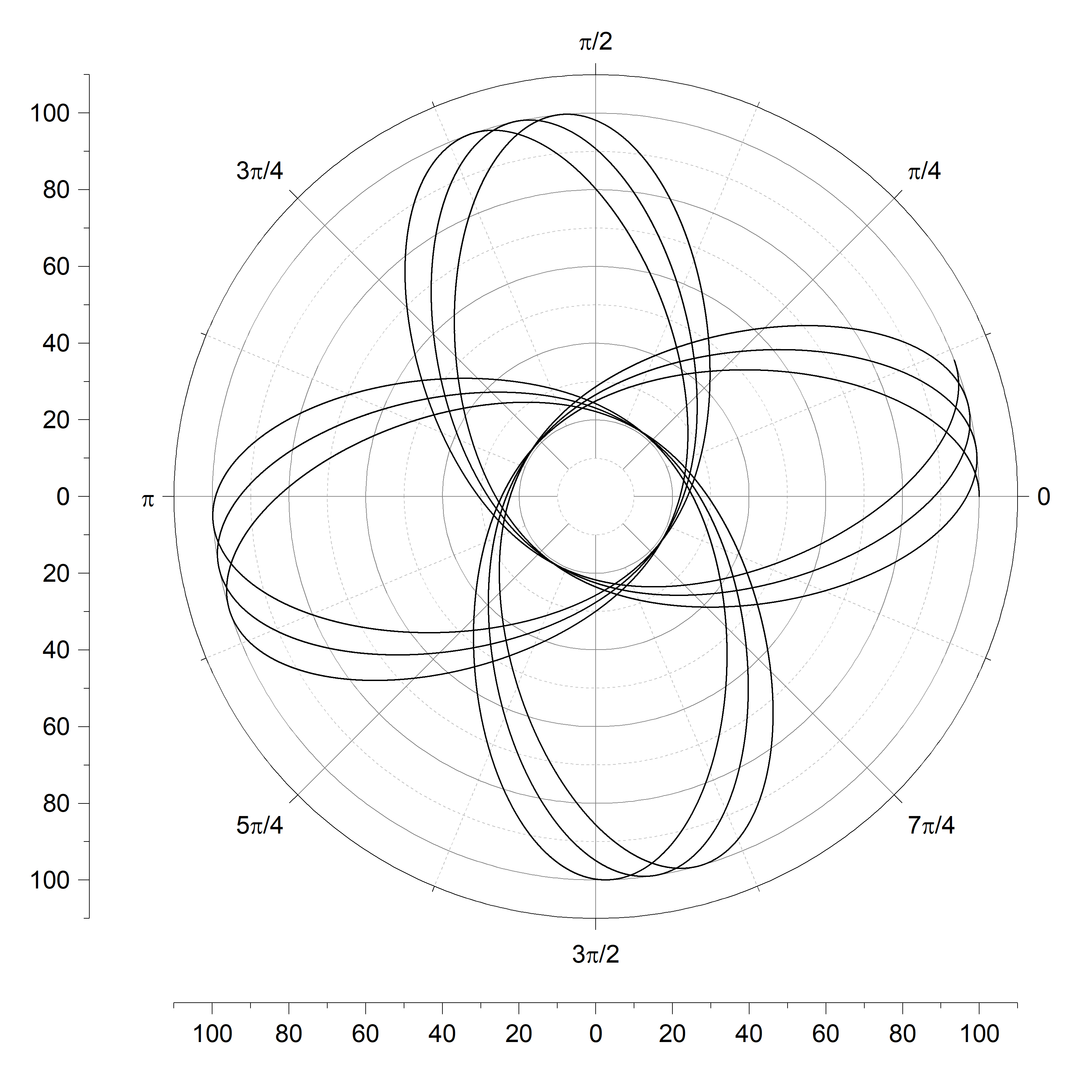}
         \caption{$(28.5,0.715,4,3,0.02556)$}
     \end{subfigure}
          \centering
     \begin{subfigure}{0.25\textwidth}
         \centering
         \includegraphics[width=\textwidth]{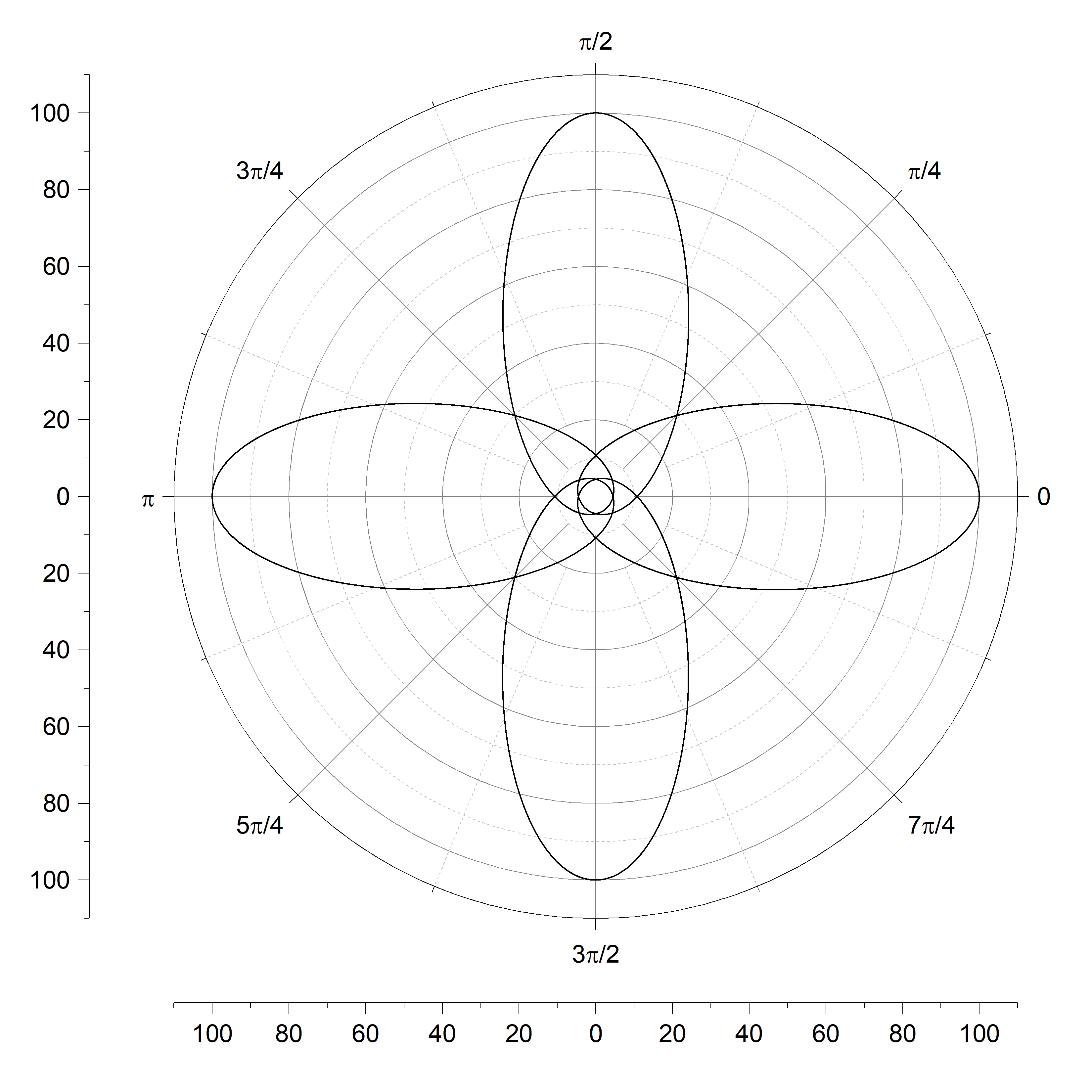}
         \caption{$(7.5,0.925,4,5,0)$}
     \end{subfigure}%
     \begin{subfigure}{0.25\textwidth}
         \centering
         \includegraphics[width=\textwidth]{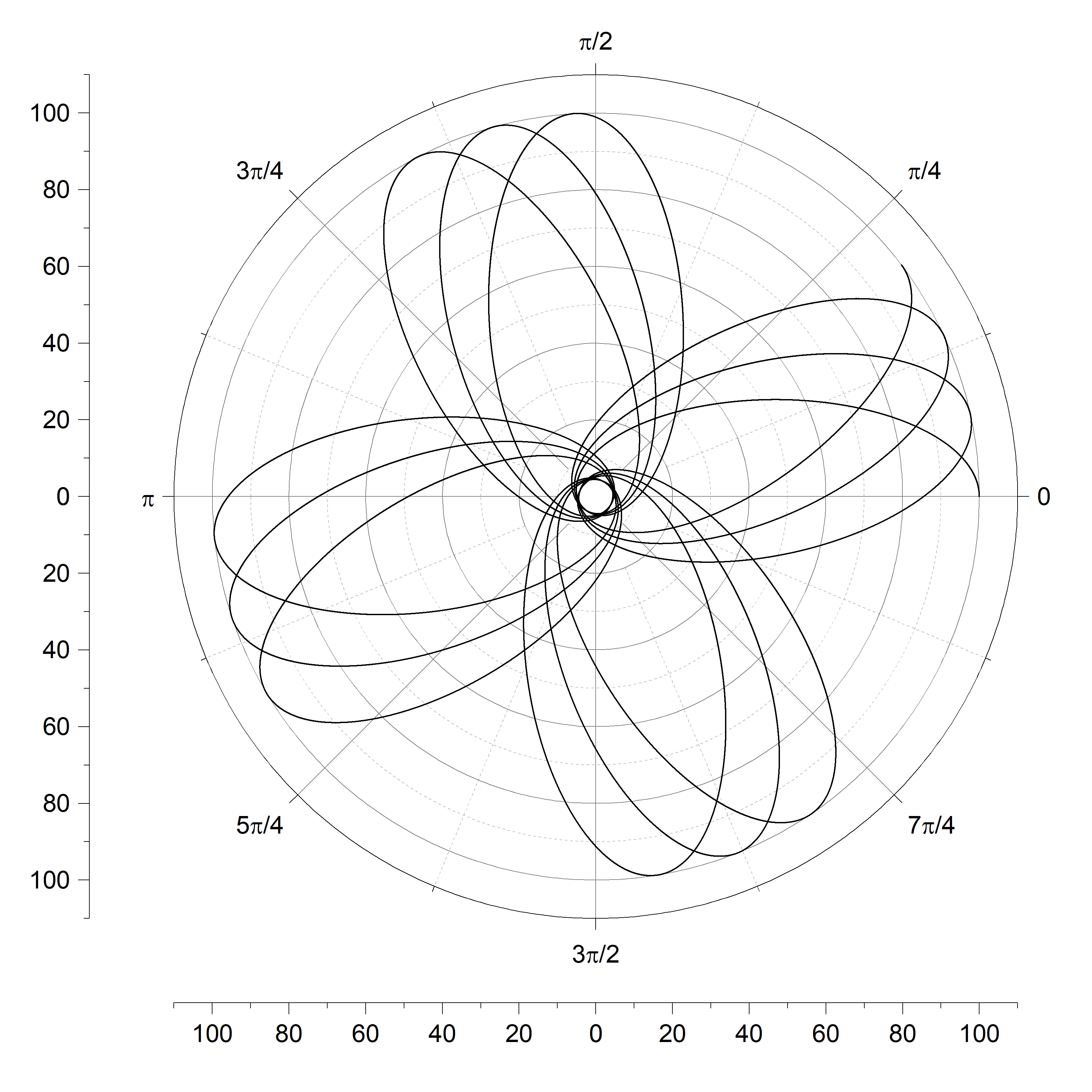}
         \caption{$(8,0.92,4,5,0.027323)$}
     \end{subfigure}%
          \begin{subfigure}{0.25\textwidth}
         \centering
         \includegraphics[width=\textwidth]{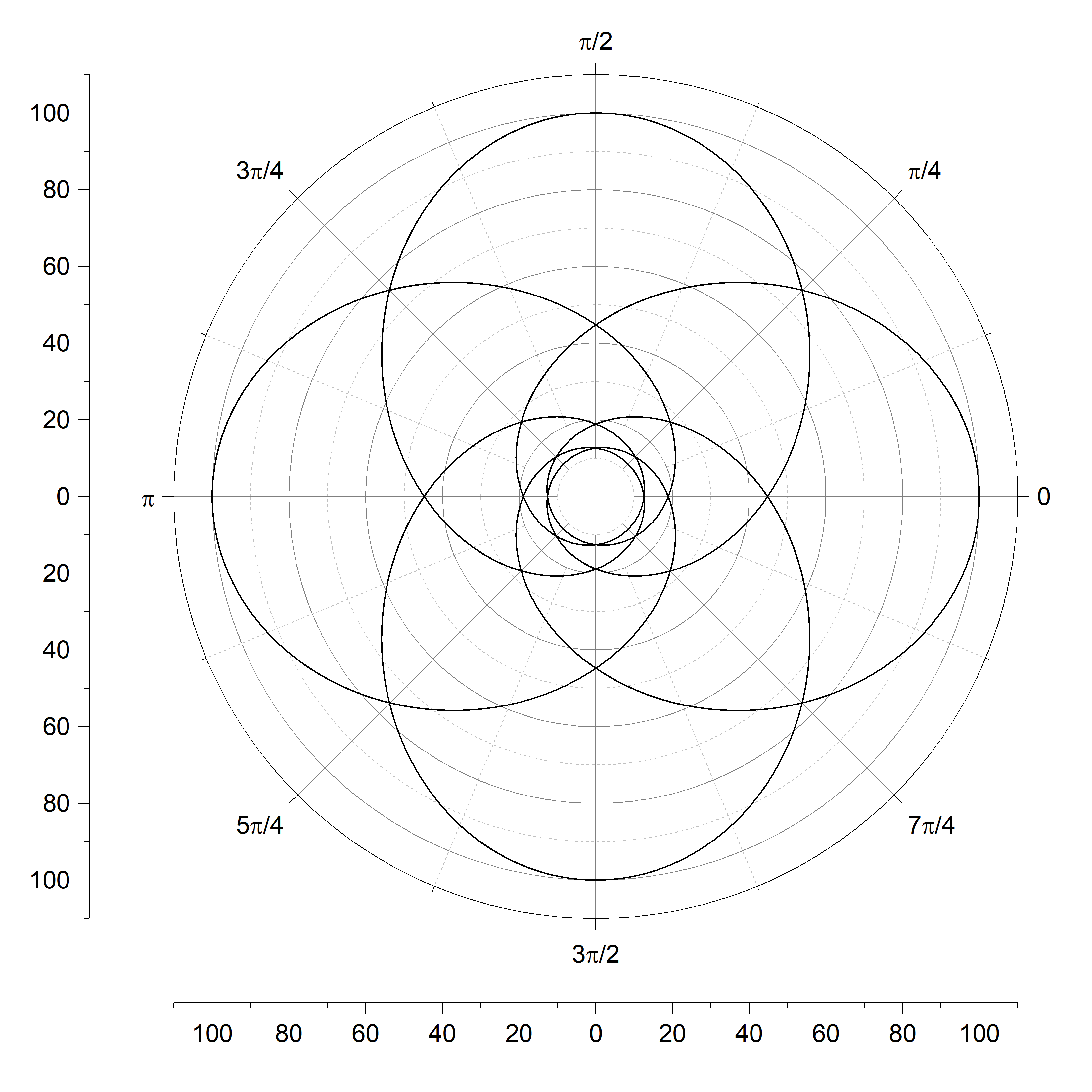}
         \caption{$(23.7,0.763,4,7,0)$}
     \end{subfigure}%
     \begin{subfigure}{0.25\textwidth}
         \centering
         \includegraphics[width=\textwidth]{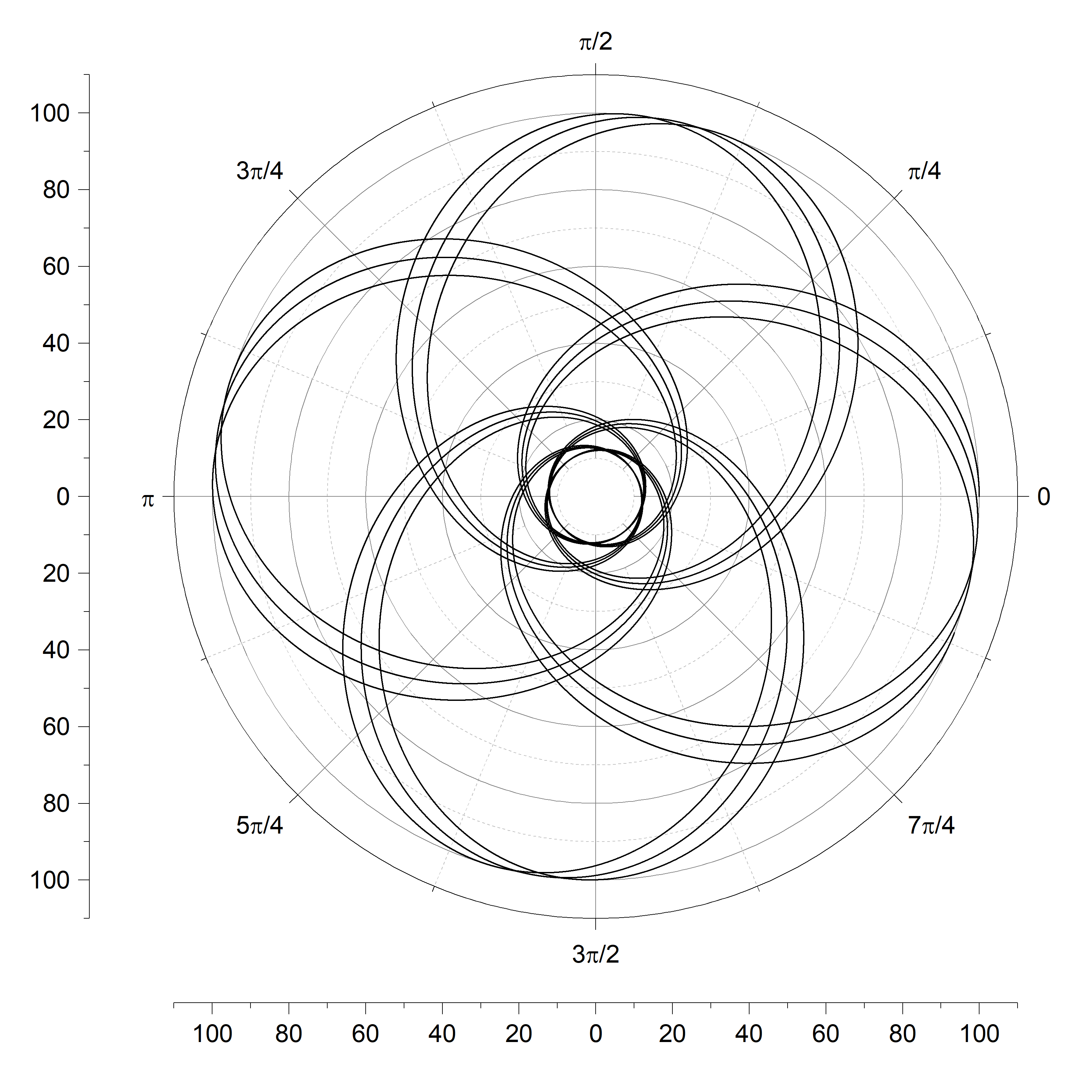}
         \caption{$(23.6,0.764,4,7,-0.01104)$}
     \end{subfigure}
          \centering
     \begin{subfigure}{0.25\textwidth}
         \centering
         \includegraphics[width=\textwidth]{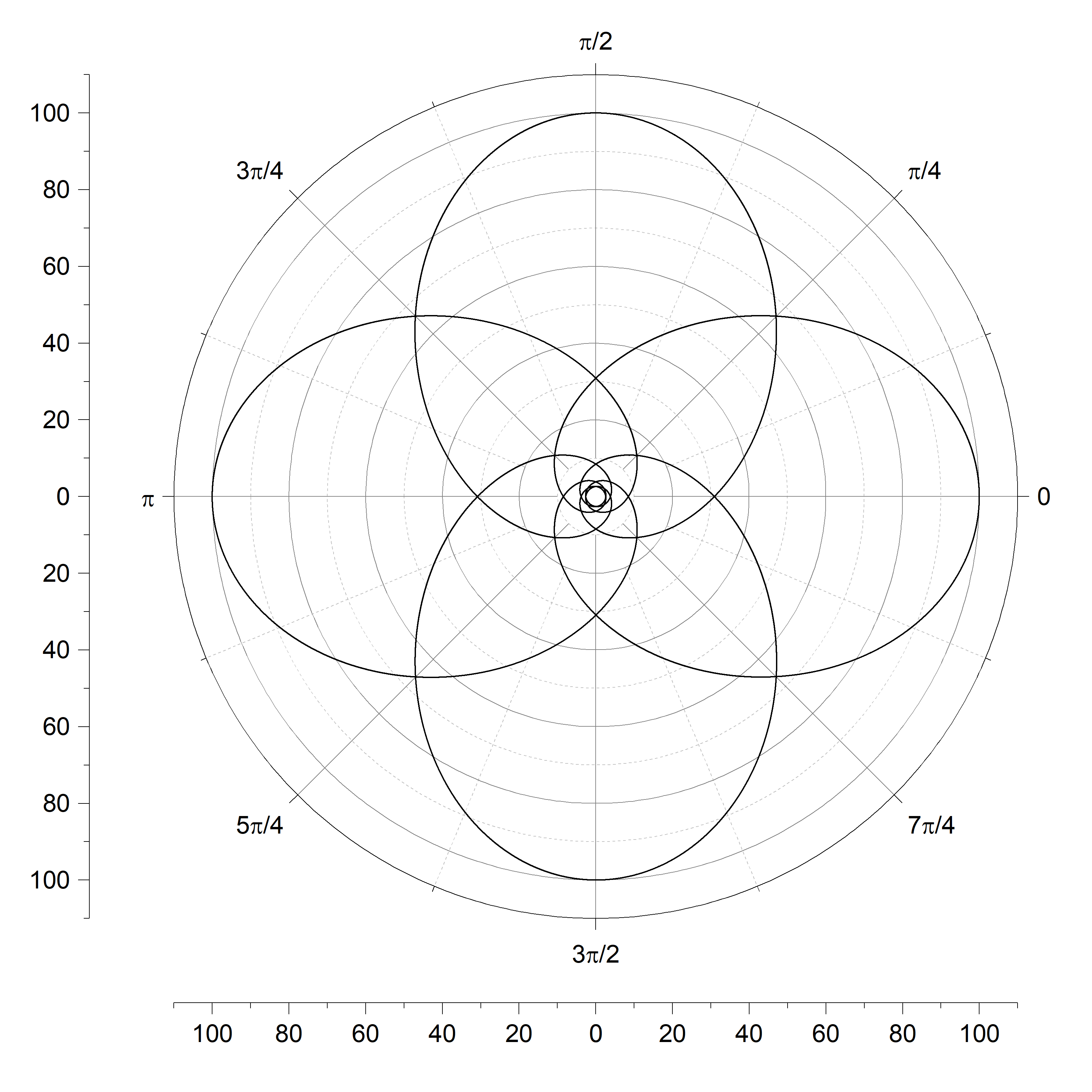}
         \caption{$(11,0.89,4,9,0)$}
     \end{subfigure}%
     \begin{subfigure}{0.25\textwidth}
         \centering
         \includegraphics[width=\textwidth]{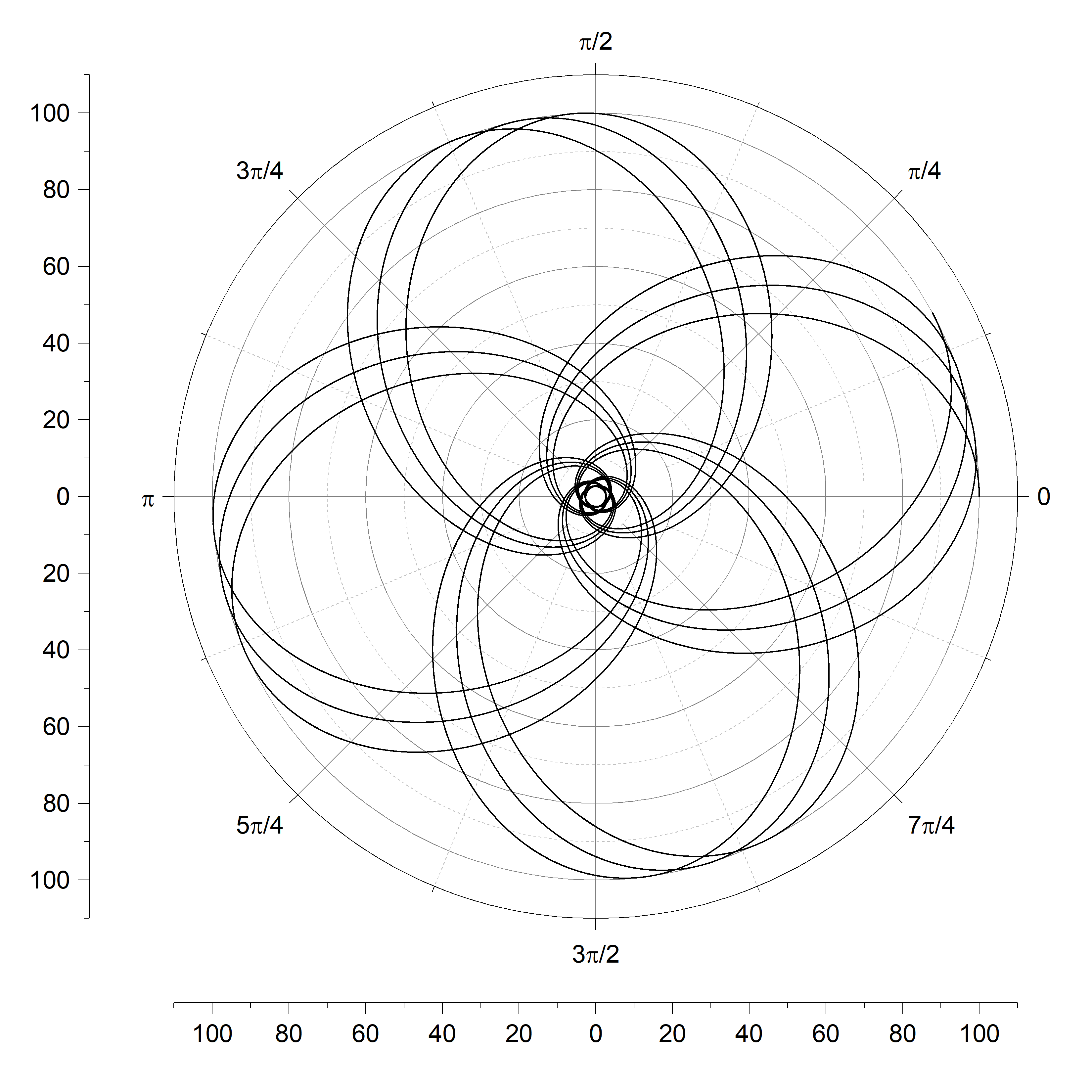}
         \caption{$(11.1,0.889,4,9,0.011752)$}
     \end{subfigure}%
          \begin{subfigure}{0.25\textwidth}
         \centering
         \includegraphics[width=\textwidth]{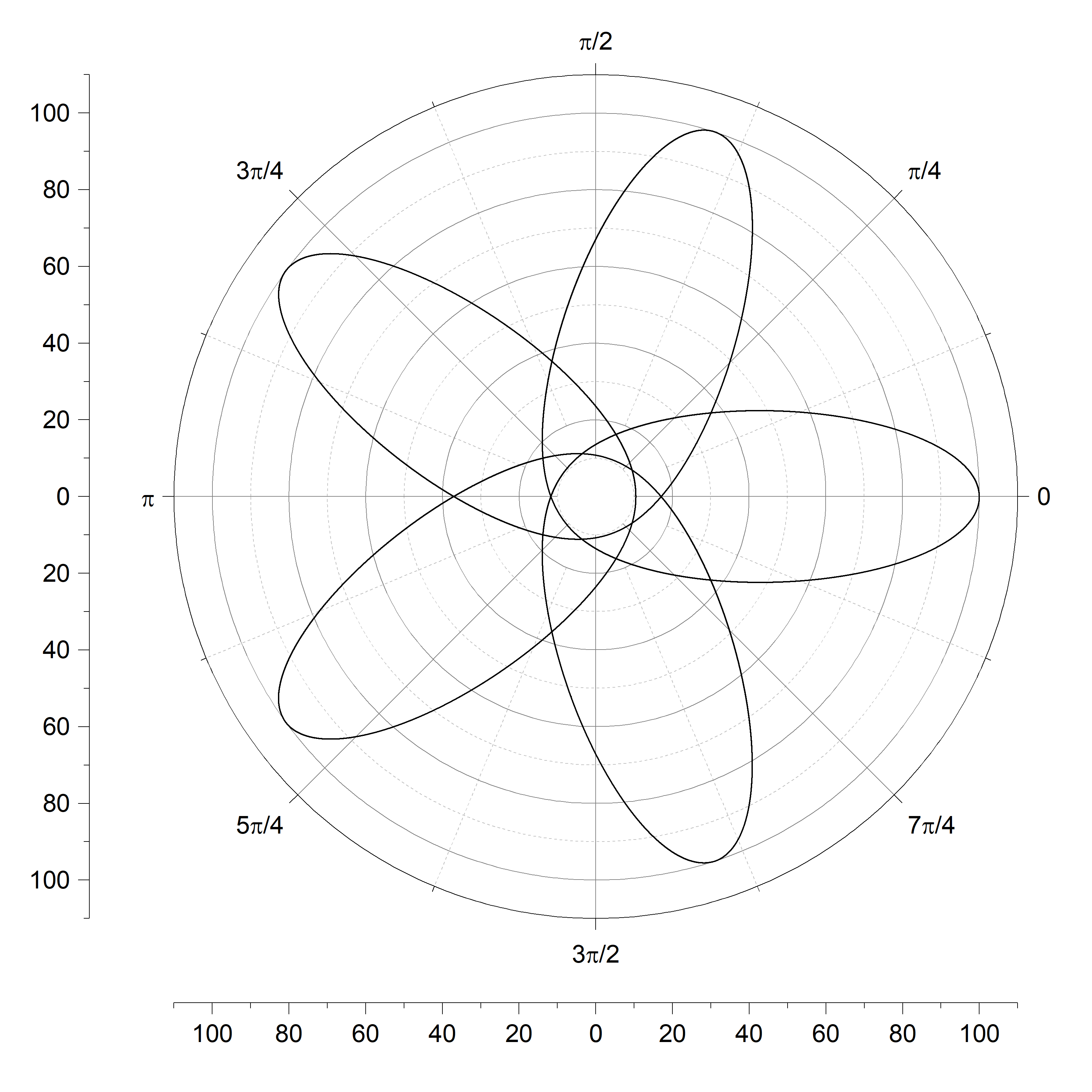}
         \caption{$(13.5,0.865,5,4,0)$}
     \end{subfigure}%
     \begin{subfigure}{0.25\textwidth}
         \centering
         \includegraphics[width=\textwidth]{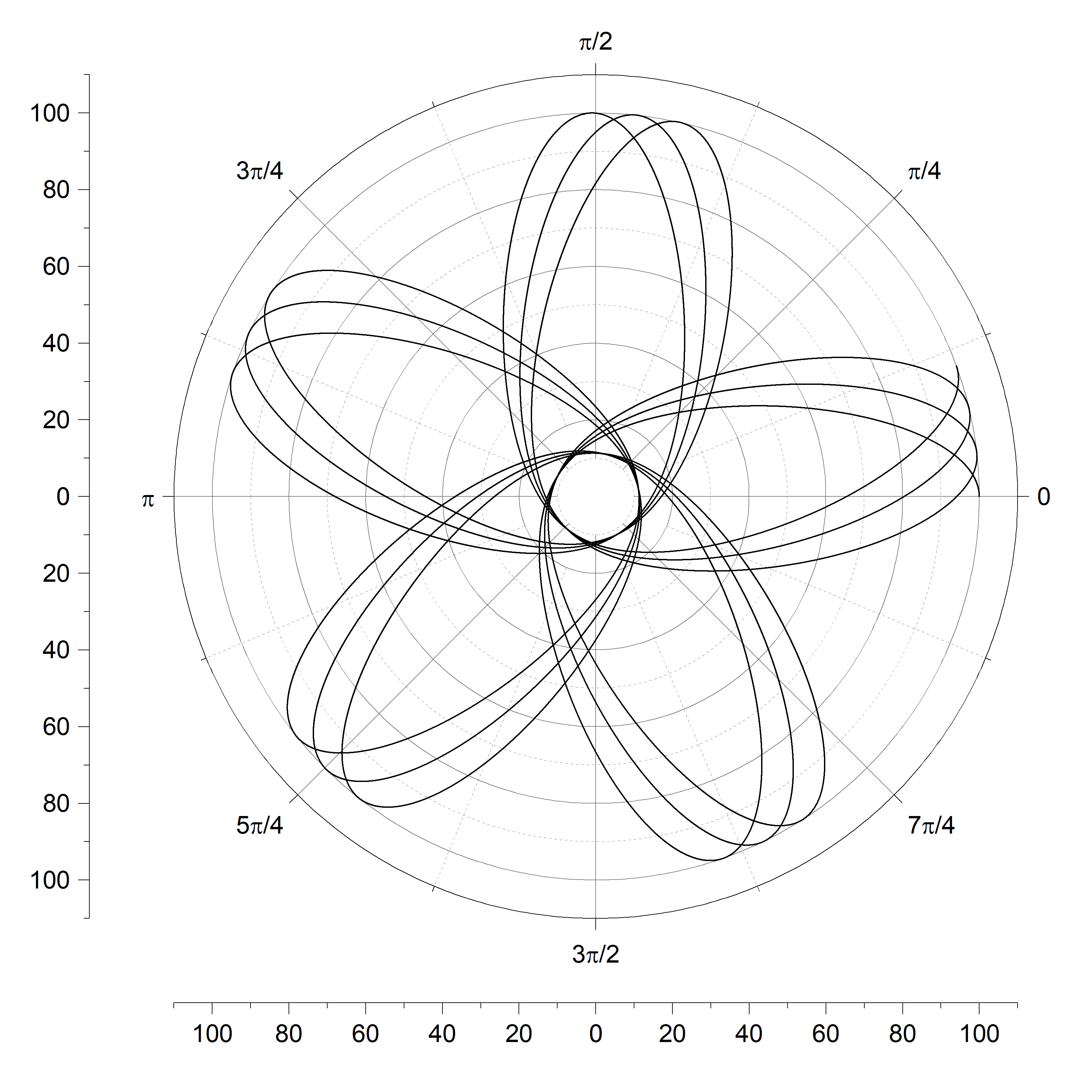}
         \caption{$(14.8,0.852,5,4,0.022983)$}
     \end{subfigure}
          \centering
    \begin{subfigure}{0.25\textwidth}
         \centering
         \includegraphics[width=\textwidth]{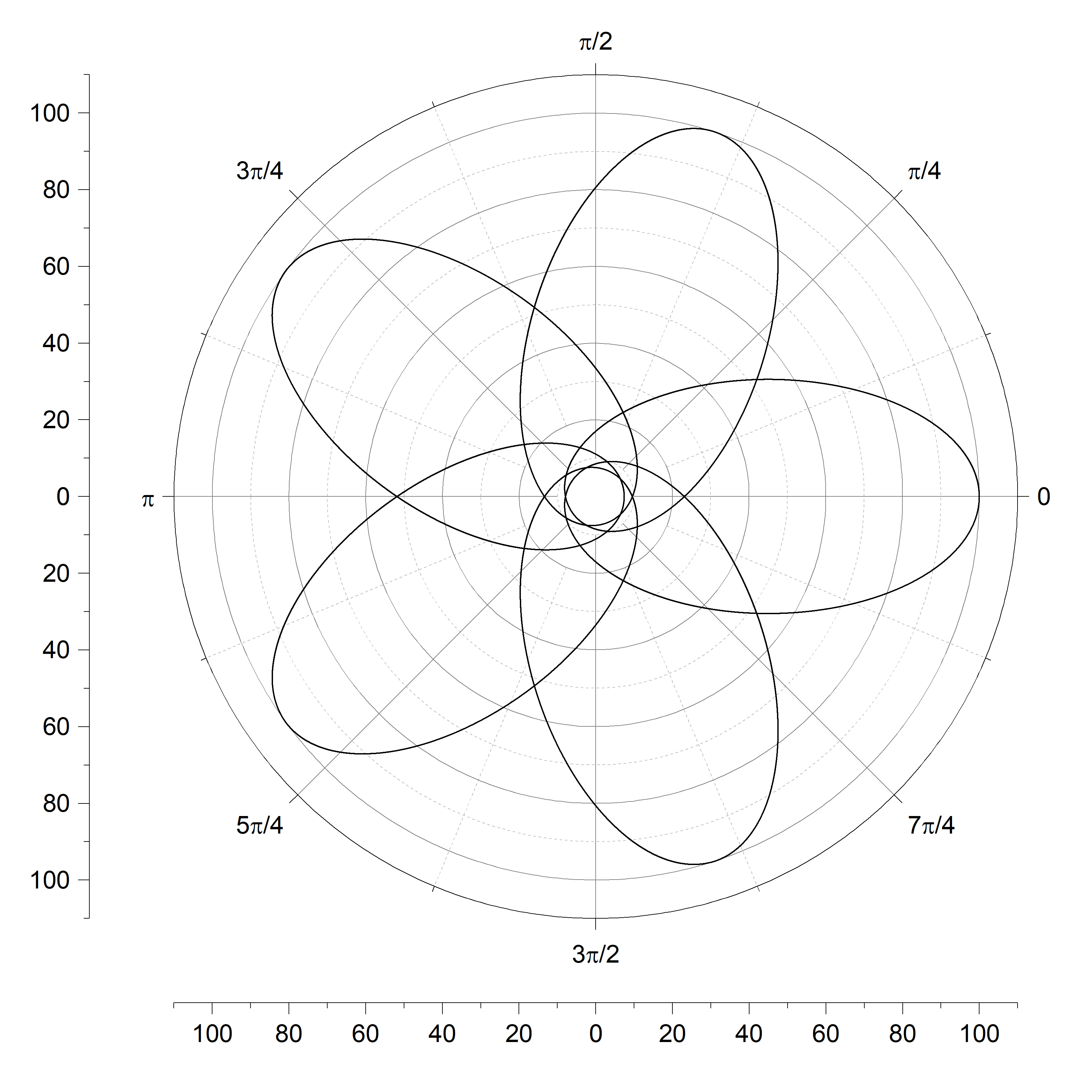}
         \caption{$(12.6,0.874,5,6,0)$}
     \end{subfigure}%
     \begin{subfigure}{0.25\textwidth}
         \centering
         \includegraphics[width=\textwidth]{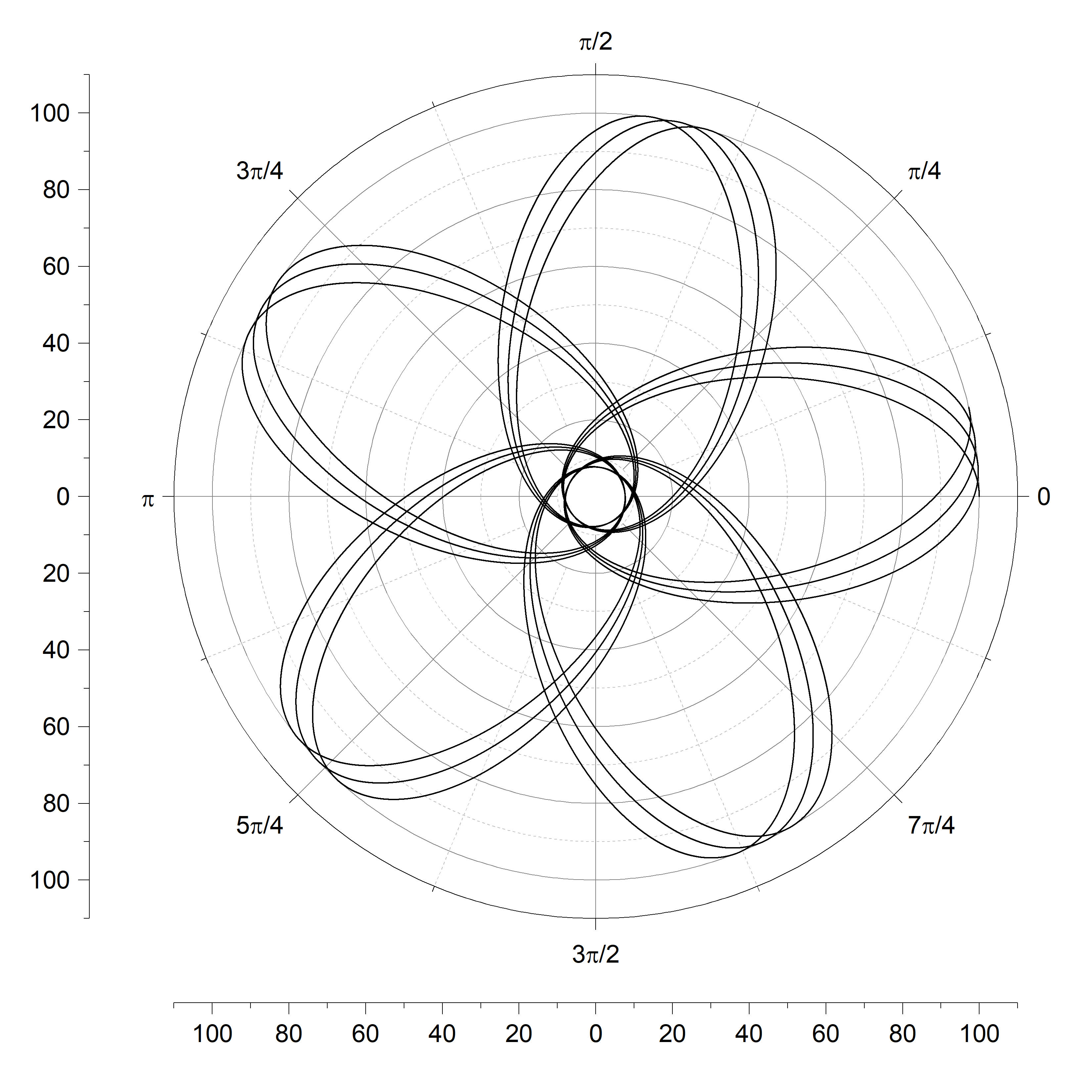}
         \caption{$(12.95,0.8705,5,6,0.01039)$}
     \end{subfigure}%
     \begin{subfigure}{0.25\textwidth}
         \centering
         \includegraphics[width=\textwidth]{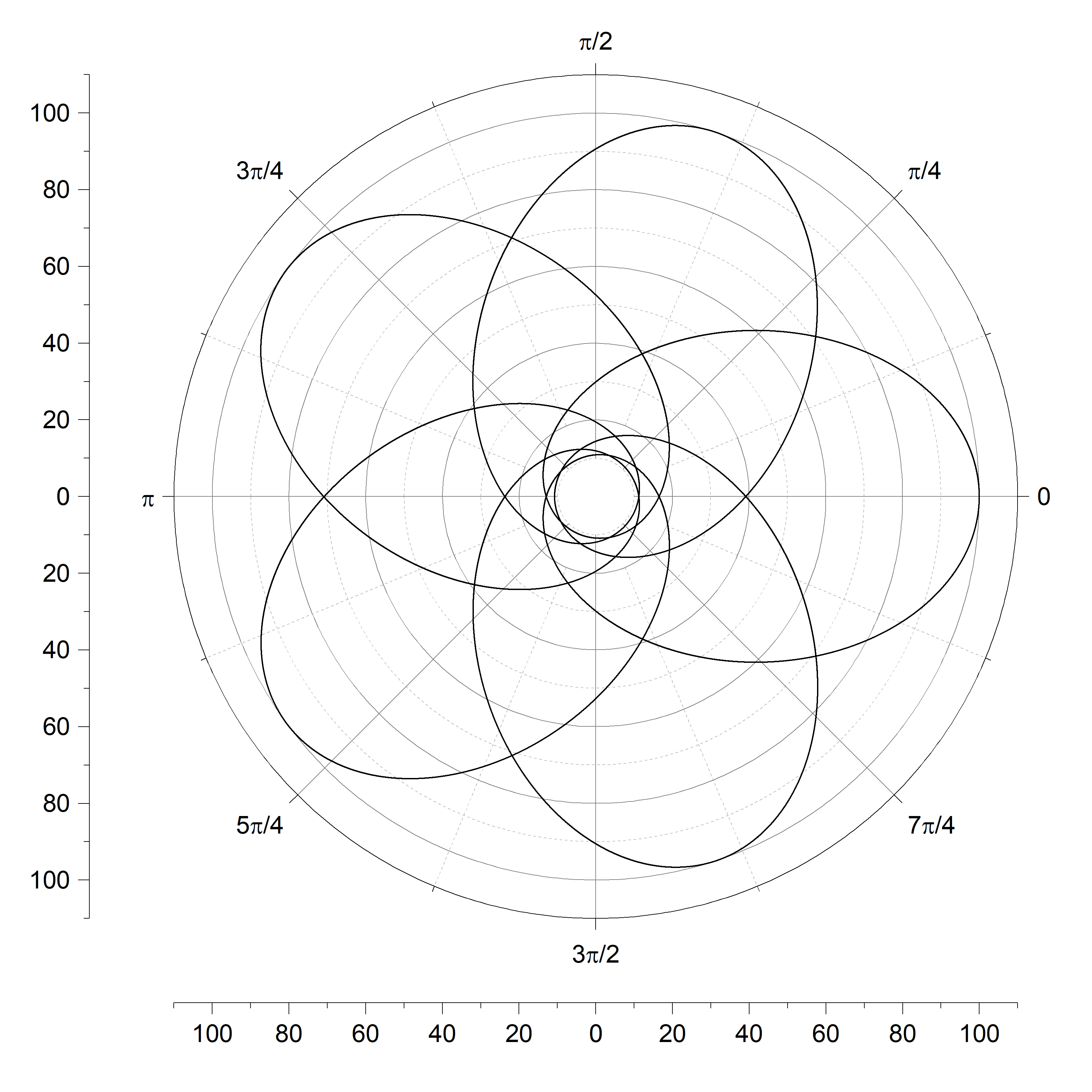}
         \caption{$(19.55,0.8045,5,7,0)$}
     \end{subfigure}%
     \begin{subfigure}{0.25\textwidth}
         \centering
         \includegraphics[width=\textwidth]{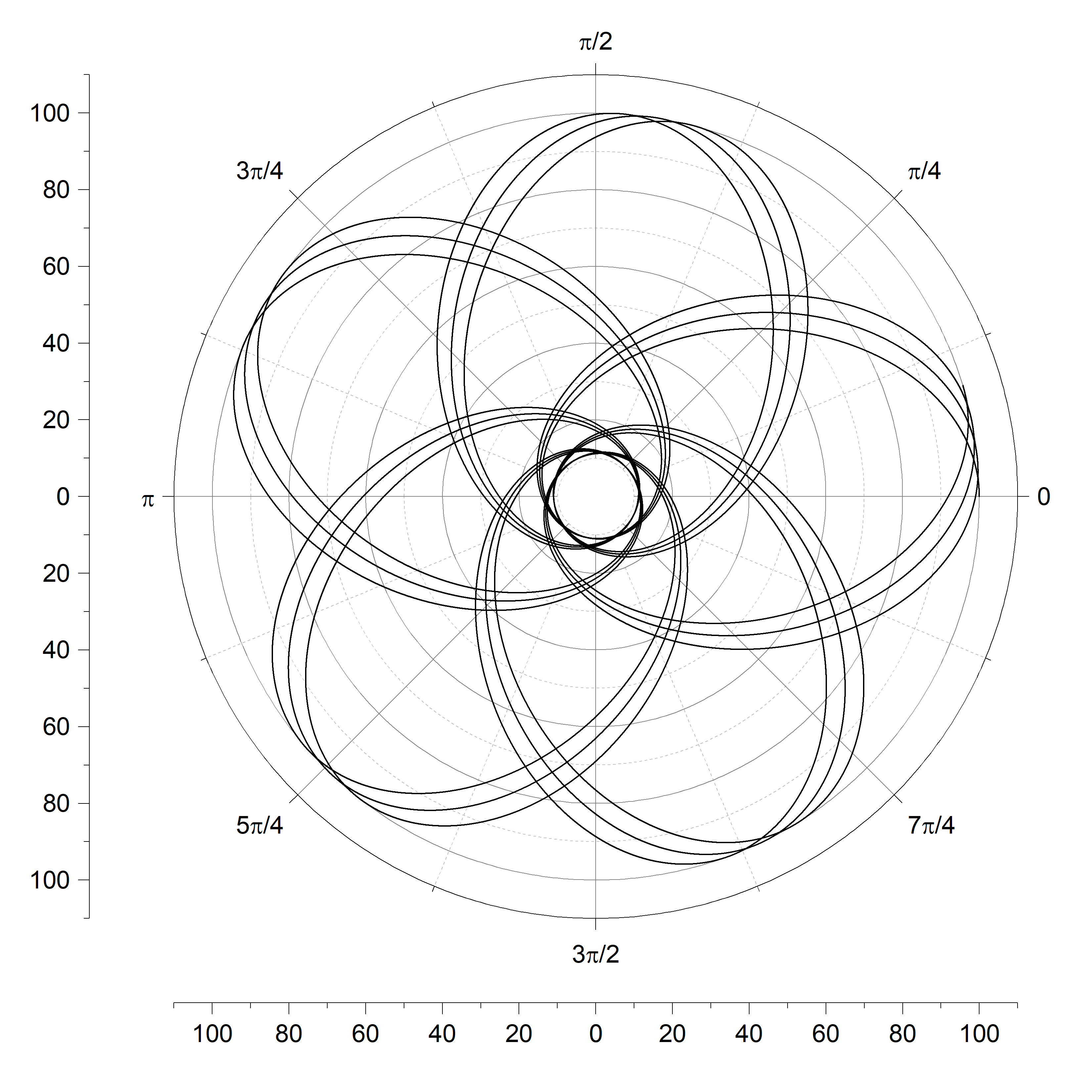}
         \caption{$(19.7,0.803,5,7,0.011116)$}
     \end{subfigure}
     \caption{Examples of massive particle orbits in Levi-Civita spacetime II}
\end{figure*}

\begin{figure*}[htb]
     \centering
     \begin{subfigure}{0.25\textwidth}
         \centering
         \includegraphics[width=\textwidth]{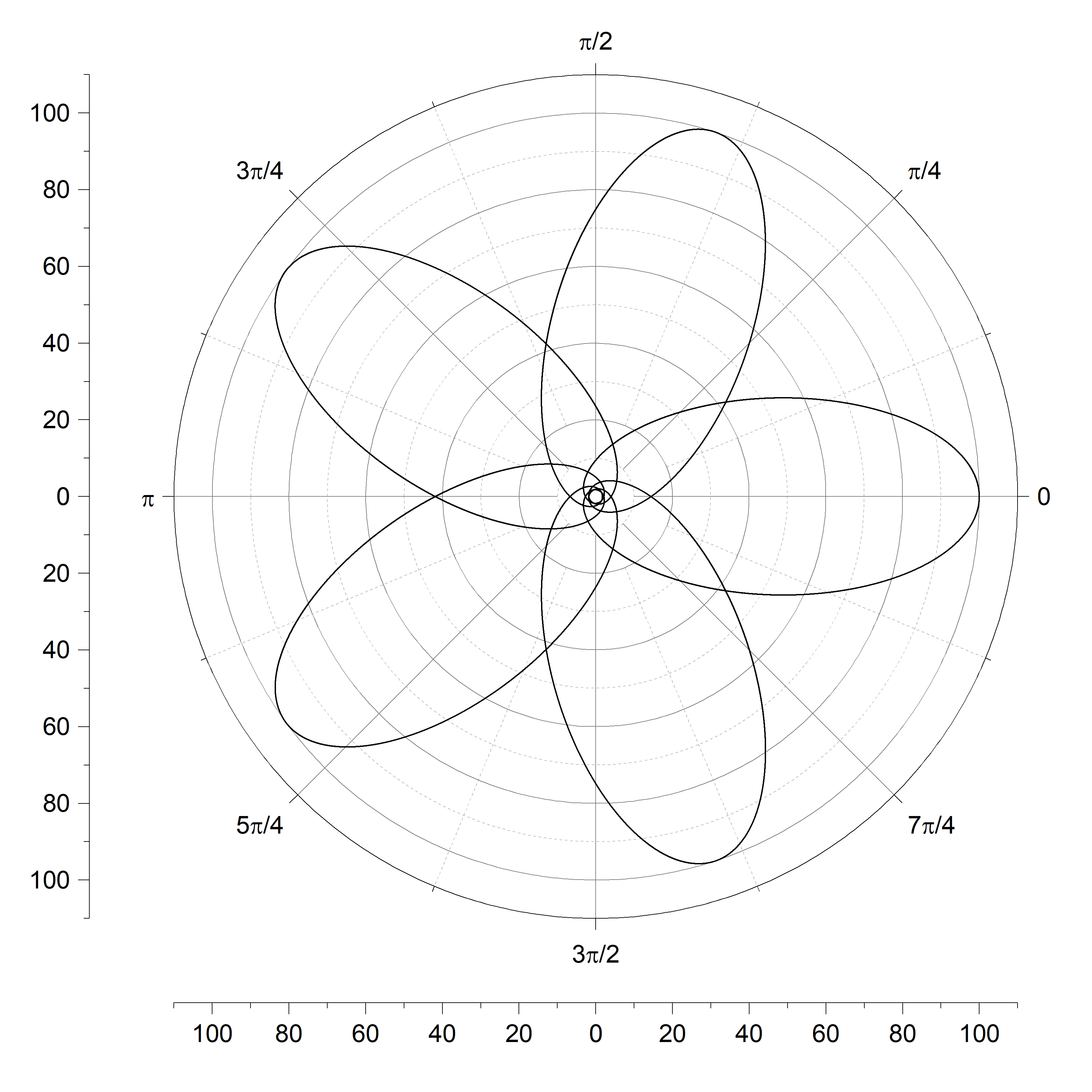}
         \caption{$(5.4,0.946,5,8,0)$}
     \end{subfigure}%
     \begin{subfigure}{0.25\textwidth}
         \centering
         \includegraphics[width=\textwidth]{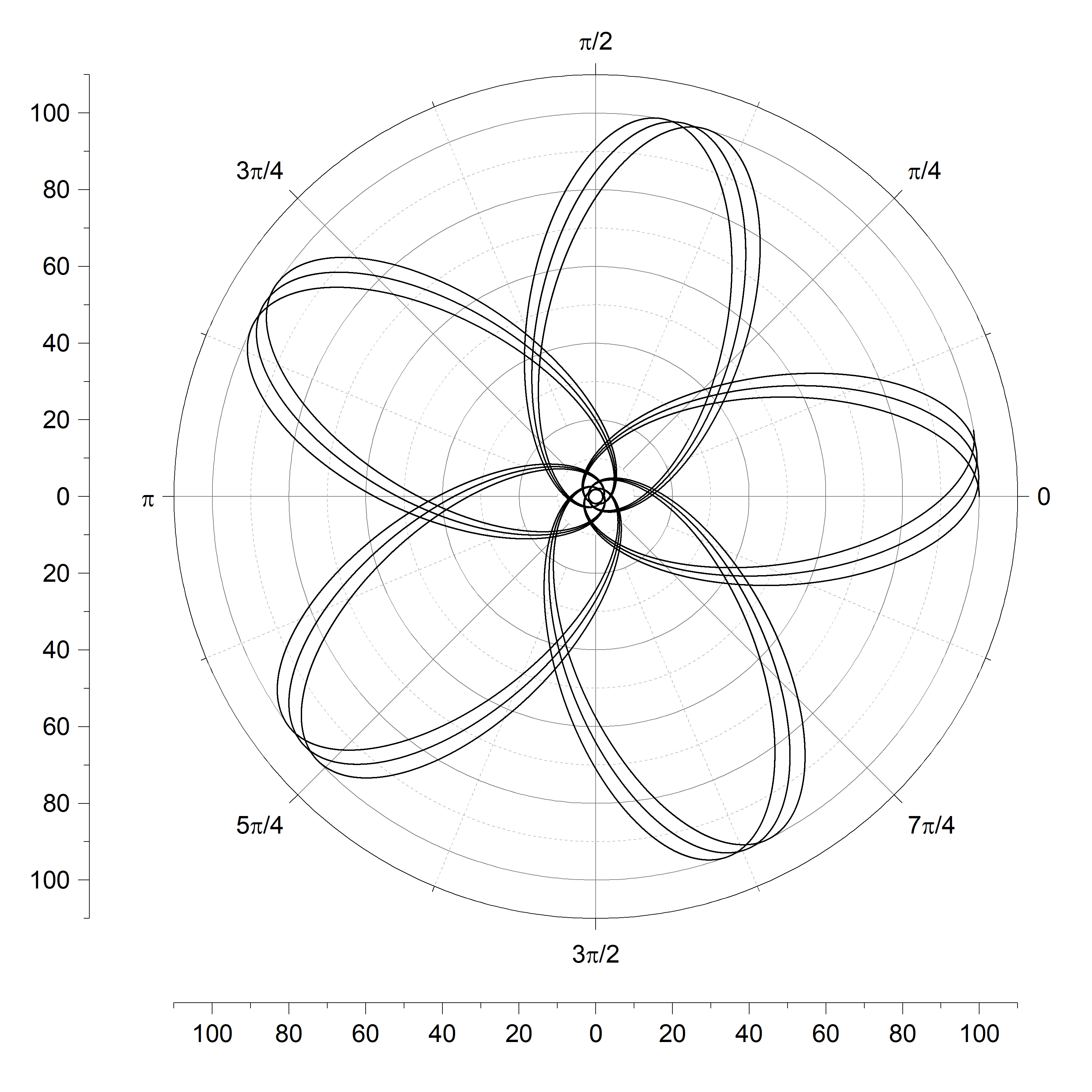}
         \caption{$(5.5,0.945,5,8,0.0057942)$}
     \end{subfigure}%
          \begin{subfigure}{0.25\textwidth}
         \centering
         \includegraphics[width=\textwidth]{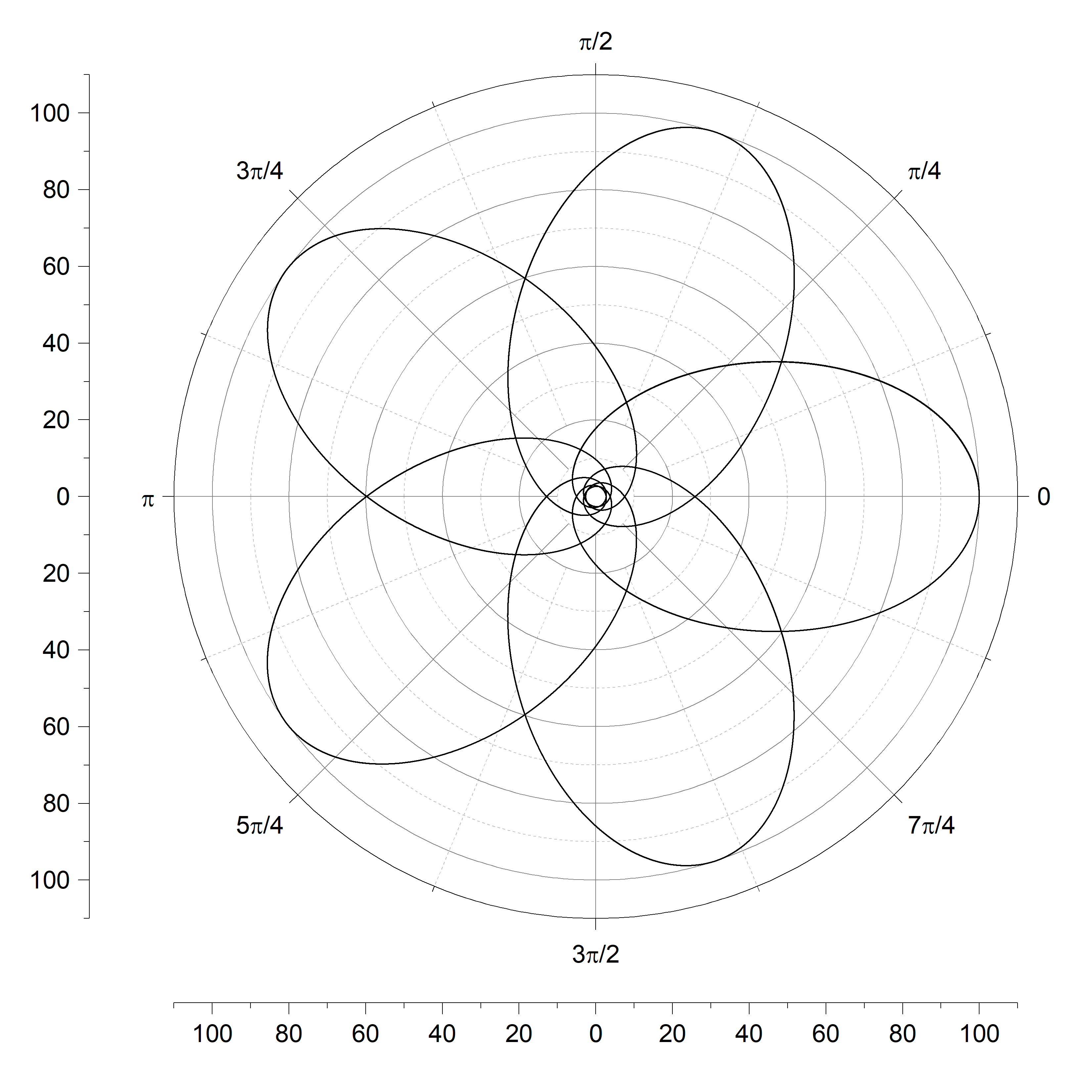}
         \caption{$(8.3,0.917,5,9,0)$}
     \end{subfigure}%
     \begin{subfigure}{0.25\textwidth}
         \centering
         \includegraphics[width=\textwidth]{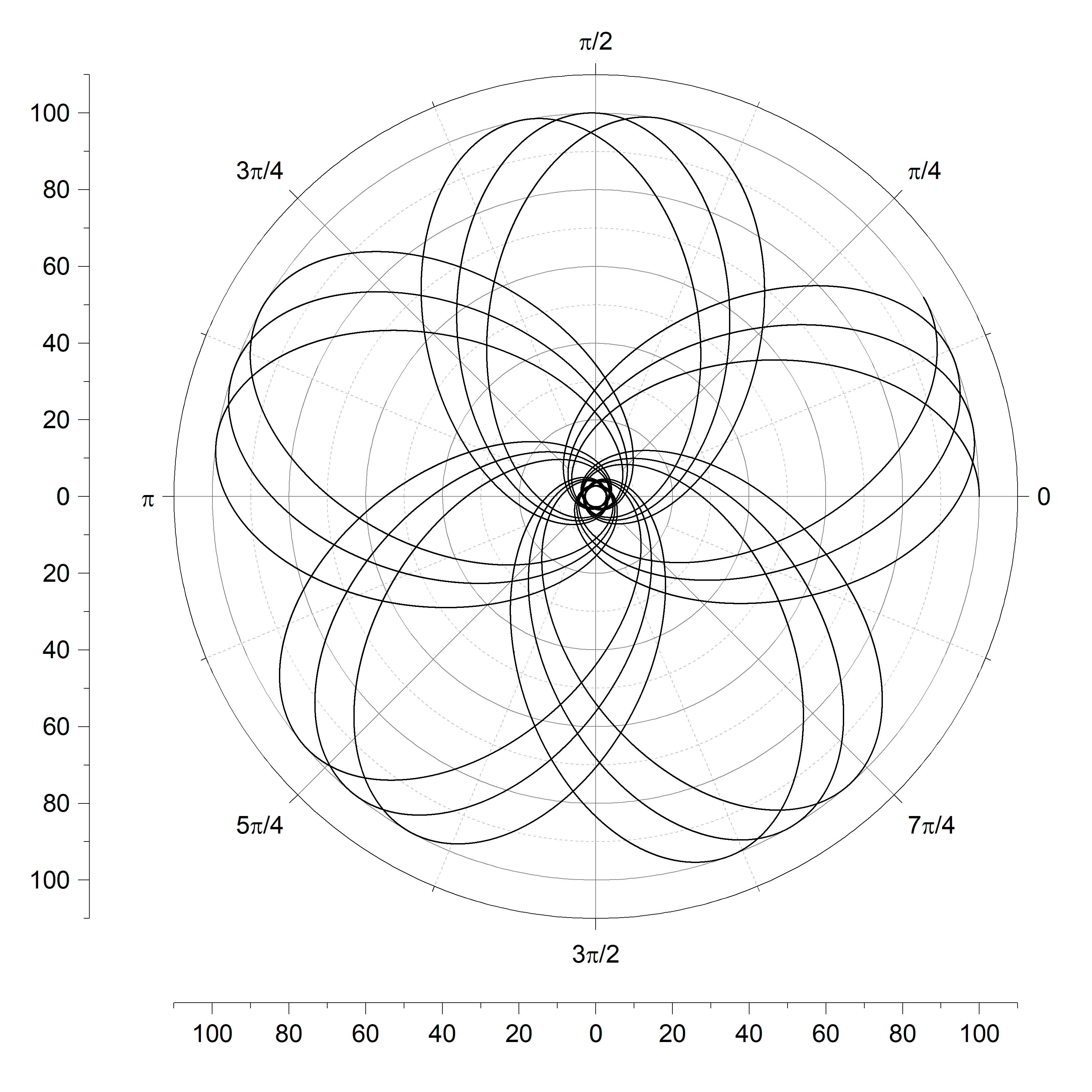}
         \caption{$(8.4,0.916,5,9,0.016061)$}
     \end{subfigure}
          \centering
     \begin{subfigure}{0.25\textwidth}
         \centering
         \includegraphics[width=\textwidth]{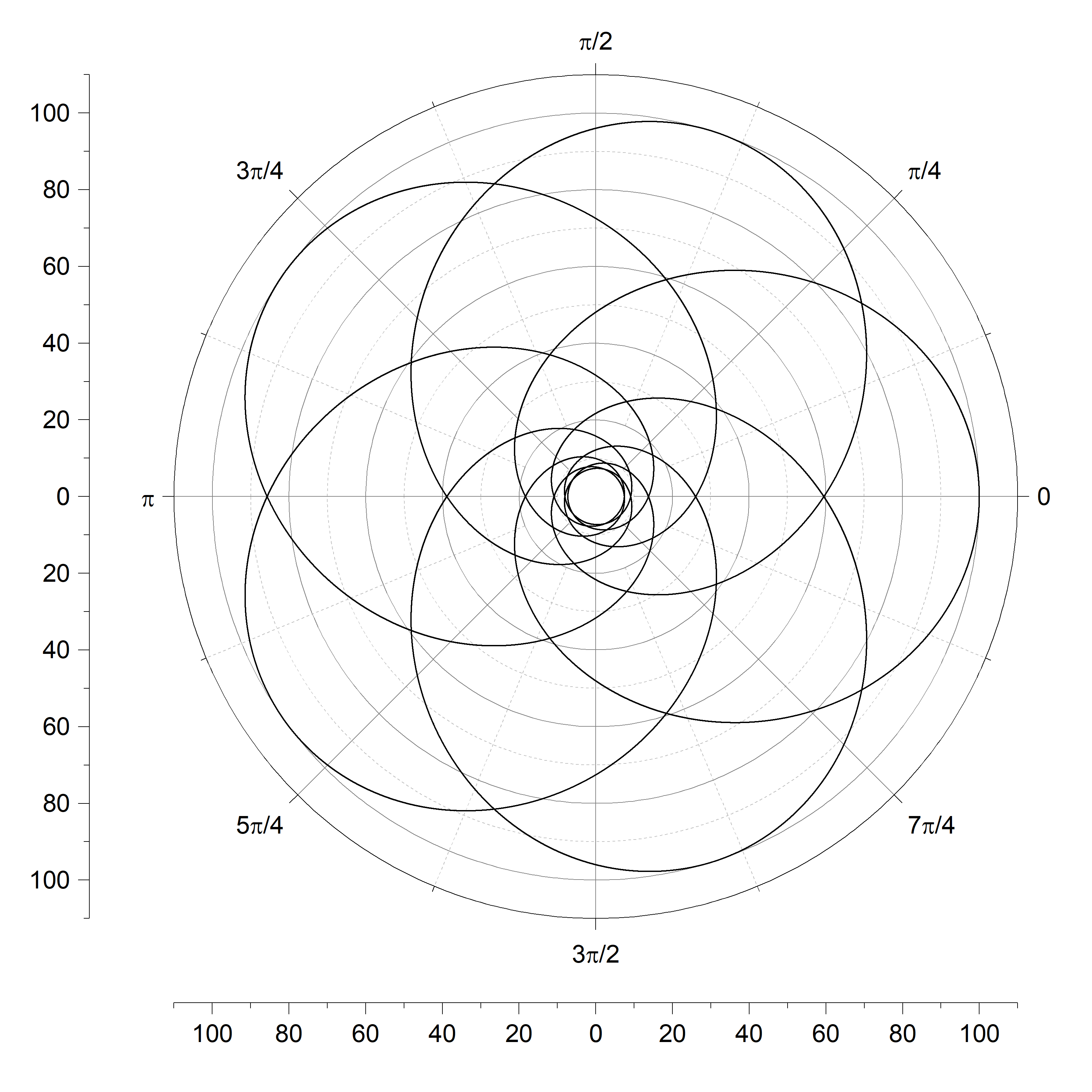}
         \caption{$(19.5,0.805,5,11,0)$}
     \end{subfigure}%
     \begin{subfigure}{0.25\textwidth}
         \centering
         \includegraphics[width=\textwidth]{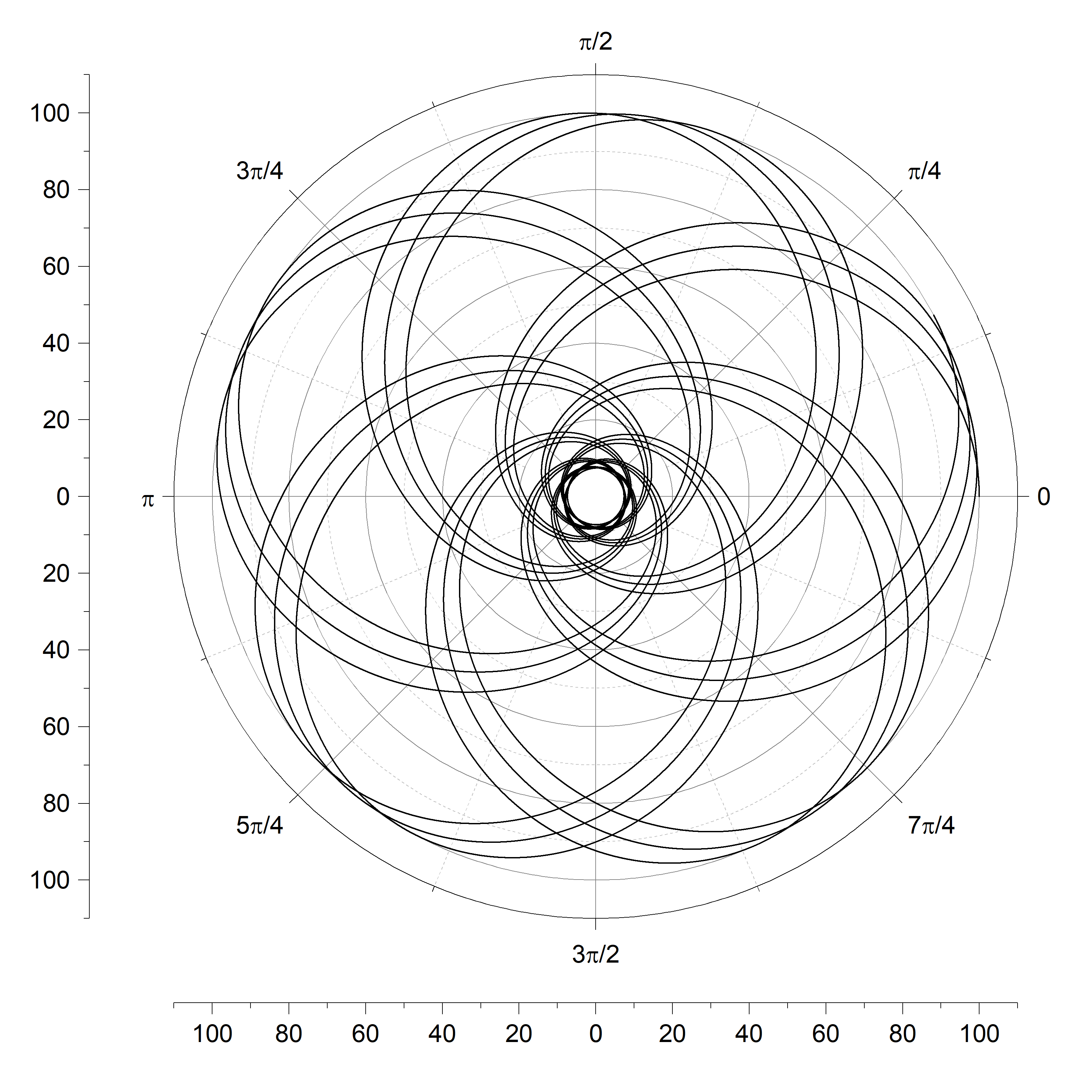}
         \caption{$(19.4,0.806,5,11,0.011862)$}
     \end{subfigure}%
          \begin{subfigure}{0.25\textwidth}
         \centering
         \includegraphics[width=\textwidth]{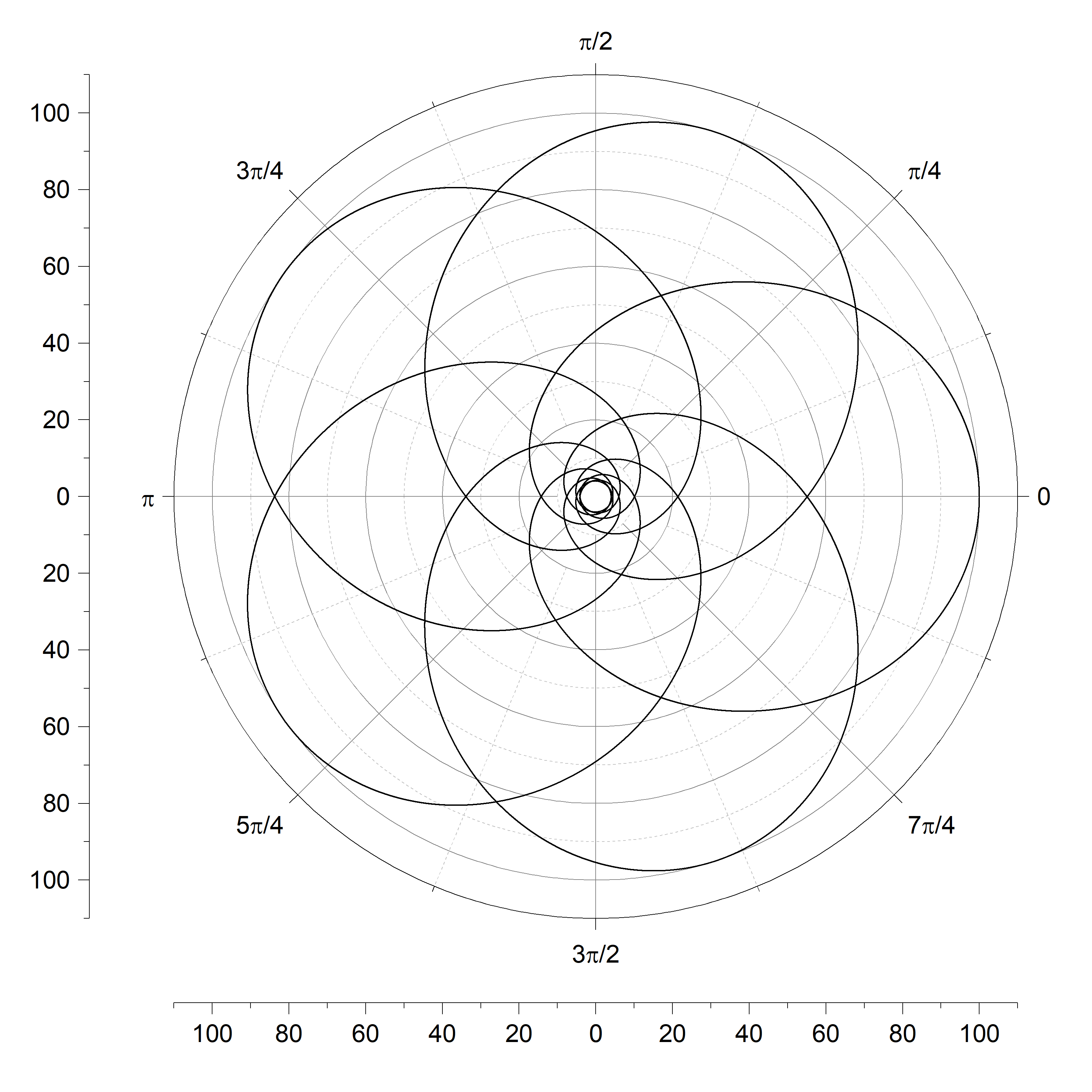}
         \caption{$(14.7,0.853,5,12,0)$}
     \end{subfigure}%
     \begin{subfigure}{0.25\textwidth}
         \centering
         \includegraphics[width=\textwidth]{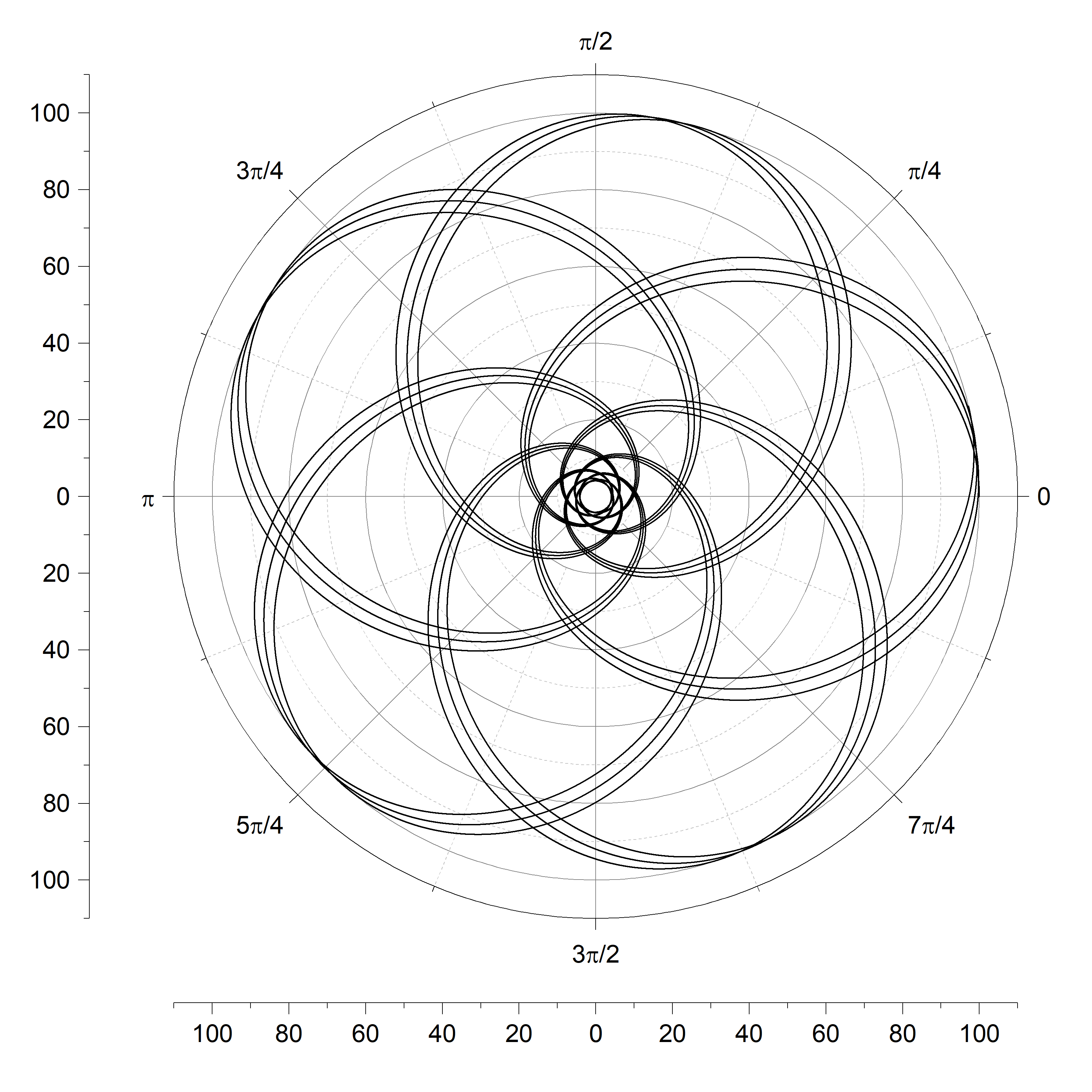}
         \caption{$(14.8,0.852,5,12,0.005258)$}
     \end{subfigure}
          \centering
     \begin{subfigure}{0.25\textwidth}
         \centering
         \includegraphics[width=\textwidth]{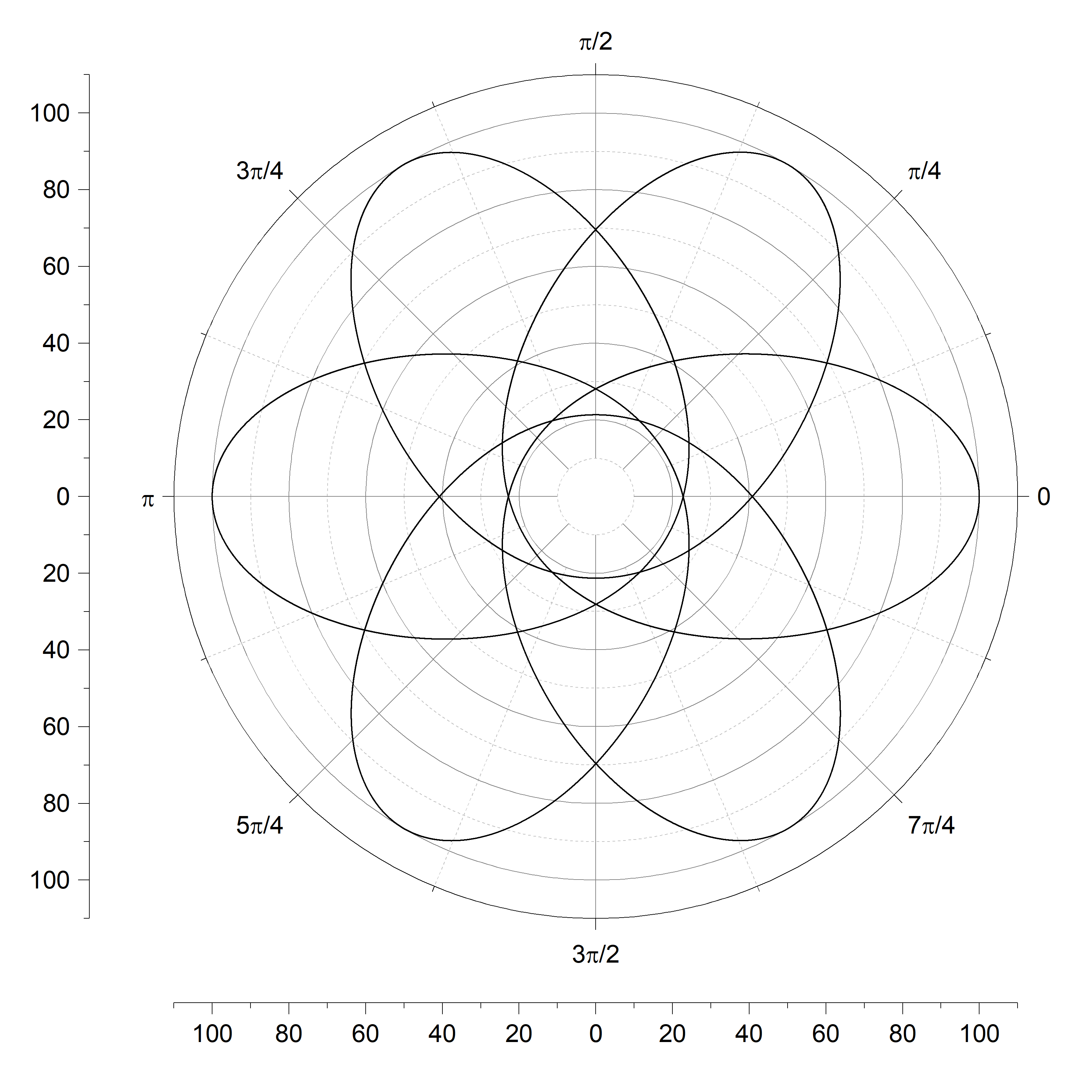}
         \caption{$(30.5,0.695,6,5,0)$}
     \end{subfigure}%
     \begin{subfigure}{0.25\textwidth}
         \centering
         \includegraphics[width=\textwidth]{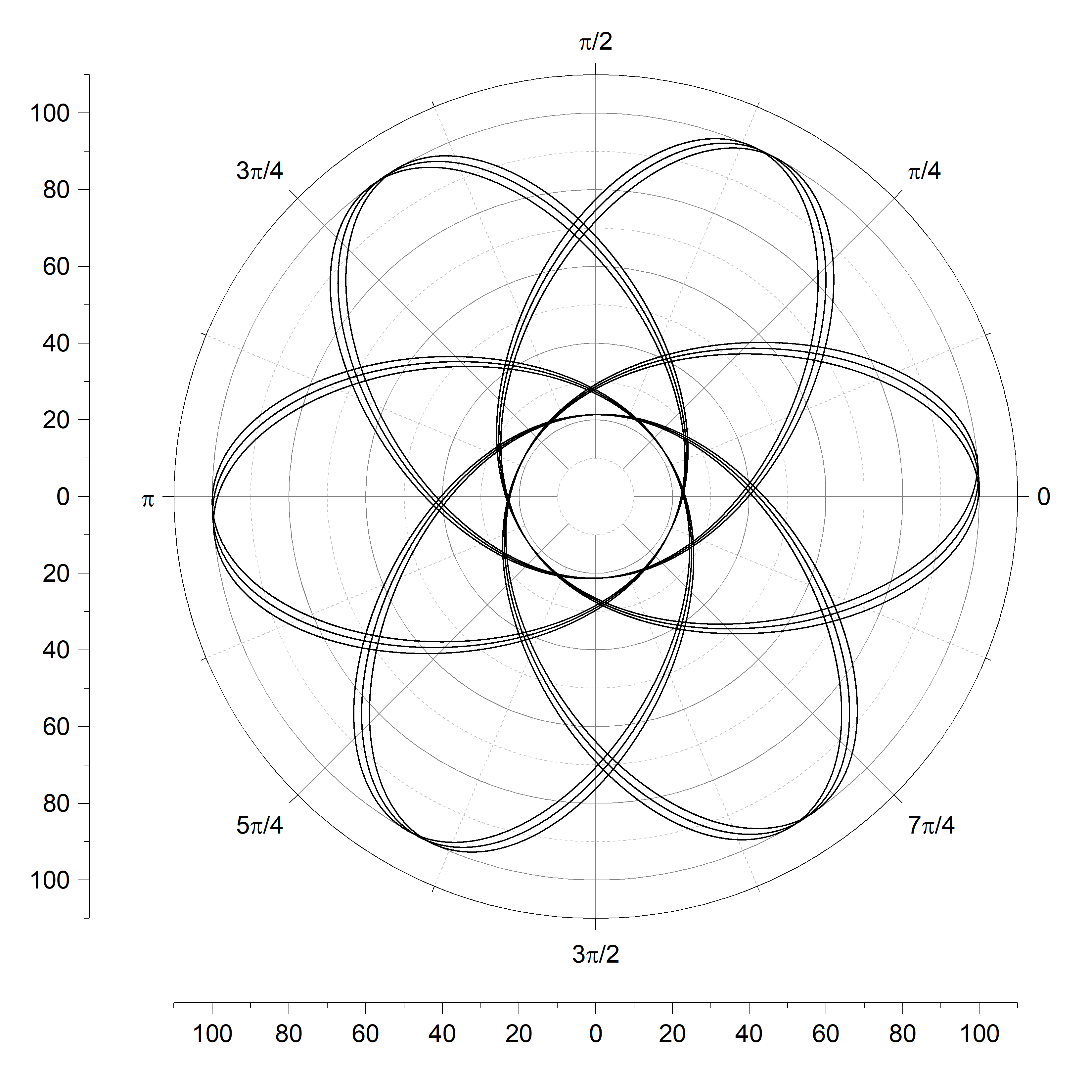}
         \caption{$(30.2,0.698,6,5,0.0057664)$}
     \end{subfigure}%
          \begin{subfigure}{0.25\textwidth}
         \centering
         \includegraphics[width=\textwidth]{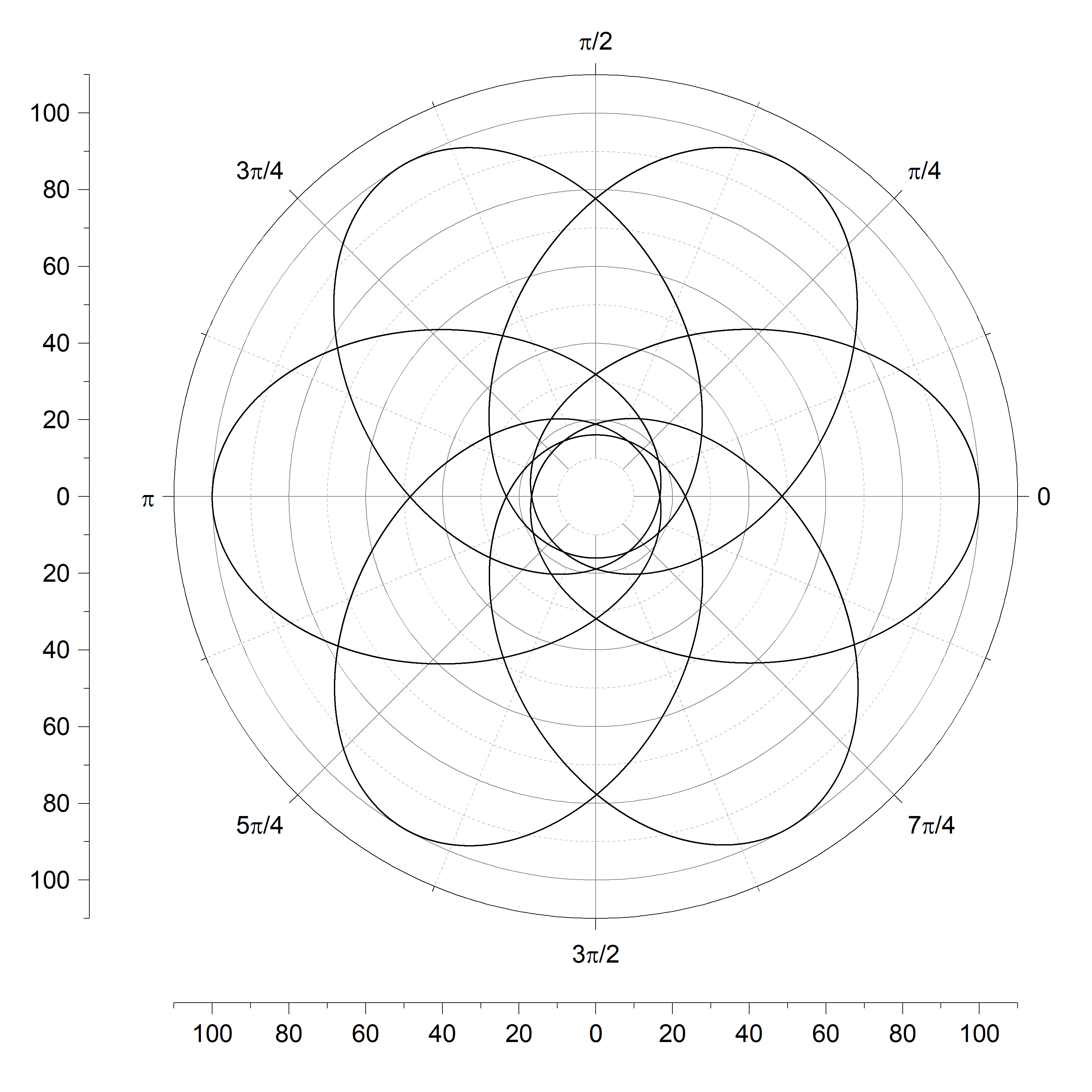}
         \caption{$(25.8,0.742,6,7,0)$}
     \end{subfigure}%
     \begin{subfigure}{0.25\textwidth}
         \centering
         \includegraphics[width=\textwidth]{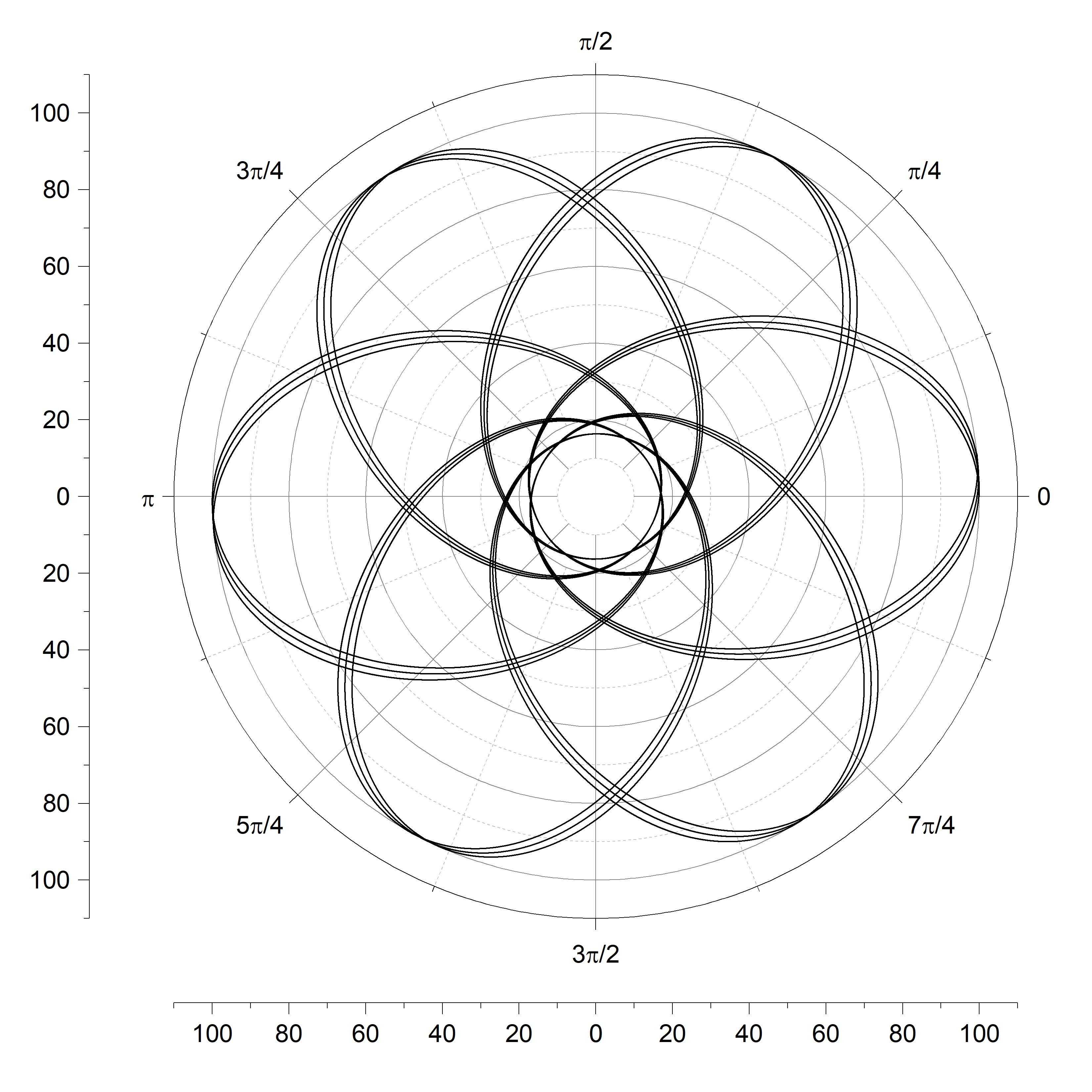}
         \caption{$(26.1,0.739,6,7,0.005057)$}
     \end{subfigure}
          \centering
     \begin{subfigure}{0.25\textwidth}
         \centering
         \includegraphics[width=\textwidth]{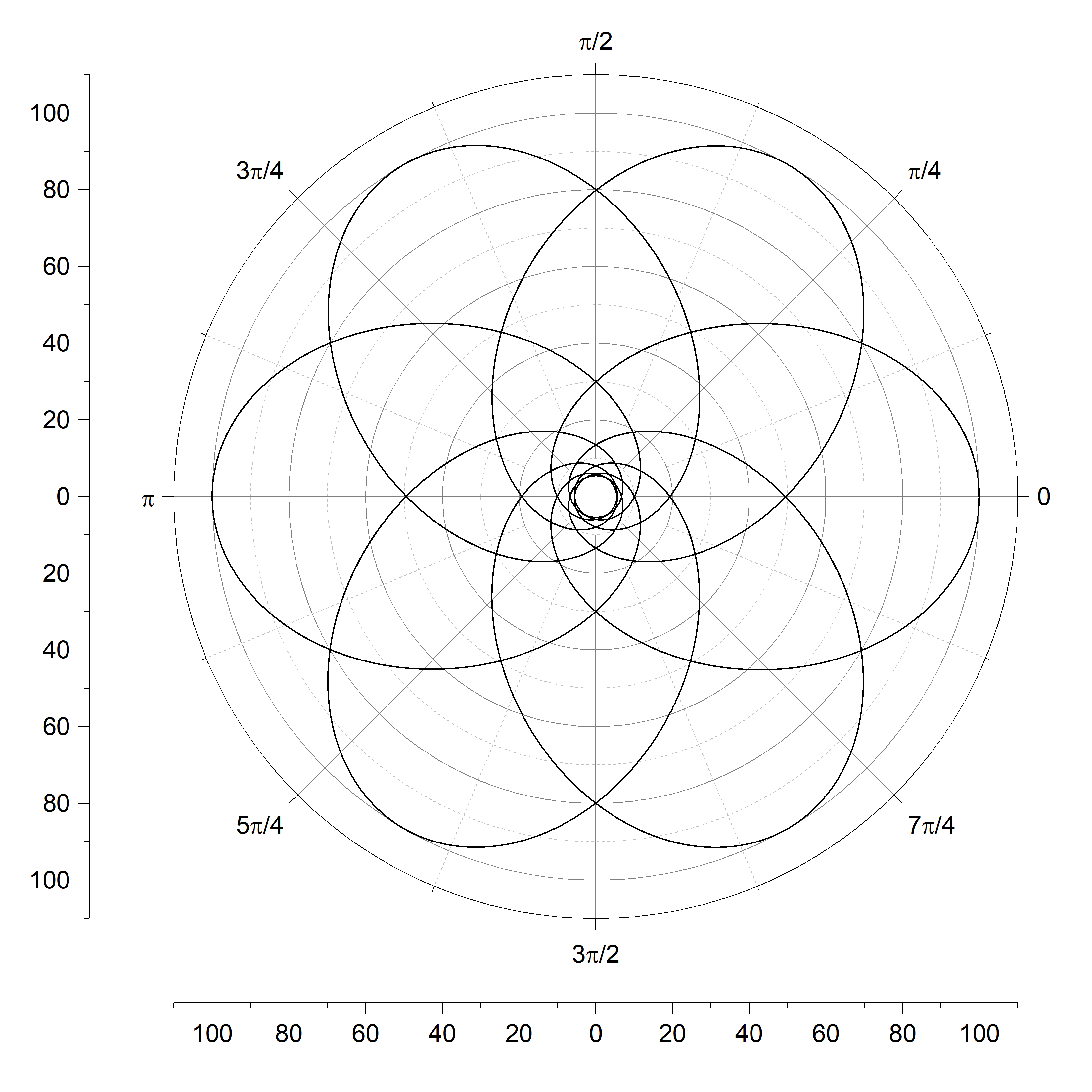}
         \caption{$(13.8,0.862,6,11,0)$}
     \end{subfigure}%
     \begin{subfigure}{0.25\textwidth}
         \centering
         \includegraphics[width=\textwidth]{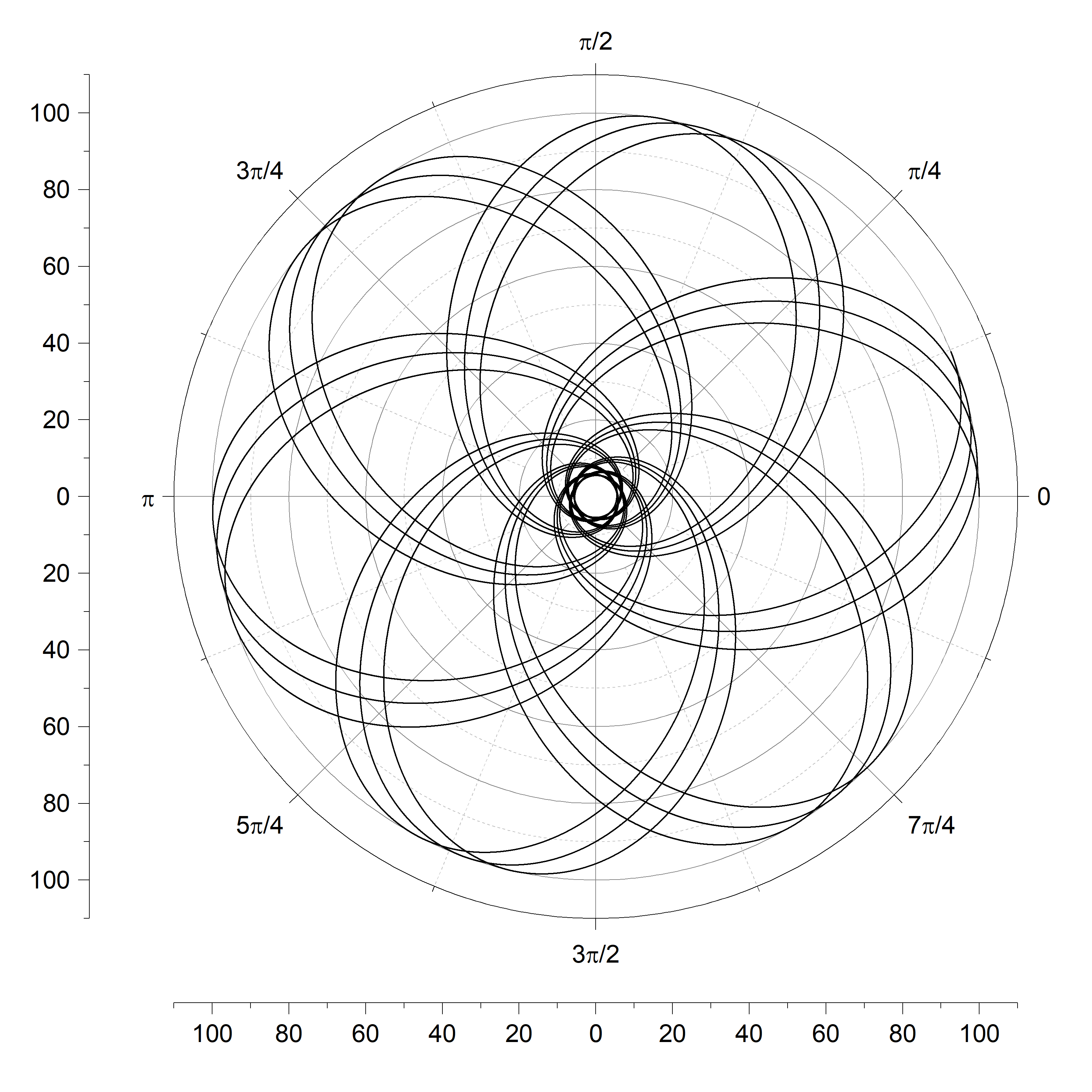}
         \caption{$(13.7,0.863,6,11,0.011192)$}
     \end{subfigure}%
          \begin{subfigure}{0.25\textwidth}
         \centering
         \includegraphics[width=\textwidth]{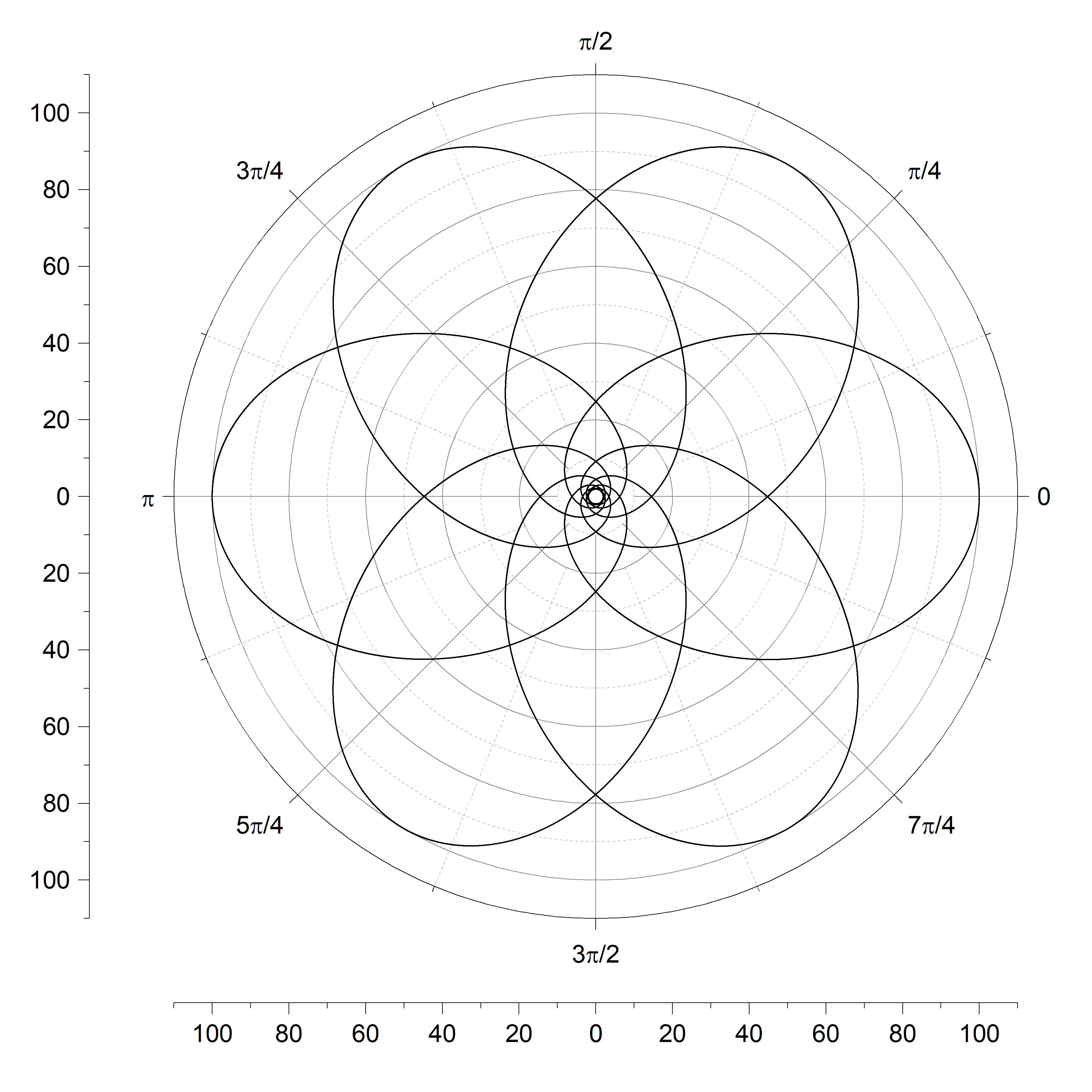}
         \caption{$(9.1,0.909,6,13,0)$}
     \end{subfigure}%
     \begin{subfigure}{0.25\textwidth}
         \centering
         \includegraphics[width=\textwidth]{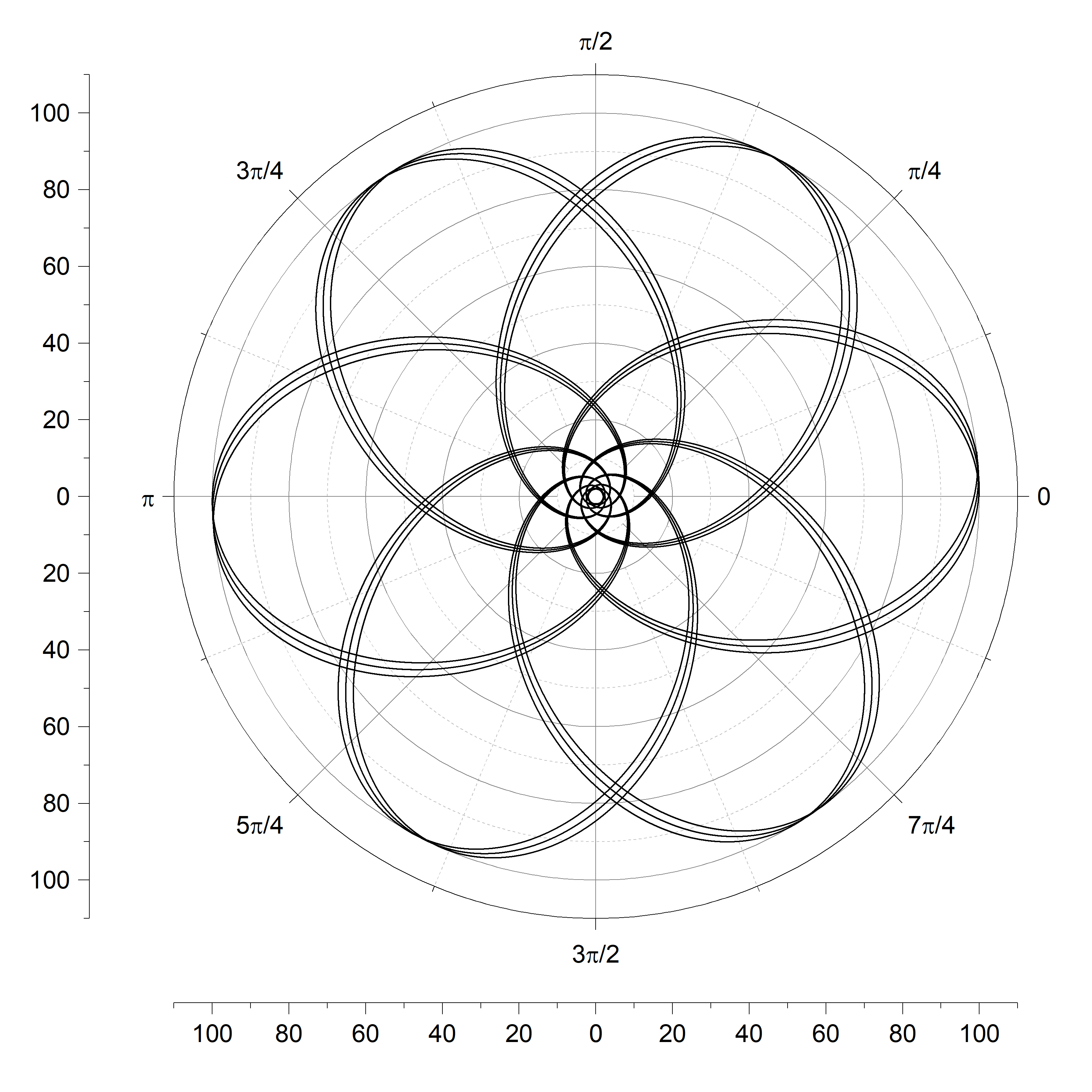}
         \caption{$(9,0.91,6,13,0.0028481)$}
     \end{subfigure}
     \caption{Examples of massive particle orbits in Levi-Civita spacetime III}
\end{figure*}

\clearpage

Observing the images in FIG. 8, FIG. 9, and FIG. 10, we make few qualitative inferences regarding the orbit parameters. We begin our discussion analysing the eccentricity of the orbits. In FIG. 8(a), FIG. 8(c), and FIG. 8(e) we see three one-petal orbits with increasing winding numbers $n$. The orbits in the mentioned images have eccentricities $0.73$, $0.738$, and $0.8$ respectively. It means, orbit in FIG. 8(c) is slightly more eccentric than the one in FIG. 8(a), and the one in FIG. 8(e) is even more eccentric. Despite of the mentioned fact, the orbit in FIG. 8(a) ``looks'' more eccentric than the orbits in FIG. 8(c) and FIG. 8(e). It is because the orbit in FIG. 8(c) and FIG. 8(e) makes one and two extra rotations before closing. Generally, orbits with same eccentricity but possessing higher winding numbers look more round-shaped than the one with lower $n$ values. This behaviour is seen in all the later orbits as well.

Another general behaviour of massive particle orbits in Levi-Civita spacetime we have observed is that the $\gamma$ parameter of an orbit decreases as the spacetime parameter $\sigma$ increases. Furthermore, increasing the initial azimuthal momentum decreases the $\gamma$ parameter as well. These two behaviours are of significant interest, as these can be used to generate orbits with desired number of petals and winding numbers.

For an orbit with a closed corolla, a small increment of parameter $\gamma$ introduces a small clockwise precession to the orbit. That is why, increasing $\sigma$ or $A$ a little bit produces small counter-clockwise precession of the orbit. One can use this property to produce closed orbits by ``fine tuning'' the precessing ones.

Raising the spacetime parameter $\sigma$ to a higher value leads to a sufficiently small value of $\gamma$, for which the winding number $n$ of an orbit is much bigger than the number of petals $m$ of that orbit. In that situation, the particle can make one or more extra rotations around the cosmic string during its course from one petal to another. These kinds of orbits are collectively known as ``zoom-whirl orbits'' for their zooming and whirling behaviour. For example, the orbits in FIG. 8(c) and FIG. 8(e) are single-petal zoom-whirl orbits. The zoom-whirl orbits are ``strong gravity phenomena'' as they require higher values of spacetime parameter $\sigma$.

\section{\label{section-6}Comparison Between Newtonian and Relativistic Orbits}

In section IIC, we demonstrated that the gravitational field generated by a cosmic string matches with the gravitational field of its Newtonian counterpart in the weak field limit, $\sigma\ll 1$. In this section we explore the orbits of massive particles moving in the Newtonian gravitational field of an infinitely-long linear mass distribution. We begin our discussion by deriving the complete, exact solution by quadrature to the equation of motion of a point particle around the Newtonian linear mass distribution. Considering the Newtonian gravitational potential of a linear mass distribution, we write the Hamilton-Jacobi equation for the system in cylindrical coordinates. Following the standard procedure \cite{jose1998-06}, we obtain the following general integral solution to the equations of motion for the point particle.

\begin{widetext}
\begin{subequations}
\begin{eqnarray}
    t & = & \int_{\rho_0}^\rho \biggr\{2\Delta - A^2\rho^{-2} - B^2 - \frac{4\sigma}{1-2\sigma+4\sigma^2} \ln{\biggr(\frac{\rho}{R} \biggr)}\biggr\}^{-\frac{1}{2}} d\rho
    \\
    \phi & = & \phi_0 + A \int_{\rho_0}^\rho \rho^{-2} \biggr\{2\Delta - A^2\rho^{-2} - B^2 - \frac{4\sigma}{1-2\sigma+4\sigma^2} \ln{\biggr(\frac{\rho}{R} \biggr)}\biggr\}^{-\frac{1}{2}} d\rho
    \\
    z & = & z_0 + B \int_{\rho_0}^\rho \biggr\{2\Delta - A^2\rho^{-2} - B^2 - \frac{4\sigma}{1-2\sigma+4\sigma^2} \ln{\biggr(\frac{\rho}{R} \biggr)}\biggr\}^{-\frac{1}{2}} d\rho
\end{eqnarray}
\end{subequations}
\end{widetext}

Similar to the case of a massive particle in Levi-Civita spacetime in equation-array (8), we have used the initial coordinates, $\rho_0$, $\phi_0$, and $z_0$, the conserved momentum conjugate to $\phi$ coordinate $A$, conserved momentum in $z$ coordinate $B$ and the total energy $\Delta$ as constants of motion in the final integral solution in the equation-array (24). However, one might want to write the integral solution in equation-array (24) terms of $\Gamma$, the initial value of the momentum conjugate to the radial coordinate, using the relation

\begin{eqnarray}
    \Gamma = \biggr\{2\Delta  - \frac{A^2}{{\rho_0}^2} - B^2 - \frac{4\sigma}{1 - 2\sigma + 4\sigma^2} \ln{\biggr(\frac{\rho_0}{R}\biggr)}
    \biggr\}^\frac{1}{2}
\end{eqnarray}

instead of $\Delta$; doing so (23) shall be useful for running numerical routines that take $\Gamma$ as an initial data, rather than $\Delta$.

The Newtonian gravitational potential, $\Phi$ for an infinitely-long linear mass distribution is unbounded from above and goes to infinity as the radial coordinate increases, implying that the motion of a point particle around it is always bounded, like its relativistic counterpart. As a result, on $z=z_0$ plane, the trajectory of a point particle around a Newtonian linear mass distribution always remain on a washer, defined by $\underline{\rho}<\rho<\bar{\rho}$, where $\underline{\rho}$ and $\bar{\rho}$ are the lower and the upper limits of the radial coordinate respectively. To see this, one starts with the Hamilton-Jacobi equation of the system in cylindrical coordinates and rewrite it in terms of conserved energy $\Delta$, and conserved momenta $A$ and $B$ of the motion. And then, setting the momentum conjugate to the radial coordinate and $B$ to zero, one gets

\begin{equation}
    \frac{4\sigma}{1-2\sigma+4\sigma^2}\ln{\biggr(\frac{\rho}{R}\biggr)}\rho^2 - 2 \Delta\rho^2+A^2=0
\end{equation}

Equation (26) is the Newtonian analog of the equation (18). The coordinate $\rho$ in equation (26) has two distinct solutions, like its relativistic counterpart. To show it graphically, we define $g(\rho;A,\Delta,\sigma,R):=\frac{4\sigma}{1-2\sigma+4\sigma^2}\ln{\biggr(\frac{\rho}{R}\biggr)}\rho^2 - 2 \Delta\rho^2+A^2$ and plot it against $g(\rho;A,\Delta,\sigma,R)$. In FIG. 11, we can see that the horizontal line $h(\rho)=0$ intersects twice the curve of $g(\rho;A,\Delta,\sigma,R)$, providing the two solutions to $\rho$. We denote the smaller solution by $\underline{\rho}$ and the larger one by $\bar{\rho}$. This establishes the fact the radial coordinate $\rho$ is bounded between $\rho=\underline{\rho}$ and $\rho=\bar{\rho}$, as mentioned earlier in section IIIA.

\begin{figure}[htb]
     \centering
     \begin{subfigure}{0.25\textwidth}
         \centering
         \includegraphics[width=\textwidth]{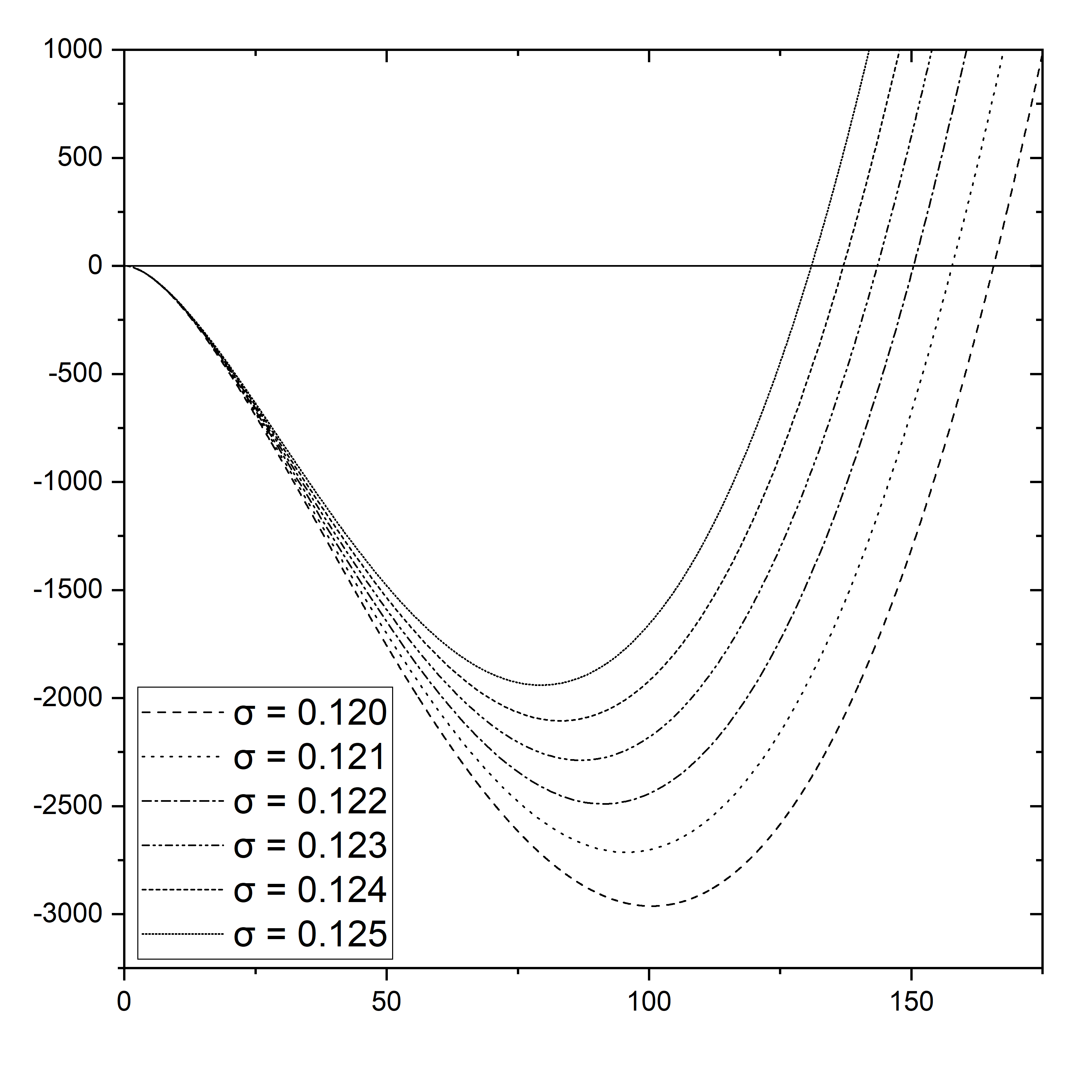}
         \caption{Varying $\sigma$ keeping $A=1$, $\Delta=1.5$, and $R=1$}
     \end{subfigure}%
     \begin{subfigure}{0.25\textwidth}
         \centering
         \includegraphics[width=\textwidth]{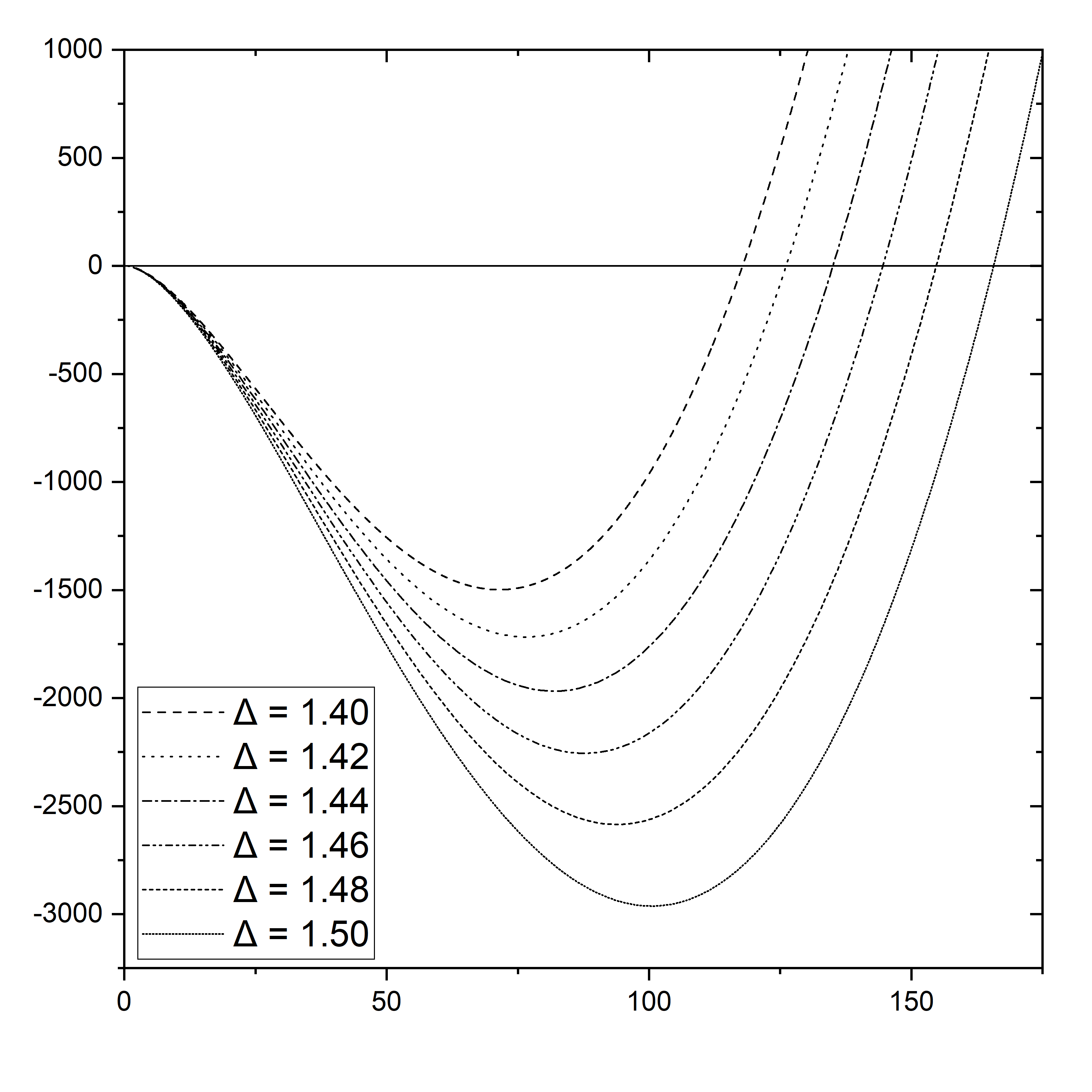}
         \caption{Varying $\Delta$ keeping $A=1$, $\sigma=0.12$, and $R=1$}
     \end{subfigure}
     \caption{$g(\rho; A, \Delta, \sigma, R)$ plots intersect $h(\rho)=0$ line twice, showing the existence of a lower limit $\underline{\rho}$ and an upper limit $\bar{\rho}$ of the radial coordinate}
\end{figure}

We have demonstrated in section IVA that, in the weak field limit, the orbit of a massive particle on $z=z_0$ plane in Levi-Civita spacetime follows approximately a precessing elliptical path. Naturally, the motion of a point particle around a Newtonian linear mass distribution is expected to be a precessing ellipse as well. To demonstrate that, we take the equation (24b) and make approximations to it considering $B=0$, $\sigma \ll 1$, and $\lvert{\underline{\rho}/\rho}\rvert < 1$, leading to the exact closed form formula as in equation (19). And the parameters of the orbit $C$, $\epsilon$, $\gamma$, and $\delta$ are given by the relations

\begin{subequations}
\begin{eqnarray}
    C = & & \frac{1}{4\sigma\underline{\rho}^2}\times \biggr[A^2 + 2\sigma\underline{\rho}^2 - 2\sigma A^2 + 4A^2\sigma^2\biggr]
    \\
    \epsilon = & & \frac{1}{4\sigma\underline{\rho}^2}\times\biggr[ 2\Delta A^2 \underline{\rho}^2- 8 \sigma \Delta A^2 \underline{\rho}^2 - 6 \sigma A^2 \underline{\rho}^2
    \nonumber\\
    & & + 4\sigma\Delta \underline\rho^4 - 4\sigma A^2\underline\rho^2\ln\biggr({\frac{\underline{\rho}}{R}}\biggr) + 24 \sigma^2 \Delta A^2\underline{\rho}^2
    \nonumber\\
    && + 12 \sigma^2 A^2 \underline{\rho}^2 - 8 \sigma^2\Delta\underline{\rho}^4 + 4 \sigma^2\underline{\rho}^4
    \nonumber\\
    && + 8 \sigma^2 A^2 \underline\rho^2\ln{\biggr(\frac{\underline{\rho}}{R}\biggr)} - 8\sigma^2 \underline\rho^4\ln{\biggr(\frac{\underline{\rho}}{R}\biggr)}\biggr]^{\frac{1}{2}} 
    \\
    \gamma = & & \biggr[1-2\sigma+4\sigma^2\biggr]^{-\frac{1}{2}}
    \nonumber\\
    & & \times\biggr[1 - 2\sigma + 2\sigma\frac{\underline\rho^2}{A^2} + 4\sigma^2\biggr]^\frac{1}{2}
    \\
    \delta = & &\phi_0-\frac{1}{\gamma}\arccos\biggr(\frac{\rho_0-C\underline{\rho}}{\epsilon \rho_0}\biggr)
\end{eqnarray}
\end{subequations}

As we have discussed earlier for the case of massive particle orbits in Levi-Civita spacetime, only the parameters $C$, $\epsilon$, and $\gamma$ are enough to uniquely determine the trajectory, whereas the parameter $\delta$ can always be set to zero by a rotation of coordinates around $z$ axis. Once again $C$, $\epsilon$, and $\gamma$ contain the information regarding the size, eccentricity, and precession rate of the orbit respectively.

However, from the equation-arrays (20) and (27), it is not apparent how the trajectory of a massive particle in Levi-Civita spacetime matches with trajectory of a point mass around a Newtonian linear mass distribution in the weak field limit. To see the correspondence, we make further approximations to the equation-arrays (20) and (27) considering $\sigma\ll 1$ and keep the terms containing the lowest two powers in $\sigma$. We compare the results side-by-side in equation-array (28),

\begin{subequations}
\begin{eqnarray}
    C_N \approx & & \frac{A^2}{4\sigma\underline{\rho}^2} + \frac{1}{2} - \frac{A^2}{2\underline{\rho}^2}\\
    C_R \approx & &  \frac{A^2}{4\sigma \underline{\rho}^2} +  \frac{1}{2} -  \frac{3A^2}{2 \underline{\rho}^2} -  \frac{ A^4}{2 \underline{\rho}^4}\\
    \epsilon_N \approx & & \biggr[  \frac{\Delta A^2}{8\sigma^2 \underline{\rho}^2} +  \frac{\Delta}{4\sigma} - \frac{3 A^2}{8\sigma \underline{\rho}^2}  - \frac{ A^2 \ln(\frac{\underline{\rho}}{R})}{4\sigma\underline{\rho}^2} \nonumber \\
    & & - \frac{\Delta A^2}{2 \sigma \underline{\rho}^2}\biggr]^{\frac{1}{2}}
    \\
    \epsilon_R \approx & & \biggr[\frac{\tilde{\Delta} A^2 }{ 8 \sigma^2 {\underline{\rho}^2}} +\frac{\tilde{\Delta}  }{ 4\sigma}   - \frac{3 A^2}{ 8 \sigma {\underline{\rho}^2}} - \frac{ A^2 \ln(\frac{\underline{\rho}}{P})}{ 4 \sigma {\underline{\rho}^2}} \nonumber \\ 
    & & - \frac{ 3 \tilde{\Delta} A^2 }{ 2 \sigma {\underline{\rho}^2}} + \frac{\tilde{\Delta}   A^2 \ln(\frac{\underline{\rho}}{P})}{ 2 \sigma {\underline{\rho}^2}} - \frac{\tilde{\Delta} A^4 }{ 2 \sigma {\underline{\rho}^4}} \nonumber \\
    & & - \frac{ \tilde{\Delta} A^2\ln(\frac{\underline{\rho}}{P}) }{\sigma {\underline{\rho}^2}}\biggr]^{\frac{1}{2}} \\   
    \gamma_N \approx & & 1 + \sigma\frac{\underline\rho^2}{A^2}\\
    \gamma_R \approx & & 1 + \sigma\frac{\underline{\rho}^2}{A^2} - 2\sigma
\end{eqnarray}
\end{subequations}

where $\tilde\Delta$ is defined by the relation $\tilde\Delta:=\frac{1}{2}(\Delta^2-1)$. In the equation-array above, parameters of particle orbit around a Newtonian linear mass density are denoted by $C_N$, $\epsilon_N$, and $\gamma_N$, whereas, $C_R$, $\epsilon_R$, and $\gamma_R$ are the parameters of massive particle orbit in Levi-Civita spacetime. It is apparent from the equation-array (28) that the massive particle orbits in Levi-Civita spacetime coincide with particle orbits around a Newtonian linear mass density in the weak gravity limit, as the corresponding orbit parameters match in the lowest power of $\sigma$ and start to deviate only when the terms with second lowest power in $\sigma$ are considered. Finally, let us compare the orbits graphically. 

\begin{figure}[htb]
     \centering
     \begin{subfigure}{0.25\textwidth}
         \centering
         \includegraphics[width=\textwidth]{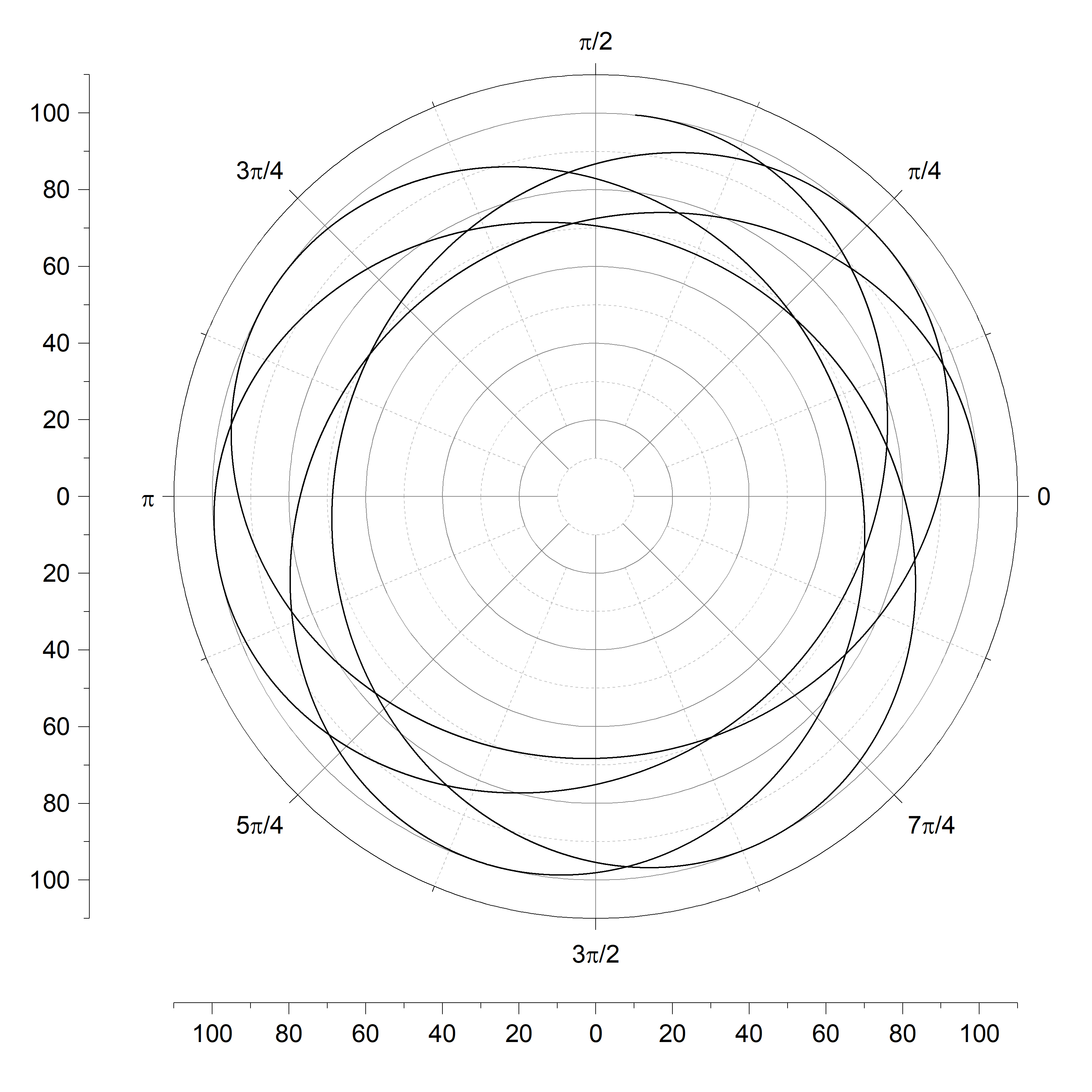}
         \caption{$(80.8,0.192,3,2,0.1645)$}
     \end{subfigure}%
     \begin{subfigure}{0.25\textwidth}
         \centering
         \includegraphics[width=\textwidth]{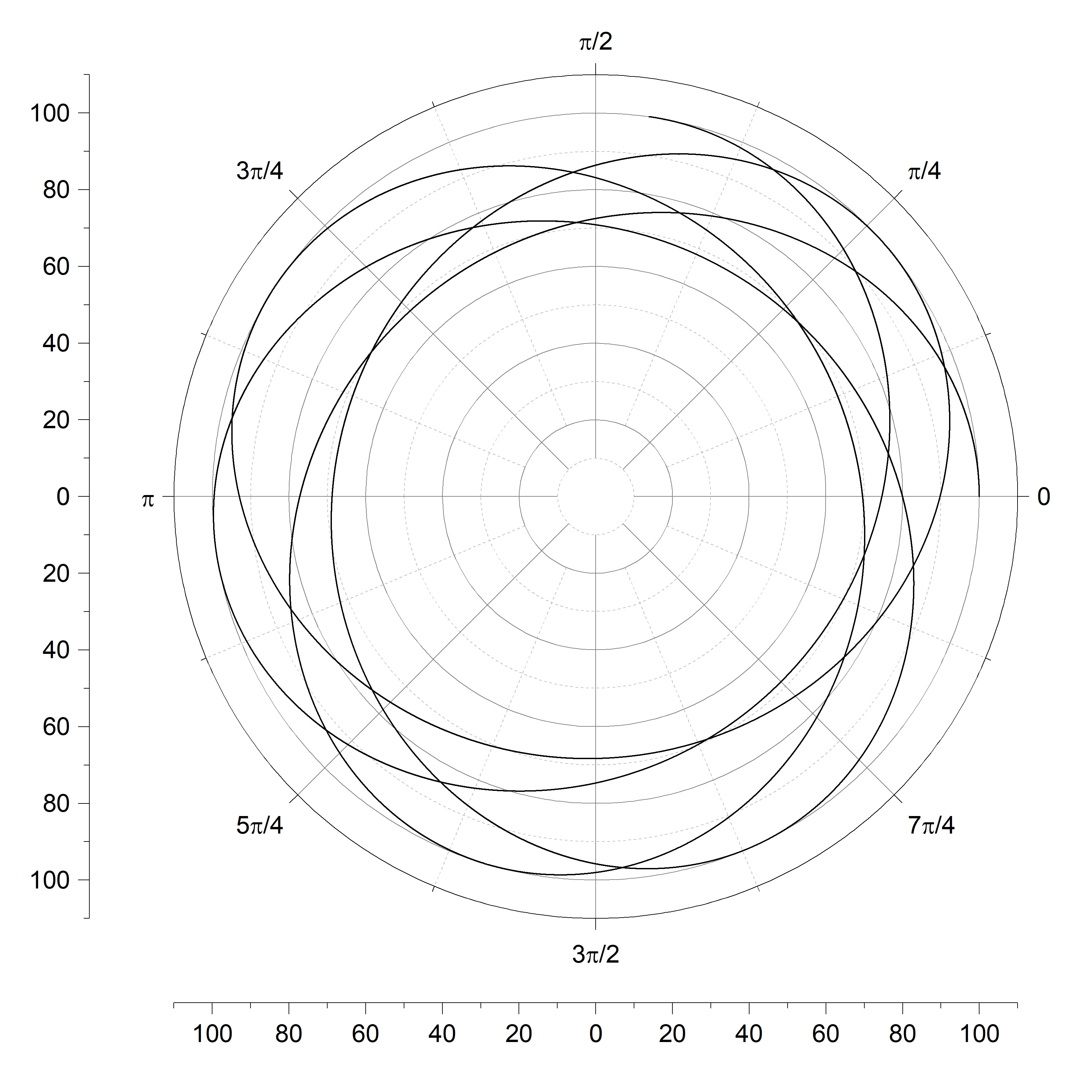}
         \caption{$(80.8,0.192,3,2,0.1612)$}
     \end{subfigure}
     \caption{Massive particle moving in Levi-Civita spacetime (left) and around a Newtonian linear mass distribution (right) with parameter $\sigma=0.0003$}
\end{figure}

\begin{figure}[htb]
     \centering
     \begin{subfigure}{0.25\textwidth}
         \centering
         \includegraphics[width=\textwidth]{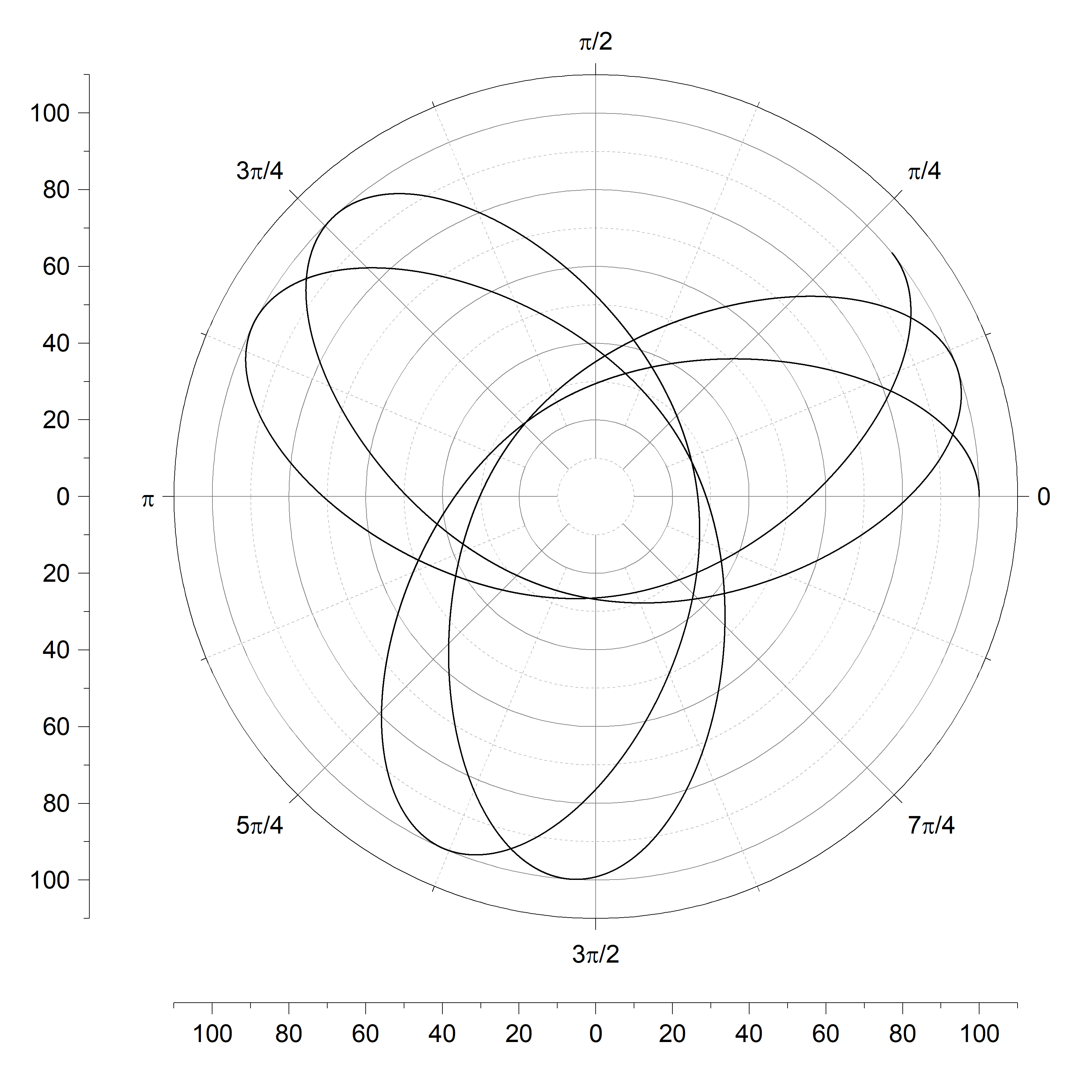}
         \caption{$(35.8,0.642,3,2,0.07999)$}
     \end{subfigure}%
     \begin{subfigure}{0.25\textwidth}
         \centering
         \includegraphics[width=\textwidth]{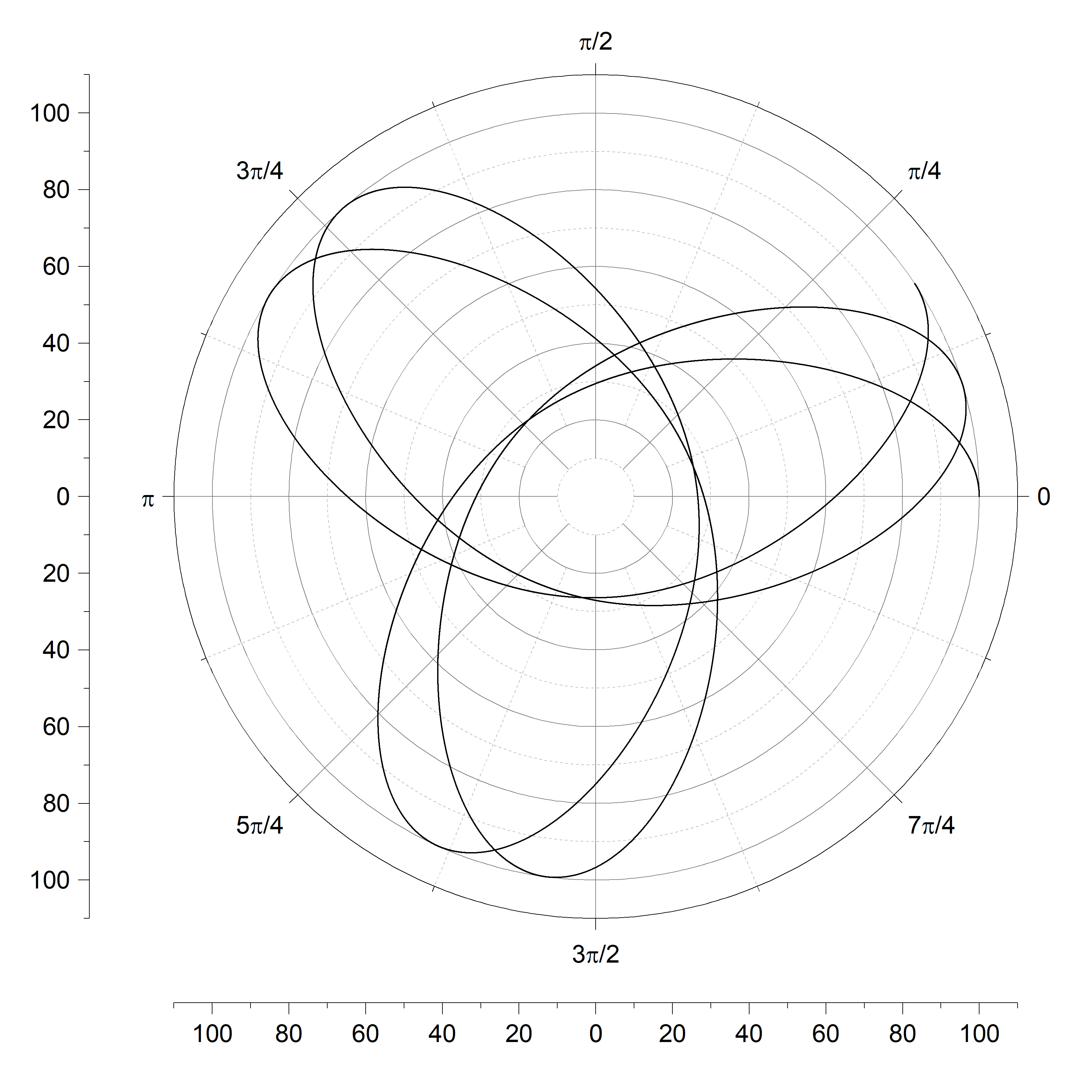}
         \caption{$(35.8,0.642,3,2,0.06868)$}
     \end{subfigure}
     \caption{Massive particle moving in Levi-Civita spacetime (left) and around a Newtonian linear mass distribution (right) with parameter $\sigma=0.001$}
\end{figure}

\begin{figure}[htb]
     \centering
     \begin{subfigure}{0.25\textwidth}
         \centering
         \includegraphics[width=\textwidth]{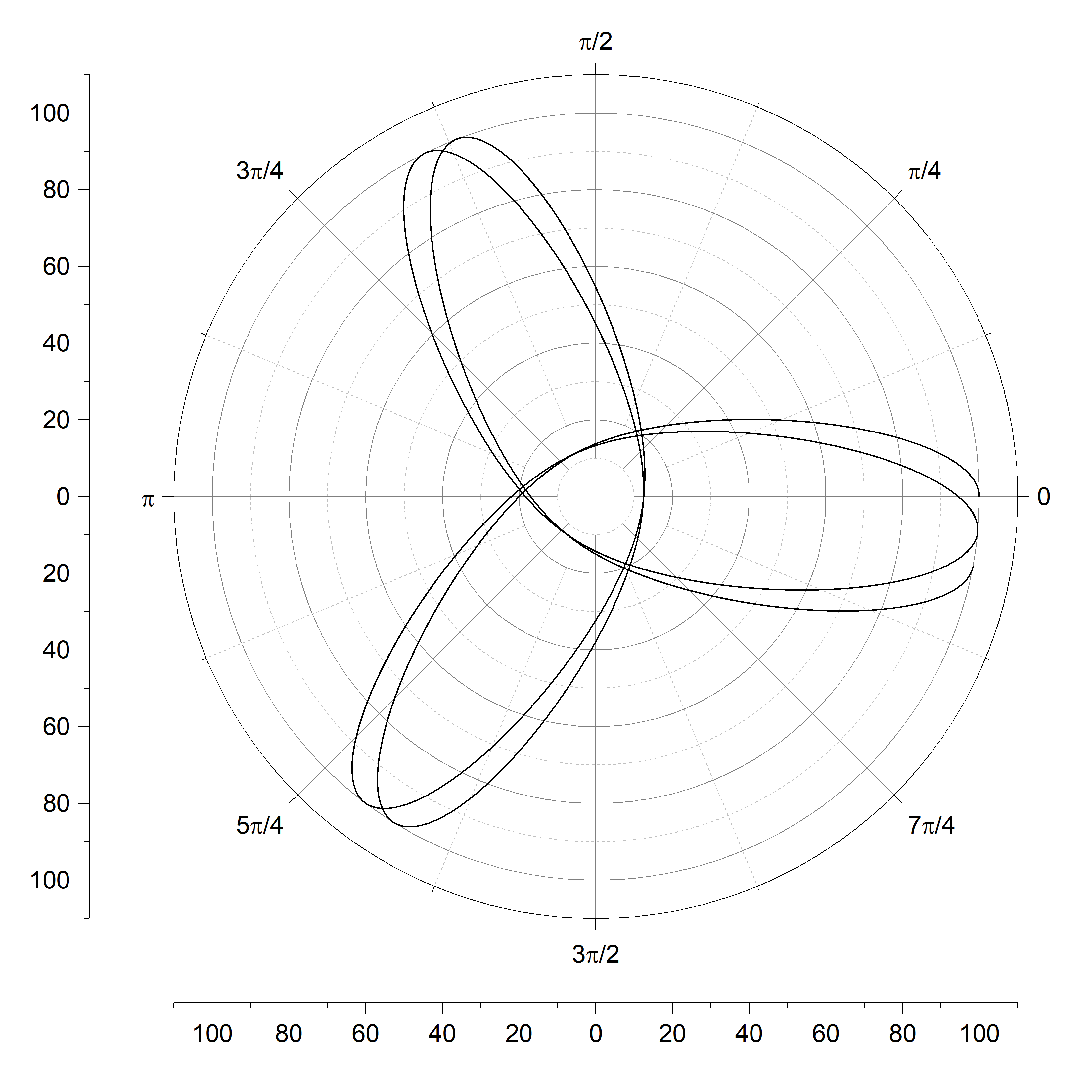}
         \caption{$(14.9,0.851,3,2,0.022021)$}
     \end{subfigure}%
     \begin{subfigure}{0.25\textwidth}
         \centering
         \includegraphics[width=\textwidth]{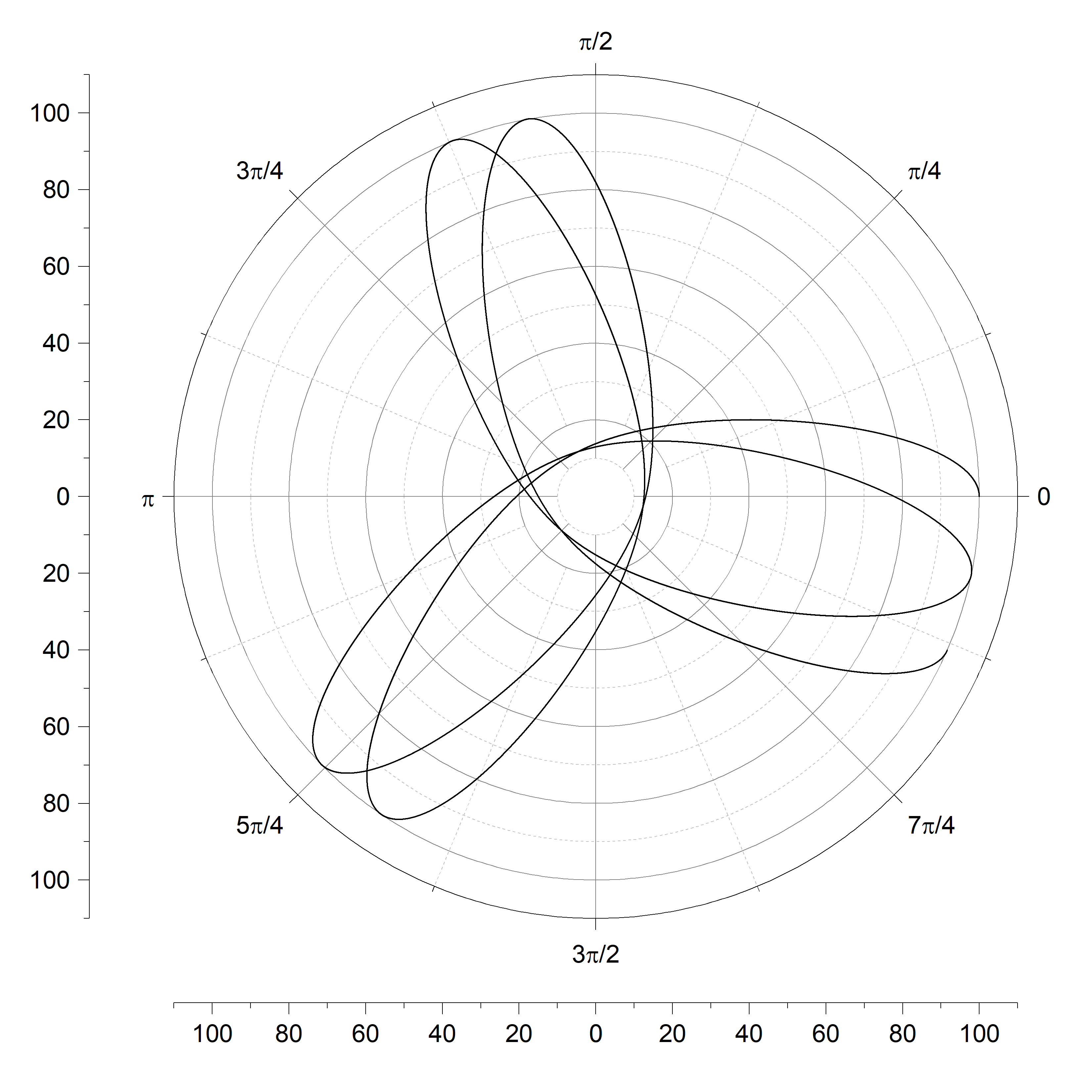}
         \caption{$(15,0.85,3,2,0.049991)$}
     \end{subfigure}
     \caption{Massive particle moving in Levi-Civita spacetime (left) and around a Newtonian linear mass distribution (right) with parameter $\sigma=0.003$}
\end{figure}

\begin{figure}[htb]
     \centering
     \begin{subfigure}{0.25\textwidth}
         \centering
         \includegraphics[width=\textwidth]{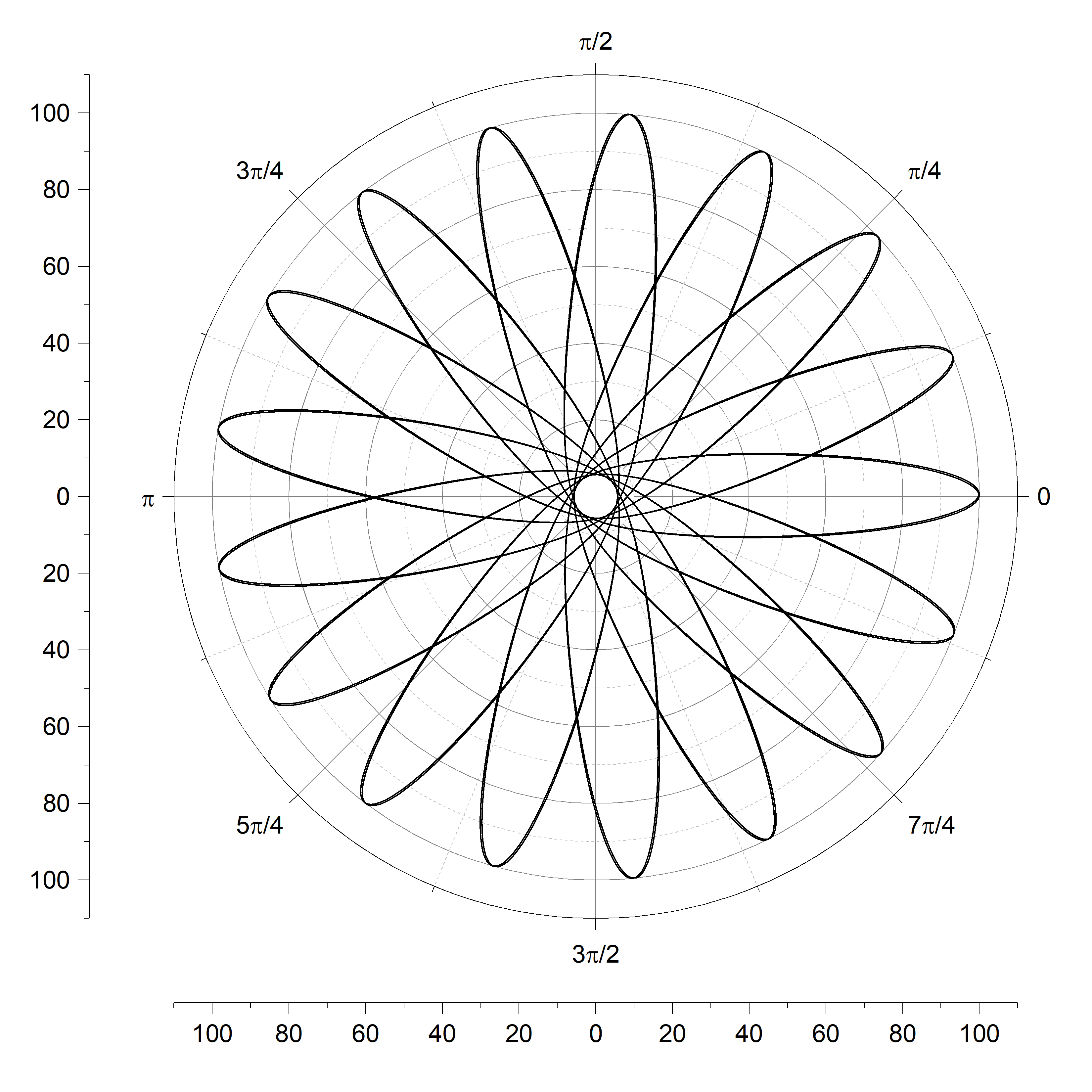}
         \caption{$(5.2,0.948,17,11,0.001294)$}
     \end{subfigure}%
     \begin{subfigure}{0.25\textwidth}
         \centering
         \includegraphics[width=\textwidth]{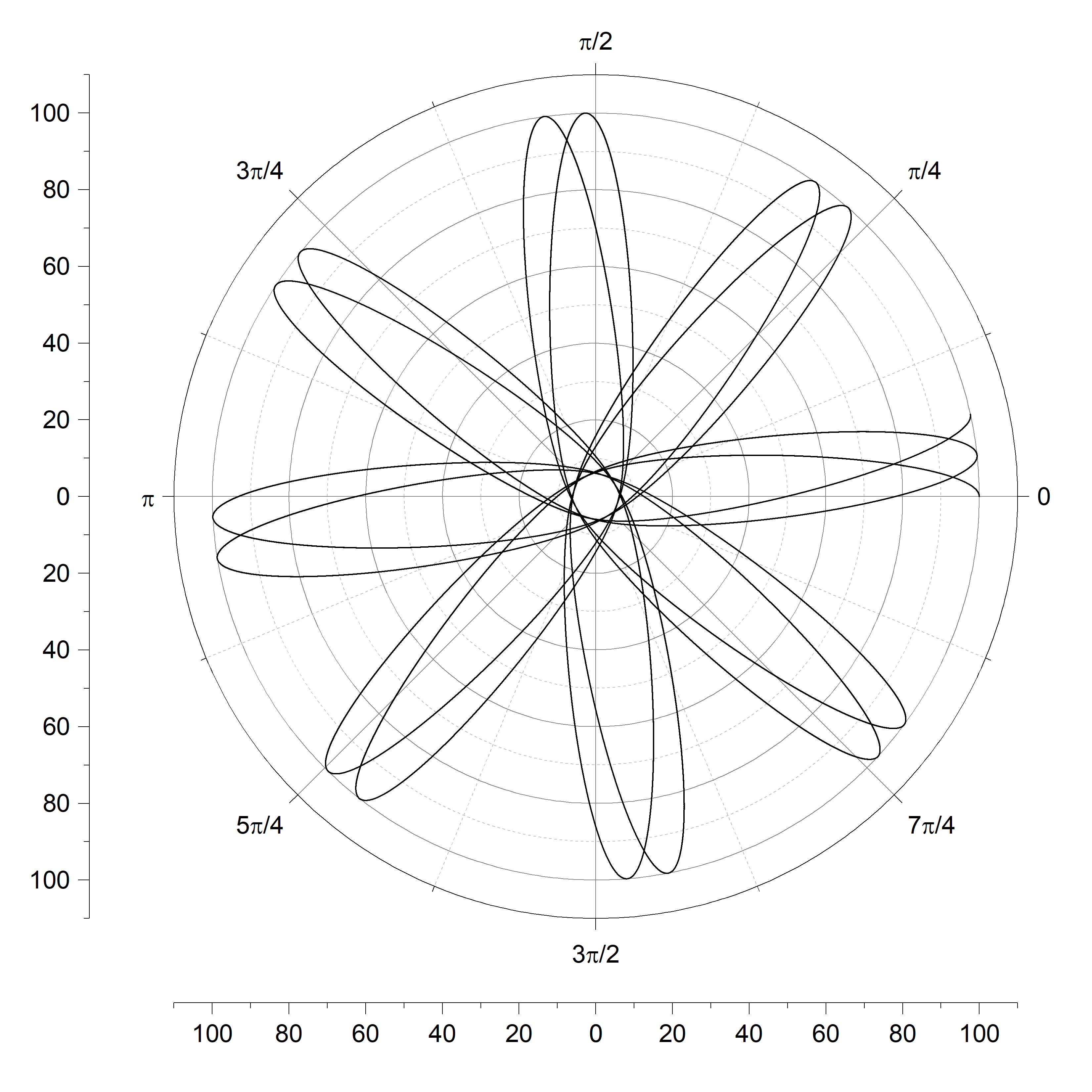}
         \caption{$(5.2,0.948,8,5,0.027502)$}
     \end{subfigure}
     \caption{Massive particle moving in Levi-Civita spacetime (left) and around a Newtonian linear mass distribution (right) with parameter $\sigma=0.01$}
\end{figure}

For any set of initial data, it is expected for the orbits of massive particles in Levi-Civita spacetime to match the orbits around a Newtonian linear mass distribution in weak gravitational field limit. FIG. 12 shows a particle moving in Levi-Civita spacetime (left) and a particle moving around a Newtonian linear mass distribution (right) where both particles have the same set of initial data $(\rho_0,\phi_0,z_0,\Gamma, A,B)=(100,0,0,0,2,0)$. For both cases the energy density parameter is set to $\sigma=0.0001$ making strings the sources of weak gravitational fields. As we can see that the both particles in FIG. 12 follow effectively the same trajectory, confirming the equation-array (28). However, the orbits have a small deviation in their precession parameter $\chi$. The difference goes down even further for the weaker gravitational sources.

Keeping the set of initial data unchanged, we increase the parameter $\sigma$ and set it to $0.001$ in FIG. 13, which makes both orbits more eccentric. However, the difference in the $\chi$ values of the orbits become larger. Apparently, the $\chi$ value for the orbit around the Newtonian linear mass density changes more rapidly than that of massive particle orbit in Levi-Civita spacetime as $\sigma$ changes. In FIG. 14, the $\sigma$ parameter is set to $0.003$, without changing the initial data. As a result, the orbits in FIG. 14 become more eccentric. However, the orbit in Levi-Civita spacetime becomes slightly more eccentric than the other one. And the $\chi$ value continues to change more rapidly in the orbit around the Newtonian linear mass density. All these qualitative observations are consistent with the equation-array (28).

Finally, in FIG. 15, we turn up the $\sigma$ parameter even further to $0.01$, keeping the initial data fixed. At this point, the parameter $\gamma$ of the orbits become different enough to break the visual similarity between the orbits.

\section{\label{section-7}Applications and Future Works}

Our work potentially unveils many doors for future works. Unsurprisingly, the results may have definite implication for corresponding observational works related to cosmic strings and networks of them. There is a room future theoretical works as well, for instance, further endeavours can investigate the geodetic motions confined to the $\phi=\phi_0$ plane, i.e. particles with no initial angular momentum, elaborately. Analysis similar to the one can be carried out for photon trajectories at sufficient depth. All of these works can be extended to the corresponding setting of Newtonian gravity.

The classification scheme developed in this article can now be applied to orbits in spacetimes with different characteristic symmetries, such as those of the Schwarzschild and Kerr spacetimes. Another possible extension to our work is to investigate the gravitational radiation for highly eccentric orbits of massive and massless particles around cosmic strings. Nonetheless, more pathways for future works will unfold as more works are carried out on relevant topics.

\section{\label{section-8}Conclusion}

In this article, we have utilised the separability of the Hamilton-Jacobi equations to derive the complete, exact integral solution to the geodesic equations for massive and massless particles in Levi-Civita spacetime. Additionally, we have expanded the resulting integral solutions into corresponding infinite series. Integrals have been performed to obtain solutions in terms of well-known functions for a few special cases. Moreover, the complete, exact integral solution to the Newtonian differential equation of motion for massive particles moving around an infinitely-long straight linear mass distribution was obtained as well.

A surprising discovery is that, both the Newtonian and relativistic orbits are always bounded, and they coincide in the weak field limit. Most importantly, we have presented a simple classification scheme for the orbits and compared to the other existing scheme. Finally, we have applied the scheme to categorize the closed and the precessing relativistic orbits of a massive particle in Levi-Civita spacetime. The relativistic massive particle trajectories were compared against their Newtonian counterparts.

\acknowledgments

We would like to thank John David Brown, Prosanto Chokroborty, and Sadman Sakib for their valuable suggestions on various occasions that have improved the article significantly. This material is based upon works supported by Community of Physics.\footnote{\href{https://www.cpbd.ac/}{www.cpbd.ac}}

\appendix

\section{\label{appendix-a}Differential Geodesic Equations}

In a four-dimensional curved spacetime, a causal geodesic is a path that a particle, whether it is massive or not, follows. During the course, a massive particle tries to maximize the proper time which, as it turns out, can be treated as the action corresponding to the motion. Therefore, extremizing the action \cite{misner2017-13}

\begin{equation}
    \mc{S}[q] =\int_{\tau_i}^{\tau_f} \sqrt{-g_{\mu\nu}\dot{q}^\mu\dot{q}^\nu} \ d\tau
\end{equation}

leads to the equations of geodesic

\begin{equation}
    \ddot{q}^\kappa + \Gamma^\kappa_{\mu\nu}\dot{q}^\mu\dot{q}^\nu = 0
\end{equation}

where, the $q^\mu$ denote the spacetime coordinates, the over-dots denote derivatives with respect to the propertime parameter, $\tau$, and $\Gamma^\kappa_{\mu\nu}$ denote the connection coefficients of the spacetime. For the metric tensor in equation (4) we can find the differential geodesic equations as follows

\begin{subequations}
\begin{eqnarray}
    \ddot{t}+\frac{4\sigma}{(1-2\sigma)\rho}\dot{\rho}\dot{t} = 0
    \\
    \ddot{\rho} + \frac{2\sigma }{(1-2\sigma)P}\biggr(\frac{\rho}{P}\biggr)^{4\sigma-1} \dot{t}^2 + \frac{4\sigma^2}{(1-2\sigma)\rho}\dot{\rho}^2
    \nonumber\\
    - \biggr(\frac{\rho}{P}\biggr)^{-\frac{8\sigma^2}{1-2\sigma}}\rho\dot{\phi}^2 + \frac{2\sigma}{P} \biggr(\frac{\rho}{P}\biggr)^{-\frac{1+2\sigma}{1-2\sigma}} \dot{z}^2 = 0\\
    \ddot{\phi}+\frac{2}{\rho}\dot{\rho}\dot{\phi}=0
    \\
    \ddot{z}-\frac{4\sigma}{\rho}\dot{\rho}\dot{z}=0
\end{eqnarray} 
\end{subequations}

The above equations are not simple enough to solve analytically. That is why we seek for another method, namely, the Hamilton-Jacobi procedure, to solve the geodesic equations.

\section{\label{appendix-b}Hamilton-Jacobi Procedure}

Considering the action in equation (A1), we can define the Lagrangian of the system as

\begin{equation}
    L(q,\dot{q};\tau) = \sqrt{-g_{\mu\nu}\dot{q}^\mu\dot{q}^\nu}
\end{equation}

and the canonical momenta corresponding to the spacetime coordinates as

\begin{equation}
    p_\kappa = \pd{L}{\dot{q}^\kappa} = -\frac{g_{\kappa\lambda}\dot{q}^\lambda}{\sqrt{-g_{\mu\nu}\dot{q}^\mu\dot{q}^\nu}}
\end{equation}

Using the above definitions of the Lagrangian and the momenta, we can construct the canonical Hamiltonian of the system as

\begin{equation}
    H(q,p;\tau)=p_\kappa \dot{q}^\kappa - L(q,\dot{q};\tau) = 0
\end{equation}

This shows that the Lagrangian of the system is degenerate, meaning the relations between the velocity and the momentum variables are not invertible. It can quite easily be shown that the determinant of the corresponding Hessian matrix is zero, confirming the degeneracy of the Lagrangian.

\begin{equation}
    \det\biggr(\frac{\partial^2L}{\partial\dot{q}^\kappa\partial\dot{q}^\lambda}\biggr)=0
\end{equation}

In such a case, Dirac reasoned that there exists a constraining relation between the momentum variables themselves \cite{henneaux1992-01}. For a massive particle, it takes the form

\begin{equation}
    1 + g^{\mu\nu}p_\mu p_\nu = 0
\end{equation}

However, for a massless particle there exists no such action principle. Rather, the constraining relation among the momenta comes from the first integral of the geodesic, which is

\begin{equation}
    g^{\mu\nu}p_\mu p_\nu = 0
\end{equation}

Together with the definition of the generating function

\begin{equation}
    p_\mu = \pd{S}{q^\mu}
\end{equation}

the equations in (B5) and (B6) can be treated as the Hamilton-Jacobi equations for the massive and massless particles, respectively \cite{gerlach1969}. The solution to the differential geodesic equations can be found using

\begin{equation}
    Q^\mu = \pd{S}{P_\mu}
\end{equation}

where, $Q^\mu$ and $P_\mu$ are the constants of motion. For massive particles, equation (B5) (and for massless particles, (B6) instead) together with equations (B7) and (B8) constitutes a complete set of equations from which a unique solution to the geodesics can be obtained.

\section{\label{appendix-c}Classification Scheme Applied to Schwarzschild Spacetime}

The classification scheme laid out in Section IV applies to other important spacetime geodesics as well. Particularly, it aptly describes the massive particle geodesics in Schwarzschild spacetime. Even though it is well-known in the literature \cite{synge1960-08}, we show it systematically here for completeness. We begin with the integral solution to the orbit equation of massive particle geodesics in Schwarzschild spacetime written in Schwarzschild coordinates $(t, r, \theta, \phi)$ (at $\theta = \pi/2$) given by the following equation:

\begin{equation}
    \phi = \phi_0 + \int_{r_0}^r \frac{l}{r^2} \biggr\{E^2-1 - \frac{2M}{r} - \frac{l^2}{r^2} + \frac{2Ml^2}{r^3} \biggr\}^{-\frac{1}{2}} dr
\end{equation}

where $E$ is the energy of a massive particle with unit mass and $l$ is the angular momentum of the particle. \cite{hackmann2008} We expand the integrand with respect to the periastron location $\underline{r}$ and keep the terms up to the second order in $|\underline{r}/r|$. We then integrate the approximate trajectory integral, giving the orbit equation exactly of the format depicted in equation (19), where the radial variable $\rho$ is replaced by the Schwarzschild radial coordinate $r$ with $\eta=0$. The orbit parameters $C$, $\epsilon$, $\gamma$, and $\delta$ are given by the following relations:

\begin{subequations}
\begin{eqnarray}
    C & = & \frac{2k_2}{2k_2-k_1}
    \\
    \epsilon & = & \sqrt{\frac{4k_0k_2+k_1^2}{(2k_2-k_1)^2}} \\
    \gamma & = & \frac{\underline{r}\sqrt{k_2}}{l}
    \\
    \delta & = & \phi_0-\frac{1} {\gamma}\arccos\biggr(\frac{r_0-C\underline{r}}{\epsilon r_0}\biggr)
\end{eqnarray}
\end{subequations}

In the above relations, $k_0$, $k_1$, and $k_2$ are defined as follows:

\begin{subequations}
\begin{eqnarray}
    k_0 & = & E^2 - 1 + \frac{l^2}{4M^2} + \frac{2M}{\underline{r}} - \frac{l^2}{\underline{r}^2} \\ \nonumber
    &&+ \frac{3l^2}{4M^2} \ln\biggr({\frac{2M}{\underline{r}}}\biggr) + \frac{9l^2}{8M^2} \ln^2\biggr({\frac{2M}{\underline{r}}}\biggr)
    \\
    k_1 & = & - \frac{3l^2}{4M^2} - \frac{2M}{\underline{r}} + \frac{2l^2}{\underline{r}^2} - \frac{9l^2}{4M^2} \ln\biggr(\frac{2M}{\underline{r}}\biggr)
    \\
    k_2 & = & - \frac{3l^2}{4M^2} + \frac{l^2}{\underline{r}^2} - \frac{9 l^2}{8M^2} \ln\biggr(\frac{2M}{\underline{r}}\biggr)
\end{eqnarray}
\end{subequations}

Having shown that the approximate geodesic in Schwarzschild follows the path of a precessing ellipse as in equation (19), one can use the classification scheme laid out in Section IV.

\bibliographystyle{apsrev4-2}
\bibliography{lct}

\end{document}